\documentclass[11pt, twoside]{article}

\usepackage{amsthm,amsmath,amsfonts,amssymb}

\usepackage{myformatting}

\usepackage[T1]{fontenc}
\usepackage[utf8]{inputenc}

\usepackage[title]{appendix}
\usepackage{array}
\usepackage{bm}
\usepackage{bbm}
\usepackage{enumitem}
\usepackage{fancyhdr}
\usepackage{float}
\usepackage[margin=3.25cm, nomarginpar]{geometry}
\usepackage{graphicx}
\usepackage[backref=page]{hyperref}

\usepackage{caption}
\usepackage{color}
\usepackage{makecell}
\usepackage{microtype}
\usepackage{multirow}
\usepackage{subcaption}

\usepackage{titlesec}

\usepackage[round, authoryear]{natbib}
\usepackage{etoolbox}
\makeatletter
\patchcmd{\BR@backref}{\newblock}{\newblock(page~}{}{}
\patchcmd{\BR@backref}{\par}{)\par}{}{}
\makeatother

\hypersetup{
  colorlinks=true,
  citecolor=blue,
  linkcolor=blue,
  urlcolor=blue
}

\fancypagestyle{plain}{ %
  \fancyhf{} 
  \fancyfoot{}
}

\title{Bayesian inference for
  group-level cortical surface image-on-scalar-regression with
  Gaussian process priors} 

\author{Andrew S. Whiteman,$^{*}$
  Timothy D. Johnson, 
  and Jian Kang \\ 
  $^{*}$\url{awhitem@umich.edu} \\ \\
  Department of Biostatistics, \\
  University of Michigan, Ann Arbor, \\
  Michigan, U.S.A.
}

\date{}

\begin{document}

\thispagestyle{plain}
\maketitle
\label{firstpage}

\begin{abstract}
In regression-based analyses of group-level neuroimage data researchers typically fit a series of marginal general linear models to image outcomes at each spatially referenced pixel. Spatial regularization of effects of interest is usually induced indirectly by applying spatial smoothing to the data during preprocessing. While this procedure often works well, resulting inference can be poorly calibrated. Spatial modeling of effects of interest leads to more powerful analyses, however the number of locations in a typical neuroimage can preclude standard computing methods in this setting. Here we contribute a Bayesian spatial regression model for group-level neuroimaging analyses. We induce regularization of spatially varying regression coefficient functions through Gaussian process priors. When combined with a simple nonstationary model for the error process, our prior hierarchy can lead to more data-adaptive smoothing than standard methods. We achieve computational tractability through a Vecchia-type approximation of our prior that retains full spatial rank and can be constructed for a wide class of spatial correlation functions. We outline several ways to work with our model in practice and compare performance against standard vertex-wise analyses and several alternatives. Finally we illustrate our methods in an analysis of cortical surface fMRI task contrast data from a large cohort of children enrolled in the Adolescent Brain Cognitive Development study.
\end{abstract}

\setcounter{page}{1}

\section{Introduction}
\label{sec:introduction}

Modern large-scale neuroimaging studies collect massive amounts of
data, often across thousands of patients, sometimes across
several years
\cite[e.g.,][]{akil2011challenges, smith2018statistical,
  vanhorn2009multi, volkow2018conception}.
Typically these studies collect multiple
structural and/or functional scans, with the aim to probe
relationships between the images and patient-level characteristics.
We focus here on an image-on-scalar regression treatment for this
general framework, where patients' images are taken to be the
response, and covariates are individual-level scalars.

Since neuroimages are a spatially referenced data type, we cast the
image-on-scalar problem as a functional regression of the form,
\begin{equation}
  \label{eqn:main}
 y_i(\spc) = \bx_i\trans \bbeta(\spc) + \omega_i(\spc) +
  \epsilon_i(\spc).
\end{equation}
In \eqref{eqn:main} we take $y_i(\spc)$ to be the imaging outcome for
patient $i$ ($i = 1, \ldots, N$) at location $\spc \in \SpatialSet$, and
treat coefficients of interest
$\bbeta(\cdot) : \SpatialSet \to \Reals^P$ as spatially varying.
Further, we decompose the error into a sum of terms
$\omega_i(\cdot)$ and $\epsilon_i(\cdot)$, where the
$\omega_i(\cdot)$ reflect individual-level deviations from the mean
with an assumed spatial structure, and the $\epsilon_i(\cdot)$
reflect a white noise process. 
Many classical analysis methods in imaging can be cast within this
framework. For example, in the typical group-level functional magnetic
resonance imaging (fMRI) analysis, the
$y_i(\cdot)$ might represent contrasts of parameter estimates from
within-participant first level time series analyses, and the
$\bx_i \in \Reals^P$ might include an intercept term along with any
relevant covariate information.
Often, in practice, marginal univariate models are fit to the data
from each location $\spc$
\cite[e.g.,][]{mumford2009simple}. 
This procedure tremendously simplifies estimation by avoiding modeling
spatial correlations in $\bbeta(\cdot)$ and $\omega_i(\cdot)$, but can
lead to poorly calibrated inference
\cite[for example, see attempts to improve the power of tests derived
from marginal models by spatially pooling variance estimates
in][]{nichols2002nonparametric, su2009modified, wang2021moderated}.

\begin{figure}[!tbh]
  \centering
  \includegraphics[width=0.8\textwidth]{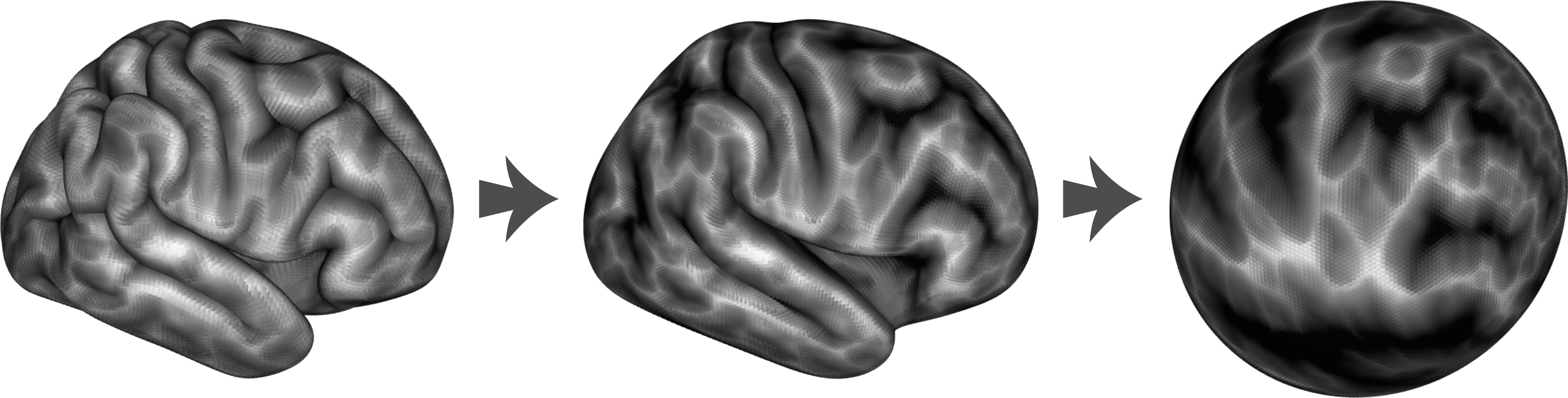}
  \caption{Example mapping of cortical surface
    coordinates onto a sphere. Left to right, the figure shows
    progressive inflation and warping of the right hemisphere of
    cortex. Gross anatomical features are highlighted
    to help visualize the mapping. This procedure was introduced to
    facilitate state-of-the-art cross-subject alignment of cortical
    features, but can also be leveraged into a mathematically
    convenient measure of geodesic distance along the cortical
    surface.}
  \label{fig:spherical-mapping}
\end{figure}

For model \eqref{eqn:main} to make sense practically, the
images must have reasonably comparable support in the spatial domain
$\SpatialSet$. Though it is still an area of active research, a 
tremendous amount of study has focused on methods to preprocess
raw neuroimage data to help coregister the images across patients and 
data collection sites
\cite[e.g.,][]{fischl1999cortical, fischl1999high, jenkinson2012fsl,
  reuter2010highly}.
In particular, certain neuroimage preprocessing tools compute
state-of-the art cross-subject alignment of cortical features by first
mapping each hemisphere of the cortex onto the surface of a sphere
with minimal distortion
\cite[][]{fischl1999cortical, fischl1999high}.
Fig. \ref{fig:spherical-mapping} gives an example of such a mapping.
This procedure standardizes the spatial support for each hemisphere of
cortex, and has been shown to lead to reduced spatial
signal contamination and result in more sensitive analyses
\cite[e.g.,][]{brodoehl2020surface}.
The inherent construction of a gray matter surface-based coordinate
system within this framework more accurately reflects the topology of
primate cortex versus simple Euclidean distance in 3D space
\cite[][]{fischl1999cortical}.
Recently, within the statistical community,
\cite{mejia2020bayesian} highlighted this preprocessing pipeline by
developing a cortical-surface-image-on-scalar regression model for
task-based fMRI data. \cite{mejia2020bayesian}
propose a joint multi-subject spatio-temporal regression
framework, model their spatial regression coefficients with Gaussian
random fields, and derive an integrated nested Laplace approximation
routine for approximate Bayesian inference. Per their data
application, Mejia \emph{et al.} develop their
model primarily for analysis of multi-subject fMRI time series
data where the number of subjects is not large.

Such joint multi-subject spatio-temporal methods are not easily
extensible to large-scale imaging studies.
Here, we consider a slightly different setting where any time
series data has already been distilled into contrast or other
summary statistic images of interest, but the number of individuals
$N$ may be quite large.
The number of spatial locations in a conventional neuroimage typically
precludes Bayesian computation in most computing environments
except by methods that either approximate (a) the spatial process by
low-rank projection or downsampling, or that approximate (b) the
posterior distribution with variational or Laplace family
approximations
\cite[see e.g.,][]{penny2005bayesian, siden2017fast,
  mejia2020bayesian}.
In general, low-rank projection methods can tend to
miss or over-smooth local features in data
\cite[e.g.,][]{stein2007spatial}, and both low-rank projection and
variational approximation can commonly underestimate posterior
variance \cite[e.g.,][]{wang2005inadequacy, rasmussen2005healing}.
Integrated nested Laplace approximation, moreover, is thought to
give accurate and scalable approximations within a wide class of
posterior distributions \cite[e.g.,][]{rue2017bayesian}, but
its accuracy can sometimes suffer when model structure is complex
\cite[see e.g.,][]{taylor2014inla}.
Here, we expand on this body of work and show how a Bayesian model
with a prior hierarchy related to that in \cite{mejia2020bayesian} can
permit estimation of coefficient functions that are realizations of a
full-rank spatial process. To be able to extend our method to
large-scale imaging studies we contribute a spatial regression model
intended primarily for group-level analyses of data indexed by
locations on the cortical surface. In the context of group-level fMRI
studies, for example, our method could simply be ``plugged in'' at the
classical second-stage analysis, with individual-level task contrast
images taken to be the response.
Our method can also be flexibly applied to analysis of
cortical thickness outcomes, or other structural indicators.
We model the probability law governing prior uncertainty in the
functions $\bbeta(\cdot)$ and $\omega_i(\cdot)$ with Gaussian
processes. Posterior computation is enabled by Vecchia approximation
of the spatial processes
\cite[][]{vecchia1988estimation, datta2016hierarchical,
  katzfuss2021general} and empirical Bayesian estimation of
associated hyperparameters.

Our model can be reasonably fit to the data from whole hemispheres of
cortex using fast optimization or scalable Markov chain Monte Carlo
(MCMC) routines without the need to downsample the original
data. Additionally, we elaborate on an approximate working model and
related Bayesian sampling scheme with computational complexity that
scales almost independently of $N$, further allowing our method to be
viable for application to large-scale neuroimaging studies.
Model computation with MCMC permits posterior inference on the
spatial extent of activation regions with simultaneous credible bands,
which also facilitate spatial inference that is inherently adjusted
for multiplicity.
Applied researchers may find this feature especially compelling:
currently, there is no broad consensus on the best way to perform
multiplicity adjustment in neuroimaging analysis.
Most popular methods are based around controlling the expected
family-wise or false discovery error rate at some level in a null
hypothesis testing paradigm. These include cluster extent-based
adjustment, false discovery rate adjustment, and permutation testing
\cite[][]{hagler2006smoothing, 
  genovese2002thresholding, nichols2002nonparametric}.
Cluster extent-based adjustment---possibly the most common in
practice---is usually implemented as a two stage procedure
requiring separate ``height'' and ``extent'' thresholds to define
regions with significant activation. In general, results can be
sensitive to the chosen combination of thresholds, and although
practitioners have endeavored to provide guides for threshold
selection \cite[][]{woo2014cluster}, the procedure may involve
high ``researcher degrees of freedom.''
In contrast, simultaneous credible bands can be expected to provide
an accurate and self-contained summary of the posterior behavior of
$\bbeta(\cdot)$ over all locations jointly.

The body of this paper contains an elaboration of our spatial
regression model hierarchy at the beginning of Section
\ref{sec:methods}. Sections \ref{sec:methods:conditional-model},
\ref{sec:methods:marginal-model}, and
\ref{sec:methods:working-model} develop schemas for using
this model and a related working model in practice.
We assess the relative accuracy and sensitivity of these
strategies against common alternative methods in a simulation study
in Section \ref{sec:simulation}.
In Section \ref{sec:analysis}, we then use our working model to
analyze n-back task contrast data ($z$-statistic images) from the
Adolescent Brain Cognitive Development (ABCD)
imaging collective data.
Section \ref{sec:discussion} concludes with a discussion of method
limitations and possible extensions.

\section{Methods}
\label{sec:methods}

Throughout this work, we assume the single hemisphere, cortical
surface-based, spherical coordinate system of
\cite{fischl1999cortical}. By isolating data from the cortical
sheet we gain anatomical specificity and a better connection to
the underlying neurobiology. 
As has been discussed by, e.g. \cite{fischl1999cortical,
  fischl1999high} and \cite{mejia2020bayesian}, 
geodesic distances along the cortical surface are
more meaningful than, say, simple Euclidean distances in the compact
3D volume. This is due to the fact that primate cortex is thought to
be topographically organized by function 
\cite[e.g.,][]{silver2009topographic}, and exhibits a folded structure 
in higher mammals to accommodate a larger cell body layer
\cite[e.g.,][]{jones2012cerebral}.

We simplify notation by considering the left and right hemispheres of
cortex as separate outcomes in separate analyses. 
Let $\SpatialSet$ denote the set of coordinates on a sphere with a
known radius $R$, and let $\mathcal{S} \subset \SpatialSet$ denote the
set of vertices for a single hemisphere of cortex at which we have
observed MRI data. For reference, the data in our application have all
been mapped to a normalized template brain space with approximately
30,000 vertices in $\mathcal{S}$. 
For any two $\spc, \spc' \in \SpatialSet$, let $d(\spc, \spc')$
measure the great-circle distance between $\spc$ and $\spc'$.
Great-circle distance is sufficient for our purpose; more generally,
however, $\SpatialSet$ might represent some topological surface, etc.,
and $d(\cdot, \cdot)$ any appropriate metric.
Beginning from \eqref{eqn:main}, we model the data likelihood as
multivariate Gaussian with a particular error structure.
We assume:
\begin{equation}
  \label{eqn:likelihood}
  y_i(\spc) \sim \Gaussian( \bx_i\trans \bbeta(\spc) + \omega_i(\spc),
  \sigma^2(\spc)), \qquad i = 1, \ldots, N, \text{ and } \spc \in
  \mathcal{S}, 
\end{equation}
where $\Gaussian(\mu, \Sigma)$ denotes the Normal distribution with
mean $\mu$ and variance $\Sigma$; $\bx_i \in \Reals^P$ are covariates;
$\bbeta(\cdot) : \SpatialSet \to \Reals^P$ are the primary effects of
interest; and $\omega_i(\cdot) : \SpatialSet \to \Reals$ reflects
individual-level deviations from $\bx_i\trans
\bbeta(\cdot)$. Conditional on $\bx_i$, $\bbeta(\cdot)$, and
$\omega_i(\cdot)$, we model the errors as a non-stationary white noise
process with spatial variances denoted by 
$\sigma^2(\cdot) : \SpatialSet \to \Reals_{>0}$.  Given the nature of
typical data in group-level functional or structural MR image
analyses, this data-level model may be sufficient for a variety of
studies.

Spatial dependence in our model arises through our prior
hierarchy on the effects $\bbeta(\cdot)$ and $\omega_i(\cdot)$.
Let $C_\btheta\{d(\spc, \spc')\}$ denote a positive
definite stationary spatial correlation function defined on
$\SpatialSet$ with parameter $\btheta$. For simplicity, we
drop the subscript $\btheta$ throughout and use $C(\cdot)$ to
represent a correlation function with implicit dependence on 
$\btheta$. We specify the prior distributions of each spatially
varying coefficient function $\beta_j(\cdot)$ as mean zero Gaussian
processes with marginal variances $\zeta_j^2 \tau^2$, i.e.,
\begin{equation}
  \label{eqn:prior-beta}
  \beta_j(\spc) \sim \GP(0, \zeta_j^2 \tau^2 C\{d(\spc, \spc')\}),
  \qquad j = 0, \ldots, P-1.
\end{equation}
This class of prior for functional regression coefficients has been
adopted by \cite{gelfand2003spatial} for general spatial regression
problems. We write the coefficient processes this way without loss of
generality: while zero mean processes are reasonable in our 
application (where outcomes are task contrast $z$-statistic images,
see Section \ref{sec:analysis}),
data from other imaging modalities could be centered
at the global mean so that zero mean priors make sense.

We treat the individual-level deviations $\omega_i(\cdot)$
as spatially varying random effects with mean zero and marginal
variance $\tau^2$,
\begin{equation}
  \label{eqn:prior-omega}
  \omega_i(\spc) \sim \GP(0, \tau^2 C\{d(\spc, \spc')\}).
\end{equation}
Next, we specify a relatively simple nonstationary process for the
error precisions, 
\begin{equation}
  \label{eqn:prior-sigma}
  \sigma^{-2}(\spc) \mid \xi \overset{\text{iid}}{\sim}
  \text{Gamma}(1/2, \xi), \qquad 
  \xi \sim \text{Gamma}(1/2, 1),
\end{equation}
using the shape-rate parameterization of the Gamma distribution. 
To complete our model hierarchy, we place weakly informative priors
on the remaining spatial precisions,
\begin{equation}
  \label{eqn:prior-tau}
  \tau^{-2} \sim \text{Gamma}(1, 1/2), \qquad
  \zeta_j^{-2} \overset{\text{iid}}{\sim} \text{Gamma}(1, 1/2).
\end{equation}

As noted above, the correlation function $C(\cdot)$ can in
general be any positive definite kernel function defined so that
$C(0) = 1$ and $C(\alpha) \leq 1$ for all
$\alpha > 0$. Given the substantial history of Gaussian smoothing in
applied MRI analysis, we will work chiefly with the two parameter
exponential radial basis function, 
\begin{equation}
  \label{eqn:rbf}
  C(\alpha) = \exp( -\psi \lvert \alpha \rvert^\nu ),
  \qquad \btheta = (\psi, \nu)\trans, \quad \psi > 0, \nu \in (0, 2],
\end{equation}
though a number of alternatives are possible.
In \ref{eqn:rbf}, $C(\cdot)$ is stationary, isotropic, and synonymous
with the Gaussian kernel when $\nu = 2$. The parameter $\psi$ is
sometimes called the bandwidth or inverse length-scale parameter and
controls how rapidly the correlations decay; $\nu$ is the kernel exponent or
smoothness parameter.
We will discuss one data-driven way
the correlation function might be selected in practice in Section
\ref{sec:methods:correlation}; the same method can also be used to
estimate the correlation parameters $\btheta$ for a given functional
family.


\subsection{Conditional model}
\label{sec:methods:conditional-model}

We outline two ways of working with model \eqref{eqn:main} in our
setting, and also study the relative behavior of an approximate
working model with connections to the standard vertex-wise analysis
framework. The regression model we have outlined is difficult to
work with without simplification for two reasons. The first and
most obvious reason is the dimension of the parameter
space. Computational strategies for spatial modeling typically involve
decomposition of a dense spatial covariance matrix.
In our case, naive decomposition of the joint covariance of 
the $\beta_j(\cdot)$ and the $\omega_i(\cdot)$ would be an
$\Order(M^3(N + P)^3)$ operation, where $M$ is number of vertices in
$\mathcal{S}$, and $N$ and $P$ are the sample size 
and number of regression predictors, respectively. In Bayesian
sampling algorithms, this decomposition often needs to be recomputed
for each sample, which would be prohibitively expensive here.
The other difficulty with the model as written is that
decomposing the error structure into the sum of spatially varying
terms (the $\omega_i(\cdot)$ and $\epsilon_i(\cdot)$)
renders the whole model at best weakly identifiable.

As we lay out in greater detail in the Supplementary Materials, we
overcome the first difficulty by using a conditional independence or
Vecchia-type approximation to the model parameters' spatial
covariance, inducing sparsity in the parameters' spatial precision.
This type of approximation can
greatly reduce the computational burden while retaining a covariance
structure with full spatial rank, leading to high accuracy and
scalability \cite[e.g.,][]{datta2016hierarchical,
  finley2019efficient}. We overcome the second difficulty in several
ways. First we introduce what we term the
``conditional'' approach to working with our model.
To explain our conditional estimation strategy, we first observe that
if we knew the correct $\omega_i(\cdot)$ the remaining terms in the
model would be relatively easy to estimate. For this approach, our
strategy will be first to obtain an approximate maximum a posteriori
(MAP) estimate of the  $\omega_i(\cdot)$, and second to condition on
those estimates, sampling the other model parameters in an Empirically
Bayesian way. To obtain these estimates, we work with an approximate
model that considers $\sigma^2(\spc) \equiv \sigma^2$ constant over
all vertices in $\mathcal{S}$, and alternate conditional maximization
of $\bbeta(\cdot)$ and 
the $\omega_i(\cdot)$ until convergence.
Once we have obtained our estimate of the
$\omega_i(\cdot)$ in this way we simply subtract the $\omega_i(\cdot)$
from the $y_i(\cdot)$, and switch to an efficient Bayesian sampling
algorithm for the remaining parameters in the model.

\subsection{Marginal model}
\label{sec:methods:marginal-model}

Alternatively, since the individual deviations $\omega_i(\cdot)$ are
not typically of direct interest, we can first integrate them out,
leading to a marginal model with respect to the $\beta_j(\cdot)$,
$\sigma^2(\cdot)$, etc. Marginalizing out the $\omega_i(\cdot)$ is
relatively straightforward given the conjugacy in our model hierarchy.
Marginalization leads to the equivalency,
\begin{equation}
  \label{eqn:marginal-likelihood}
  y_i(\spc) = \bx_i\trans \bbeta(\spc) + \epsilon_i^*(\spc), \qquad
  \epsilon_i^*(\spc) \sim \GP(0, H\{d(\spc, \spc')\}),
\end{equation}
where $H\{d(\spc, \spc')\} = \tau^2 C\{d(\spc, \spc')\} +
\sigma^2(\spc) \I\{d(\spc, \spc') = 0\}$, and $\I(\mathcal{A})$ is the
event indicator function ($\I(\mathcal{A}) = 1$ if event $\mathcal{A}$
occurs, and 0 otherwise). A computational approach to working with
model \eqref{eqn:marginal-likelihood} can then follow by 
application of Vecchia approximation directly to
the covariance of the $\epsilon_i^*(\cdot)$.
In the Supplementary Materials, we outline a
means of computing with model \eqref{eqn:marginal-likelihood} based on
estimating $\btheta$, $\tau^2$, and $\sigma^2(\cdot)$ in an
Empirically Bayesian way. Briefly, we take a two stage approach to
computation, first obtaining approximate (up to
optimization tolerance) MAP estimates of
$\bbeta(\cdot)$, $\btheta$, $\tau^2$, and $\sigma^2(\cdot)$. Second,
we fix the covariance parameters at their MAP estimates and switch to
an efficient MCMC routine to sample $\bbeta(\cdot)$.

\subsection{Working model}
\label{sec:methods:working-model}

We also introduce a third, working model to obtain
approximate inference on the $\beta_j(\cdot)$. In general, including
the $\omega_i(\cdot)$ as a separate correlated error component will
not greatly influence standard estimators of the center of the
posterior of the $\beta_j(\cdot)$, such as the posterior mean.
If out of sample prediction of imaging outcomes is not a goal of the
analysis, then the primary reason to include a spatially correlated
error component is to improve modeling the posterior variance of the
$\beta_j(\cdot)$. In a large data setting, differences in efficiency
resulting from including a correlated error component can be minimal
to negligible. If we replace the likelihood in \eqref{eqn:likelihood}
with the approximation, 
\begin{equation}
  \label{eqn:working-likelihood}
  y_i(\spc) = \bx_i\trans \bbeta^{w}(\spc) + \epsilon_i^{w}(\spc), \qquad
  \epsilon_i^{w}(\spc) \sim \Gaussian(0, \sigma^2(\spc)),  
\end{equation}
and keep the prior structure on the $\beta_j^{w}(\cdot)$ and 
$\sigma^2(\cdot)$ the same as in \eqref{eqn:prior-beta} and
\eqref{eqn:prior-sigma} above, it is natural to ask how well the
resulting model performs.
We term this approximation our ``working'' model,
and note that it can be viewed as a generalization of the
standard vertex-wise general linear model (GLM) paradigm
in a spatial Bayesian context. The model implied by fitting
vertex-wise marginal GLMs is a limiting case of our working model as
$\tau^2 \to \infty$ for select 
choices of the correlation function,
$C(\alpha) = \I(\alpha = 0)$, and (improper) prior on the
$\sigma^{-2}(\cdot) \sim \text{Gamma}(1, 0)$.
Comparing our suite of methods, we will show in simulation that, for
moderate to large sample size, posterior inference for the regression
coefficients can be quite similar across the working and marginal
variants.

\subsection{Posterior computation}
\label{sec:methods:posterior-computation}

The Supplementary Materials provide a detailed description of our
approach to posterior computation.
Briefly, we follow work on ``Nearest Neighbor Gaussian Processes''
\cite[][]{datta2016hierarchical, finley2019efficient}
to develop sparse Vecchia-type approximations to the inverses of key
spatial covariance matrices. These approximations require some
definition of a spatial neighborhood within which the spatial
precisions are non-sparse. For both our simulations and applied
analysis, we have used discs with 8 mm radii to constitute these
neighborhoods. Sensitivity analyses over this choice are presented in
the Supplementary Materials.
We combine Vecchia approximation with a quasi-Newton
Hamiltonian Monte Carlo (HMC) algorithm to sample from the conditional
posterior of the regression coefficients supported on the fixed
spatial domain $\mathcal{S}$. Each gradient step of our HMC algorithm
is scaled by a sparse preconditioning matrix related to the prior
precision of the regression parameters on $\mathcal{S}$.

This algorithm can be used
for efficient posterior computation in very large data sets.
In fact, given sufficient statistics that can be computed
with a single pass through the outcome images, all of the parameter
updates in our working model can be performed without reference to the
original data. This leads to computational time complexity that, save
for an initial data streaming step, is independent of $N$.
In large data regimes the advantage of this is obvious.
A common applied fMRI use-case when working
with task contrast images is to use a simple set of predictors:
practitioners often fit an intercept-only model, or perhaps
additionally control for select covariates like age and sex. We
benchmarked our working model software for these use cases, analyzing
right hemisphere data from over 3,000 participants
($\approx$ 30,000 vertices; see Section \ref{sec:analysis}).
Streaming the images typically took around 100 ms or less per image
\cite[CIFTI/NIFTI-2 file format; ][]{cifti}.
After streaming, analysis with HMC took around 3.3 min per 1,000
iterations for the intercept-only model, or around 18.8 min per 1,000
iterations for the three predictor model (intercept, age, and
sex). Each analysis used less than 300 Mb of free RAM,
demonstrating the scalability of our approach.
This comparison was performed on a Dell PowerEdge R440 server
(2.1 GHz Intel\textsuperscript{\textregistered} 
Xeon\textsuperscript{\textregistered} Gold 6230 processors),
with processes limited to use eight cores each.

\subsection{Estimation of $\btheta$ and $C(\cdot)$}
\label{sec:methods:correlation}

Commonly used methods to estimate spatial correlation parameters
include variogram or covariogram estimation
\cite[e.g.,][]{armstrong1984improving, cressie1984median},
and maximum marginal likelihood methods
\cite[e.g.,][]{mardia1984maximum}. These methods can also be used to
select the correlation function itself by considering a set and
retaining the candidate with the best fit to the variogram or the
highest marginal likelihood. Here, we have used a maximum marginal
likelihood for a surrogate model to estimate the
correlation function and corresponding parameters in the spirit of
Empirical Bayes. The Supplementary Materials provide a full description
of this selection method for interested readers. In our analysis
of the ABCD study data (Section \ref{sec:analysis}), we estimated
$\btheta = (0.17, 1.38)\trans$, which corresponds to a sub-Gaussian
correlation function with $5.57$ mm full-width-at-half-maximum
(FWHM).

In addition to likelihood- or (co)variogram-based estimation,
experience can guide practitioners selecting $C(\cdot)$ and $\btheta$
to a large extent.
In applied imaging it is common to apply
a Gaussian smoothing kernel to data prior to analyses.
Commonly applied smoothing kernels are specified by their
FWHMs, which are often chosen to be
within a 4--12 mm range \cite[e.g.,][]{mikl2008effects}.
A 6 mm FWHM Gaussian 
kernel, e.g., is nominally equivalent to the radial basis
function \eqref{eqn:rbf} with bandwidth parameter
$\psi = 0.077$ and exponent parameter $\nu = 2$. In our setting with
moderate to large $N$, we find posterior inference is
not overly sensitive to the choice of $\btheta$.

\section{Simulation study}
\label{sec:simulation}

\begin{figure}[!htb]
  \centering
  \includegraphics[width=0.75\textwidth]{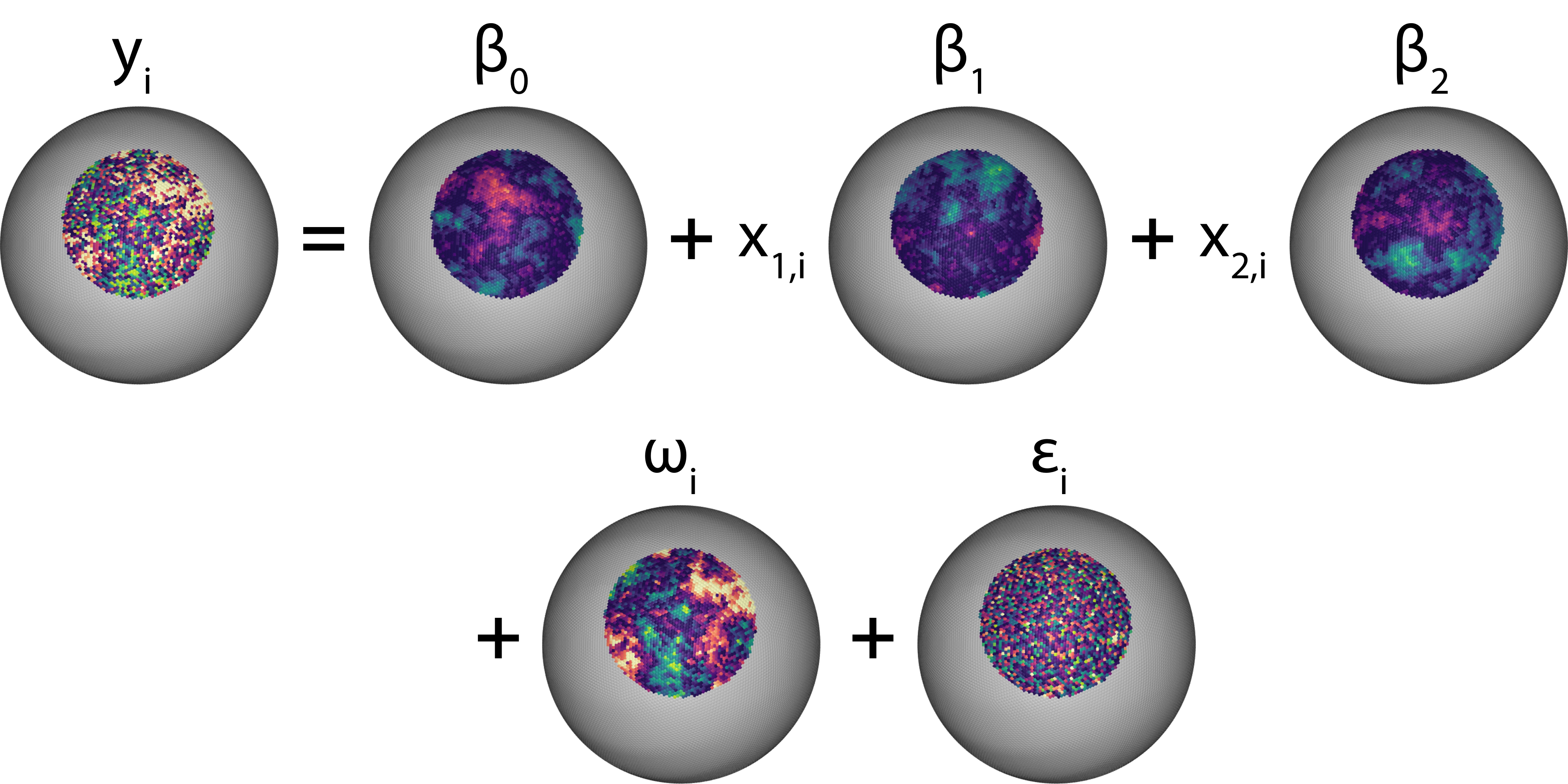}
  \caption{Simulation design. Data were simulated over a disc
    $\mathcal{D}$ of 2,000 vertices on a spherical surface. Effects of
    interest $\bbeta_j = [\beta_j(\spc)]_{\spc \in \mathcal{D}}$, 
    $j = 0, 1, 2$ were simulated as hard-thresholded Gaussian fields
    each with approximate 30\% sparsity. Error terms
    $\bomega_i = [\omega_i(\spc)]_{\spc \in \mathcal{D}}$ and
    $\bepsilon_i = [\epsilon_i(\spc)]_{\spc \in \mathcal{D}}$ were
    drawn from larger variance spatial processes 
    and dominate the spatial signals of interest.
    We have enhanced the contrast of the $\bbeta_j$ images for visual
    clarity.
  }
  \label{fig:sim-design}
\end{figure}

Our goal in simulation was to compare our suite of methods against
several common alternatives. 
In all cases, data were simulated on
a disc $\mathcal{D}$ of 2,000 vertices on the cortical surface.
See Fig. \ref{fig:sim-design} for a schematic illustration of this
design.
For each simulated data set, we generated spatially
correlated and sparse true
$\bbeta_j = [\beta_j(\spc)]_{\spc \in \mathcal{D}}$, $j = 0, 1, 2$,
by hard thresholding draws from independent Gaussian
processes with 6 mm FWHM exponential correlation functions.
To illustrate, let $\beta_j(\cdot) = \tilde{\beta}_j(\cdot)
\I\big\{ |\tilde{\beta}_j(\cdot)| > T \big\}$ for some threshold $T > 0$ and
non-sparse spatial process $\tilde{\beta}_j(\cdot)$.
Below, we will also use $\bbeta$ without the subscript to refer to the
vector of concatenated random fields
$\bbeta = (\bbeta_0\trans, \ldots, \bbeta_{P-1}\trans)\trans$.

Since our prior model in \eqref{eqn:prior-beta} is  
non-sparse, simulating the $\bbeta_j$ in this way reflects a
setting with slight model misspecification.
We made this choice to facilitate evaluating methods using
measures of inferential accuracy.
We set the marginal variance of the $\tilde{\beta}_j(\cdot)$ 
to 0.04 and threshold $T = 0.08$ so that each resulting $\bbeta_j$
image would be approximately 30\% sparse on average. 
This level of sparsity roughly matches pseudo-sparsity in the real
data we analyze in Section \ref{sec:analysis}: applying standard
vertex-wise GLM methods with a Bonferroni correction-based $p$-value
threshold to these data resulted in significant findings over about
70\% of the cortical surface.
We treated the field $\bbeta_0$ as a spatially varying intercept
parameter and paired $\bbeta_1$ and $\bbeta_2$ with covariates:
for individual $i = 1, \ldots, N$, corresponding covariates
$(x_{1,i}, x_{2,i})\trans$
were drawn from multivariate Gaussians with
mean zero, unit marginal variances, and correlation 0.5.

We developed two different settings to control the signal-to-noise
ratio (SNR) in simulation: one ``low'' SNR setting with the spatial
SNR set to 4\% (equivalently $R^2 = 3.8\%$), and one ``high'' SNR
setting with the spatial SNR set to 40\% (equivalently
$R^2 = 28.6\%$). In these low and high SNR settings, we took the
the individual-level deviation variance $\tau^2 = 1.75$ and
$0.23$, respectively; and the white noise variances
$\sigma^2(\spc) = 1.25$ and $0.07$ for all $\spc$.
Parameters in the \emph{low} SNR setting were designed to
roughly mimic those estimated from our data application in Section
\ref{sec:analysis}.
Within this design, we studied the behavior of our various comparison
methods for increasing sample size, replicating the simulation 50
times per setting.

We then fit our suite of methods conditioning on
the ``true'' correlation parameters $\btheta$ (that is, 
$\btheta$ used to generate the non-sparse
$\tilde{\beta}_j(\cdot)$), and assessed performance against:
\begin{itemize}[itemsep=0.5em]
\item An ``\textbf{oracle}'' comparator: our marginal model but
  with perfect knowledge of all of the covariance parameters
  $\tau^2$, the $\zeta_j^2$, $\sigma^2(\cdot)$, and
  $\btheta$. With this knowledge, the posterior of $\bbeta$ has a
  closed form, and the problem is small enough that no
  approximation of the effective prior on $\bbeta$ is necessary.

\item A \textbf{reduced rank} version of the oracle
  comparator. For ease of implementation, we used a truncated
  Eigenbasis representation of the prior covariance of
  $\bbeta$. Here, we retained the minimal set of
  Eigenvalues that explained at least 80\% of the prior variance
  of $\bbeta$.

\item An \textbf{INLA}-based implementation of our working
  model
  \cite[using the \texttt{R} package \texttt{inlabru}; ][]{inlabru}.
  Comparison to INLA is not entirely fair: 
  it is the only method that could use, but is not provided 
  the true $\btheta$ in the simulation.

\item Vertex-wise general linear model
  comparators: one version fit to the data directly
  (\textbf{GLM}), and another fit to pre-smoothed data
  (\textbf{GLM-PS}). These methods are comparable to those used in
  common applied practice. For GLM-PS, the outcome data were
  pre-smoothed by convolution with a 6 mm FWHM exponential
  correlation function.
\end{itemize}
For comparison, we give the standard vertex-wise GLM analysis a 
Bayesian treatment by replacing our priors on the $\beta_j(\cdot)$ and
$\sigma^2(\cdot)$ with independent Jeffreys priors (as alluded to in
Section \ref{sec:methods:working-model}) and drawing from their
posterior using Gibbs sampling.

\subsection{Results of simulation comparisons}
\label{sec:simulation:results}

Results of our simulation study are summarized in Table
\ref{tab:sim-results} and in further detail in the Supplementary
Materials.
For each method in the table, we report the mean relative squared
error of the posterior mean of $\bbeta$ as a measure of estimation
accuracy (lower is better), alongside several measures of
inferential quality: interval coverage and the Matthews correlation
coefficient \cite[MCC; ][]{matthews1975comparison}.
The column ``95\% CIs'' shows the average coverage of pointwise 95\%
posterior credible intervals. Ideally, every method would have near
nominal 95\% coverage (within reason: again this design
represents a slight model misspecification).
We also use posterior credible bands as a way to summarize the joint
uncertainty in the $\beta_j(\cdot)$ over all vertices simultaneously.
In a spatial modeling context, posterior
credible bands are a natural, fully Bayesian approach to
inference and can be estimated from MCMC samples
\cite[see e.g.,][]{ruppert2003ch6}.
Example inferential decisions were constructed based on 80\% credible
bands: this threshold was chosen to represent a selection that might
reasonably be applied in practice rather than by optimizing some
inferential criterion. 
Note from the ``80\% CBs'' column in Table \ref{tab:sim-results}
that the credible bands can have very high coverage of the truth
and so can be a conservative approach to inference.
Active (inactive) decisions were made based on absence (presence)
of zero in the credible band for $\beta_j(\spc)$. The overall
quality of these decisions against the true sparsity of 
$\bbeta$ is summarized using the MCC.
Higher MCC values are indicative of better average decision making:
MCC zero indicates chance performance; one, perfect
concordance with the truth.

When the SNR is high, most methods do quite well; when the SNR
is low, the methods we develop here may be expected to improve
estimation accuracy and inferential sensitivity over alternative
methods. Even in the low SNR setting, both our marginal
and working model variants approach oracle model estimation accuracy
with near nominal credible interval coverage.
Several methods appear to have a tendency to underestimate the
variance of the $\beta_j(\cdot)$, including our
conditional model variant, INLA, and the vertex-wise GLM with
pre-smoothed data (Table \ref{tab:sim-results}; column ``95\%
CIs'').
Given this pattern of results, we might suggest the working model
variant as an economical approximation to the full marginal form when
the sample size is around 100 individuals or greater.
For completeness, note that INLA failed in the low
SNR setting for sample sizes over 50 due to overflow
errors evaluating the model likelihood.

\begin{table}[h]
  \centering
  \caption{
    Simulation results focusing on estimation of and inferential
    accuracy for the spatially varying regression coefficients.
    Results are presented as \emph{mean}
    (\emph{simulation standard error}). Interpret standard errors
    reported as 0 to be less than $5 \times 10^{-2}$.
    Methods developed in the main text are
    denoted by asterisks. MRSE--mean relative squared error;
    MCC--Matthews correlation coefficient; CIs--posterior credible
    intervals (pointwise); CBs--posterior credible bands
    (simultaneous). \protect{\label{tab:sim-results}}}
  \begin{footnotesize}
  \begin{tabular}{ c r l c c c c }
    & & & & \multicolumn{2}{c}{\emph{Interval Coverage}} & \\
    \emph{SNR} & \multicolumn{1}{c}{$N$} & \multicolumn{1}{c}{\emph{Method}}
      & \emph{MRSE} & \emph{95\% CIs}
    & \emph{80\% CBs} & \emph{MCC} \\ 
    \hline
    \multirow{24}{*}{4\%} 
    & \multirow{8}{*}{20}
      & *Conditional & 275.7\% (11.0\%) & 61.7\% (0.7\%) & 91.1\% (0.5\%) & 0.08 (0) \\ 
   & & *Marginal & 327.8\% (11.7\%) & 91.8\% (0.3\%) & 99.9\% (0\%) & 0.02 (0) \\ 
   & & *Working & 308.7\% (12.8\%) & 83.9\% (0.4\%) & 99.5\% (0.1\%) & 0.04 (0) \\ 
   & & GLM & 576.0\% (20.9\%) & 93.6\% (0.1\%) & 100.0\% (0\%) & 0 (0) \\ 
   & & GLM-PS & 229.1\% (9.8\%) & 35.2\% (0.5\%) & 59.3\% (0.7\%) & 0.06 (0) \\ 
   & & INLA & 257.6\% (13.6\%) & 54.0\% (0.9\%) & 85.0\% (0.8\%) & 0.08 (0) \\ 
   & & Low rank (80\%) & 359.4\% (11.1\%) & 95.1\% (0.1\%) & 100.0\% (0\%) & 0 (0) \\ 
   & & Oracle & 75.4\% (1.1\%) & 95.3\% (0.3\%) & 100.0\% (0\%) & 0 (0) \\ 
    \cline{2-7}
    & \multirow{8}{*}{100}
      & *Conditional & 63.2\% (2.0\%) & 67.5\% (0.5\%) & 94.7\% (0.2\%) & 0.25 (0) \\ 
  & & *Marginal & 81.2\% (2.6\%) & 96.1\% (0.2\%) & 100.0\% (0\%) & 0.05 (0) \\ 
  & & *Working & 64.2\% (2.0\%) & 87.2\% (0.3\%) & 99.7\% (0\%) & 0.14 (0) \\ 
  & & GLM & 100.2\% (2.7\%) & 94.7\% (0.1\%) & 100.0\% (0\%) & 0.06 (0) \\ 
  & & GLM-PS & 63.9\% (1.9\%) & 27.6\% (0.3\%) & 48.2\% (0.4\%) & 0.16 (0) \\ 
    & & INLA & -- & -- & -- & -- \\
  & & Low rank (80\%) & 76.9\% (1.7\%) & 95.3\% (0.1\%) & 100.0\% (0\%) & 0.03 (0) \\ 
  & & Oracle & 39.9\% (0.8\%) & 95.3\% (0.2\%) & 100.0\% (0\%) & 0.03 (0) \\ 
    \cline{2-7}
    & \multirow{8}{*}{500}
      & *Conditional & 15.7\% (0.5\%) & 75.8\% (0.4\%) & 97.8\% (0.1\%) & 0.49 (0) \\ 
  & & *Marginal & 18.5\% (0.5\%) & 96.7\% (0.1\%) & 100.0\% (0\%) & 0.22 (0) \\ 
  & & *Working & 15.8\% (0.5\%) & 90.6\% (0.3\%) & 99.9\% (0\%) & 0.34 (0) \\ 
  & & GLM & 19.7\% (0.5\%) & 94.8\% (0.2\%) & 100.0\% (0\%) & 0.25 (0) \\ 
  & & GLM-PS & 33.8\% (0.8\%) & 17.0\% (0.1\%) & 30.6\% (0.2\%) & 0.23 (0) \\ 
    & & INLA & -- & -- & -- & -- \\
   & & Low rank (80\%) & 18.5\% (0.5\%) & 94.9\% (0.2\%) & 100.0\% (0\%) & 0.24 (0) \\ 
  & & Oracle & 14.6\% (0.4\%) & 94.7\% (0.2\%) & 100.0\% (0\%) & 0.26 (0) \\ 
    \hline
    \multirow{24}{*}{40\%} 
    & \multirow{8}{*}{20}
      & *Conditional & 47.9\% (1.7\%) & 50.7\% (0.4\%) & 84.7\% (0.4\%) & 0.35 (0) \\ 
   & & *Marginal & 51.9\% (1.8\%) & 93.8\% (0.2\%) & 99.9\% (0\%) & 0.13 (0) \\ 
   & & *Working & 44.6\% (1.5\%) & 84.5\% (0.4\%) & 99.5\% (0\%) & 0.21 (0) \\ 
   & & GLM & 57.2\% (2.1\%) & 93.6\% (0.2\%) & 100.0\% (0\%) & 0.09 (0) \\ 
   & & GLM-PS & 52.6\% (1.4\%) & 23.4\% (0.3\%) & 41.4\% (0.4\%) & 0.18 (0) \\ 
   & & INLA & 46.5\% (1.5\%) & 60.5\% (0.5\%) & 90.3\% (0.3\%) & 0.30 (0) \\ 
   & & Low rank (80\%) & 44.3\% (1.4\%) & 95.3\% (0.2\%) & 100.0\% (0\%) & 0.08 (0) \\ 
   & & Oracle & 31.6\% (0.9\%) & 95.3\% (0.2\%) & 100.0\% (0\%) & 0.08 (0) \\ 
    \cline{2-7}
    & \multirow{8}{*}{100}
   & *Conditional & 9.6\% (0.3\%) & 53.2\% (0.4\%) & 84.9\% (0.3\%) & 0.69 (0) \\ 
  & & *Marginal & 9.8\% (0.3\%) & 96.7\% (0.1\%) & 100.0\% (0\%) & 0.36 (0) \\ 
  & & *Working & 9.6\% (0.3\%) & 90.4\% (0.2\%) & 99.9\% (0\%) & 0.46 (0) \\ 
  & & GLM & 10.0\% (0.3\%) & 94.7\% (0.2\%) & 100.0\% (0\%) & 0.39 (0) \\ 
  & & GLM-PS & 30.8\% (0.7\%) & 12.7\% (0.1\%) & 23.0\% (0.1\%) & 0.21 (0) \\ 
  & & INLA & 10.9\% (0.3\%) & 80.6\% (0.3\%) & 98.9\% (0.1\%) & 0.52 (0) \\ 
  & & Low rank (80\%) & 9.4\% (0.3\%) & 95.1\% (0.2\%) & 100.0\% (0\%) & 0.40 (0) \\ 
  & & Oracle & 8.7\% (0.2\%) & 95.1\% (0.2\%) & 100.0\% (0\%) & 0.40 (0) \\ 
    \cline{2-7}
   & \multirow{8}{*}{500}
   & *Conditional & 2.0\% (0.1\%) & 53.9\% (0.4\%) & 85.0\% (0.4\%) & 0.89 (0) \\ 
  & & *Marginal & 2.0\% (0.1\%) & 97.3\% (0.1\%) & 100.0\% (0\%) & 0.75 (0) \\ 
  & & *Working & 2.0\% (0.1\%) & 93.4\% (0.2\%) & 100.0\% (0\%) & 0.83 (0) \\ 
  & & GLM & 2.0\% (0.1\%) & 94.8\% (0.2\%) & 100.0\% (0\%) & 0.81 (0) \\ 
  & & GLM-PS & 27.7\% (0.5\%) & 6.0\% (0.1\%) & 11.0\% (0.1\%) & 0.17 (0) \\ 
  & & INLA & 2.1\% (0.1\%) & 92.0\% (0.3\%) & 100.0\% (0\%) & 0.84 (0) \\ 
  & & Low rank (80\%) & 2.0\% (0.1\%) & 94.9\% (0.2\%) & 100.0\% (0\%) & 0.82 (0) \\ 
  & & Oracle & 1.9\% (0.1\%) & 94.9\% (0.2\%) & 100.0\% (0\%) & 0.82 (0) \\ 
    \hline
  \end{tabular}
  \end{footnotesize}
\end{table}

\section{Data application}
\label{sec:analysis}

\subsection{Description of the data and model terms}
\label{sec:analysis:data}

To illustrate use of our methods, we applied our working model to
analyze n-back task contrast data from a subset
of 3,267 children enrolled in the ABCD study
\cite[][]{abcd201}.
Brief comparisons of estimation differences between our three model
variants are available for these data
in the Supplementary Materials, as is a comparison to results from
a popular neuroimaging software package \cite[AFNI;][]{cox1996afni}.
The ABCD study is the product of a large
collaborative effort to study longitudinal changes in the developing
brain through childhood and adolescence, and to track biological and
environmental correlates of development
\cite[][]{feldstein2018adolescent}.

We focus our analysis on relationships between task-related
activation and individual-level task accuracy.
We took 2- vs 0-back task contrast images ($z$-statistic scale) as our
primary outcome and modeled them as functions of 2-back task accuracy;
child fluid intelligence; child age (months); child gender (binary);
parental education (five levels); parental marital status (binary);
and family income (three levels). We included
first-order interactions between child gender and parental education;
child age and parental education; child age and child gender; child
age and 2-back accuracy; and child gender and 2-back accuracy. 
Predictors were mean-centered so that the spatial intercept in our
regression can be interpreted as the expected task contrast image for
a ten year old female child of average fluid intelligence that
scored 80\% correct on the 2-back task condition (married household,
at least one parent with a post graduate degree, and household income
greater than \$100,000 USD/year).

Covariates were chosen largely on the basis of known
associations with general n-back task accuracy
\cite[][]{pelegrina2015normative}. In addition, we performed 
exploratory analyses in the classic vertex-wise framework without any
spatial smoothing to help us visualize and understand important
aspects of the data (not shown; see Supplementary Materials for
details).

\subsection{Summary of primary results}
\label{sec:analysis:results}

\begin{figure}[!htb]
  \centering
  \includegraphics[width=0.95\textwidth]{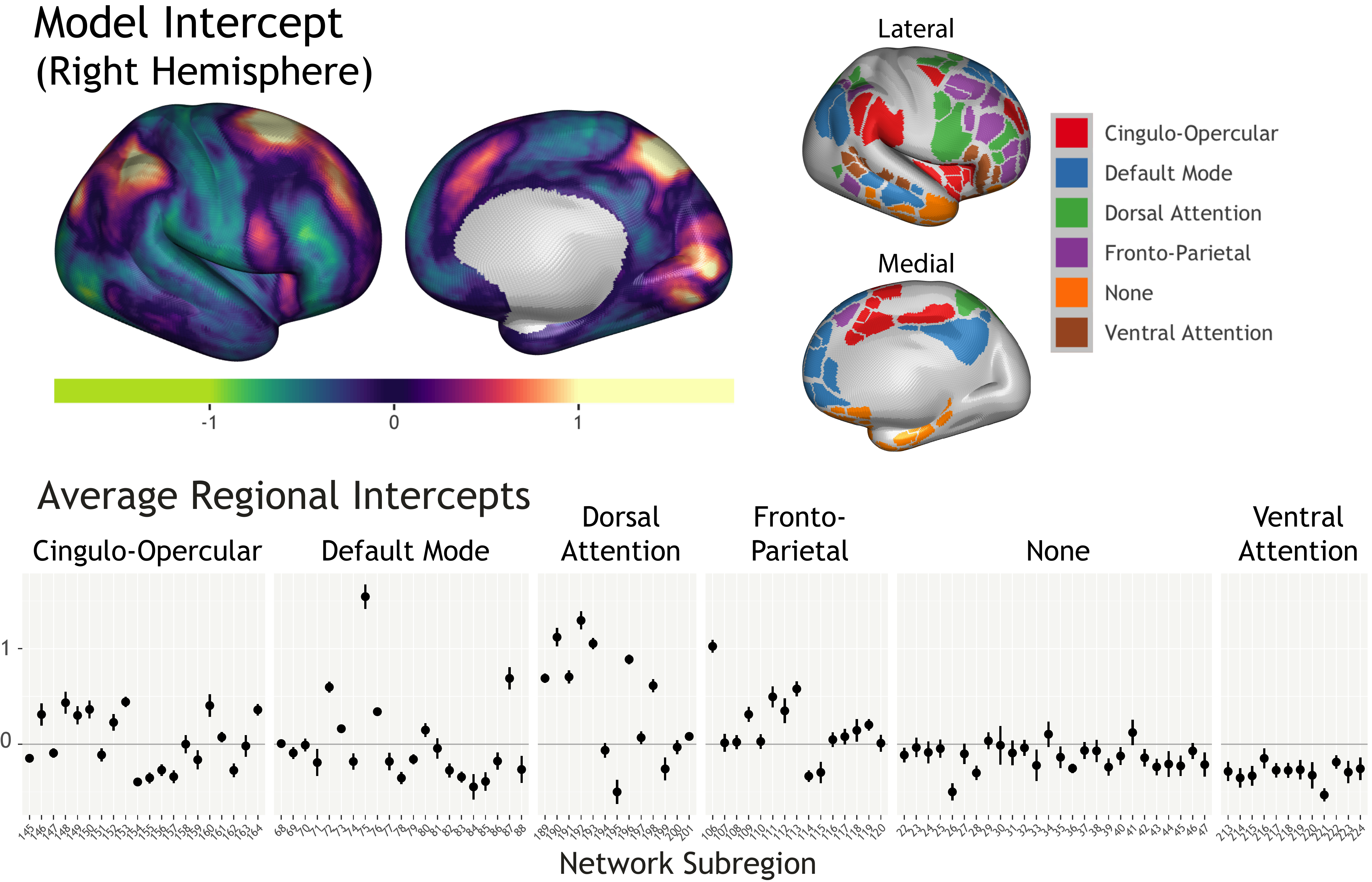}
  \caption{Model intercept coefficients summary. The upper left corner
    of the figure shows the posterior mean estimate of the intercept,
    which can be interpreted as a one-sample $z$-statistic for a 2-
    vs 0-back contrast, controlling for demographic information (see the
    main text for details). Forest plots in the bottom row of the
    figure summarize the intercept parameters in terms of region-level
    averages, with regions taken from the
    \protect\cite{gordon2016generation} 
    parcellation. Error bars
    in the forest plots correspond to Bayesian 95\% intervals
    that have been widened to be multiple-comparisons consistent
    (Bonferroni-type adjustment). The upper right panel of the figure
    shows the brain regions represented on the $x$-axis in the bottom
    row forest plots. Region numbers correspond to 
    labels from the Gordon atlas. Left to right the region
    labels read, Cingulo--Opercular: 145--164;
    Default Mode: 68--88;
    Dorsal Attention: 189--201;
    Fronto-Parietal: 106--120;
    None: 22--47;
    Ventral Attention: 213--224.
  }
  \label{fig:results-intercept}
\end{figure}

We consolidate analysis output by focusing on the right
hemisphere and noting results in the left hemisphere are highly
symmetric.
Fig. \ref{fig:results-intercept} shows the posterior
mean estimate of our model intercept ($\bbeta_0$), and 
gives a region of interest-level summary of this term. 
Regions of interest were taken from the
\cite{gordon2016generation} cortical surface 
atlas, which was created in part from resting state
functional connectivity maps and groups brain regions within
network ``communities.'' The atlas
delimits 172 brain regions in the right hemisphere (161 in the left),
each grouped within one of 13 functional communities.

To produce brain region-level summaries
we fit a series of mixed models to MCMC samples
of $\bbeta_0$, taking advantage of the Gordon atlas's hierarchically
grouped structure.
Using this natural hierarchy, we obtained
region-level averages of our spatial intercept, shrinking the 
average in each region towards its network community mean.
The bottom panel of Fig. \ref{fig:results-intercept}
displays point and multiple comparisons
consistent 95\% interval estimates for a subset of regions in
the atlas. Although we include all
172 atlas regions in this summary, we only show
estimates for those belonging to select communities
(see Fig. \ref{fig:results-intercept}). Results suggest
the largest activations generally occur in areas associated
with the Dorsal attention, Fronto-Parietal, and Cingulo-Opercular
networks. Similar conclusions were reached
by \cite{casey2018adolescent} in a smaller, preliminary subset of
ABCD data ($N = 517$), and by \cite{li2021neural} in a large study of
22--37 year old adults ($N = 949$).

\begin{figure}[!htb]
  \centering
  \begin{subfigure}[c]{0.5\linewidth}
    \includegraphics[width=\linewidth]{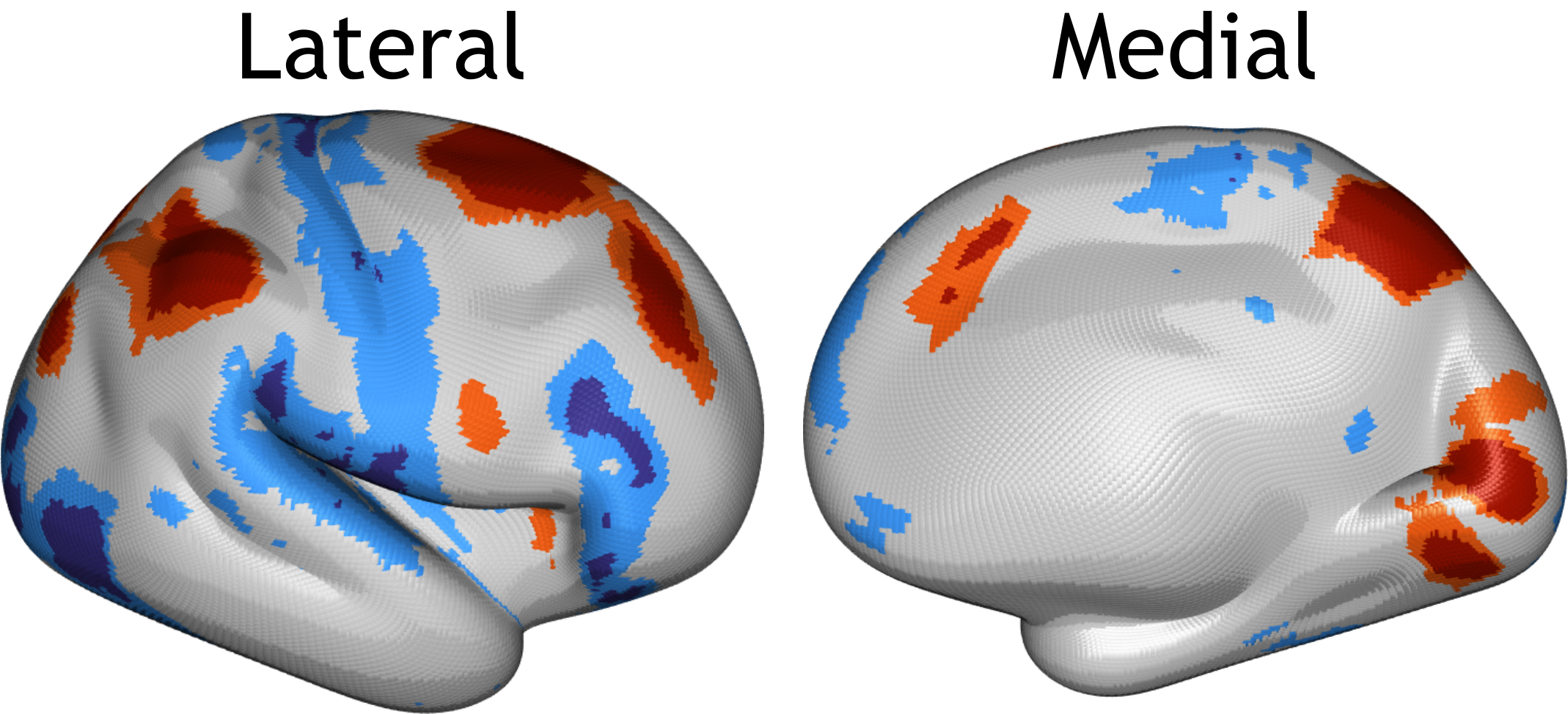}
    \caption{Spatial inference for the model intercept
      (posterior mean shown in Fig. \ref{fig:results-intercept}).}
    \label{fig:spatial-inference}
  \end{subfigure}
  \vspace*{0.5em}
  
  \begin{subfigure}[c]{0.5\linewidth}
    \includegraphics[width=\linewidth]{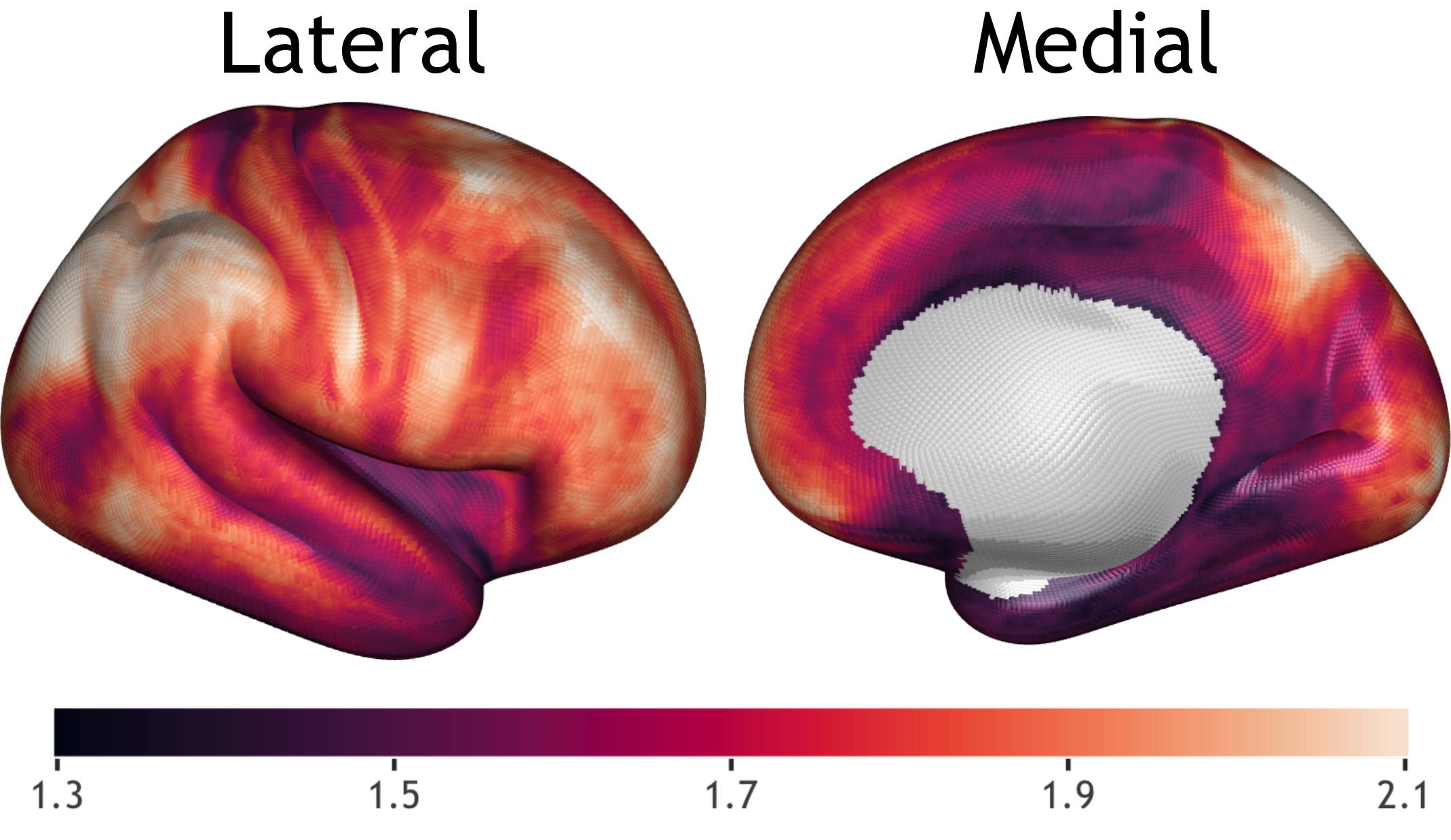}
    \caption{Residual standard deviation.}
    \label{fig:results-sigma}
  \end{subfigure}

  \caption{
    (\ref{fig:spatial-inference})
    Model intercept for the right hemisphere: example signed
    discoveries using an 80\% posterior simultaneous credible band to
    infer locations where $|\beta_0(\cdot)| > 0.4$. Red regions
    correspond to functional activations and blue regions correspond
    to deactivations. Darker colors indicate regions of simultaneous
    posterior confidence that $|\beta_0(\spc)|$ is greater than $0.4$
    for all vertices $\spc$ in those regions.
    Lighter colors can be thought of as reflecting the
    spatial uncertainty in that claim of posterior credibility.
    (\ref{fig:results-sigma})
    Residual standard deviation for the right
    hemisphere. Areas of high residual variation generally overlap
    with activation areas in the 2- vs 0-back contrast (confer with
    Fig. \ref{fig:results-intercept}).
  }
  \label{fig:results-analysis-main}
\end{figure}

Fig. \ref{fig:spatial-inference} depicts example spatial inference for
the intercept in the right hemisphere, thresholding at what
we might consider a small to medium effect size. In the figure,
colored regions denote areas where the posterior mean estimate of
$|\beta_0(\spc)|$ is greater than $0.4$.
Since we are modeling $z$-statistic outcomes, $\beta_0(\spc) > 0.4$
can be interpreted to mean, roughly, that we expect at least 2 out of
every 3 ``average'' children to show task-related
activation at location $\spc$ (versus 1 in 3 showing deactivation; the
statement can be reversed for $\beta_0(\spc) < 0.4$).
Areas of darker color in Fig. \ref{fig:spatial-inference} mark core
regions where the posterior probability that $|\beta_0(\spc)| > 0.4$
is at least 80\% simultaneously for all included vertices $\spc$.
This interpretation is similar to the notion of ``upper
confidence sets'' from \cite[][]{bowring2021confidence}.
Residual standard deviations for the right hemisphere are shown
in Fig. \ref{fig:results-sigma}. In general, areas with the highest
residual variance overlap with areas activated in the 2- vs 0-back
contrast (confer from Figs. \ref{fig:results-intercept} and
\ref{fig:results-sigma}). This result indicates substantial 
variability in individual responses in these regions. Our
fitted model explained about 6.2\% of the total variance in
the task contrast images.

\subsection{Goodness-of-fit evaluation}
\label{sec:analysis:diagnostics}

\begin{figure}[!htb]
  \begin{minipage}{\textwidth}
    \centering
    \begin{tabular}{ c c c }
      Mean & 10\textsuperscript{th} Quantile & 90\textsuperscript{th} Quantile \\
      \includegraphics[width=0.18\textwidth]{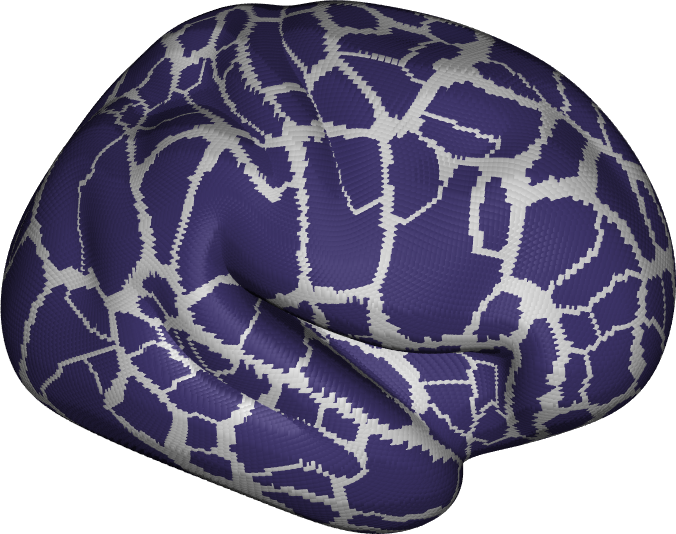} &
    \includegraphics[width=0.18\textwidth]{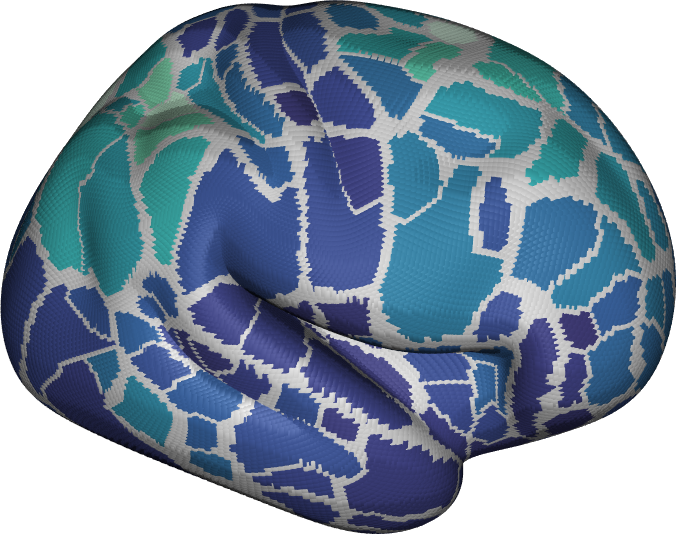} &
    \includegraphics[width=0.18\textwidth]{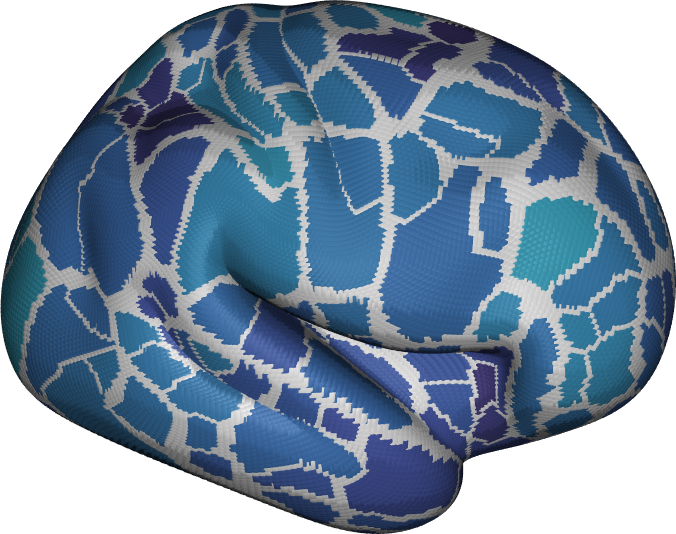}
      \\
    \multicolumn{3}{c}{\includegraphics[width=0.5\textwidth]{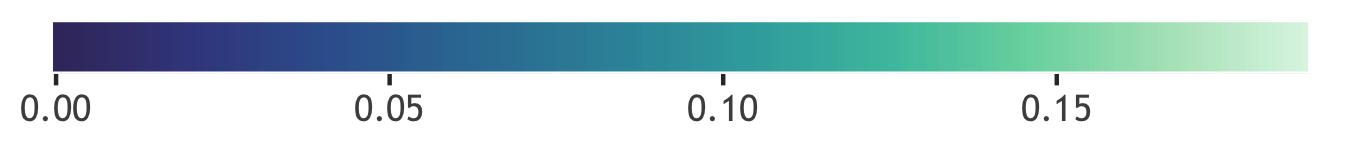}}
    \end{tabular}
  \end{minipage}
  \vspace*{2em}
  
  \begin{minipage}{\textwidth}
    \centering
    \begin{tabular}{ c c c }
      Best Case & Median Case & Worst Case \\
      \includegraphics[width=0.25\textwidth]{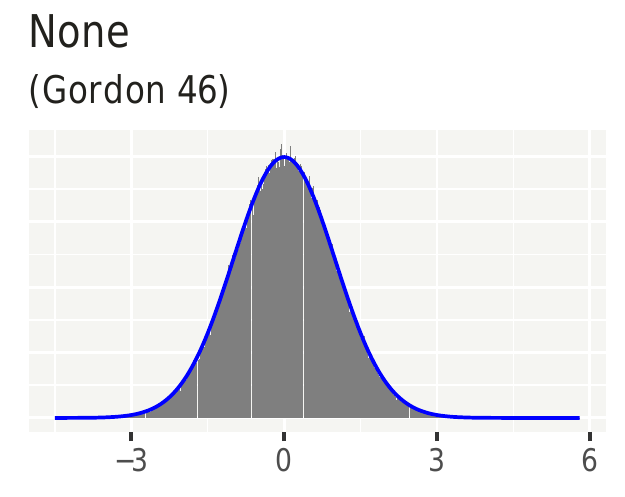} & 
    \includegraphics[width=0.25\textwidth]{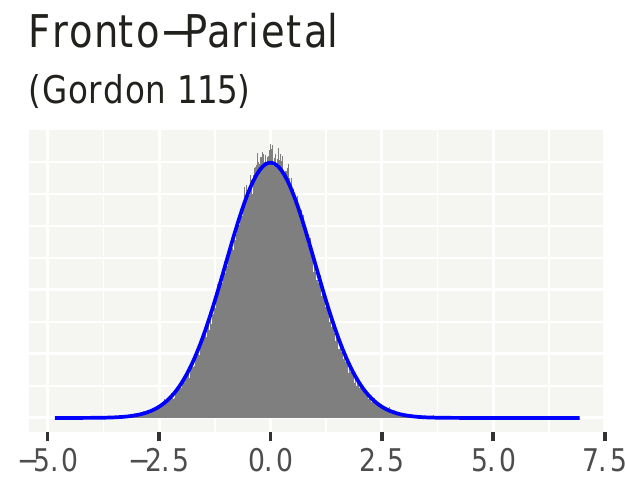} & 
    \includegraphics[width=0.25\textwidth]{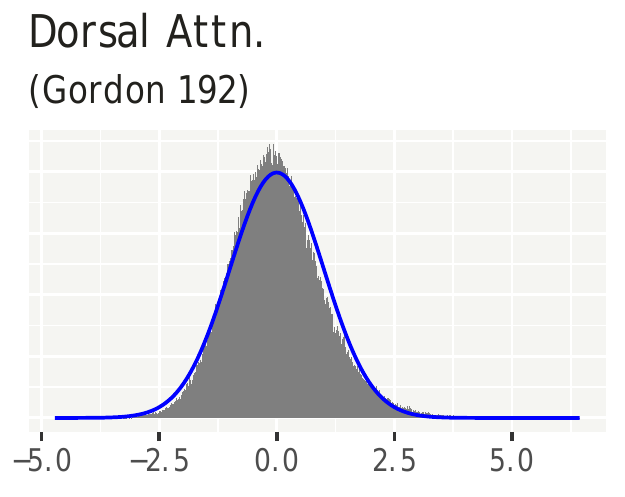} 
    \end{tabular}
  \end{minipage}
  \vspace*{1em}
  \caption{Goodness of fit checking.
    (\emph{Top}) 
    Absolute differences in the observed and mean posterior
    predictive value for three different test statistics computed
    across all participants and vertices for each brain region in the
    \protect\cite{gordon2016generation} atlas.
    Test statistics shown are the regional mean and
    10\textsuperscript{th} and 90\textsuperscript{th} quantiles.
    In the figure, predictive checking is homogeneous within each brain
    region; the gray shows a boundary area not assigned to any
    particular region. 
    (\emph{Bottom}) Histograms of standardized residuals from
    three different brain regions. Blue lines show the fit of model at
    the posterior mean. The three different regions chosen show the
    best, the median, and the worst-case scenarios for the model's 
    goodness-of-fit in these areas.
  }
  \label{fig:results-fit}
\end{figure}

Finally, we assess the fit of our model using posterior predictive
simulation  
and analysis of model residuals. Selected results of these comparisons
are presented in Fig. \ref{fig:results-fit}.
In the figure, we summarize discrepancies in the predictive and
empirical data distributions based on measures of central tendency and
spread.
To do this, we again used the \cite{gordon2016generation} 
parcellation, computed test descriptive
statistics across participants for each brain region, and compared
against the same statistics computed over model-simulated data of the
same size.
Absolute differences in empirical values and the
posterior predictive mean are shown for three such test statistics in
each brain region (Fig. \ref{fig:results-fit}).
For scale, the largest regional difference in the
figure is $< 0.2$ (10\textsuperscript{th} Quantile 
panel), whereas the range of the data is approximately $-13.7$ to
$17.1$. In general, discrepancies were extremely low for summaries of
central tendency.
Fig. \ref{fig:results-fit} also shows histograms of 
standardized residuals.
We ranked each brain region by discrepancy with a normal model and
show residual histograms for the best, median, and worst-case regions
(Fig. \ref{fig:results-fit}, lower panel). In general,
evidence is again of good model fit. Interestingly, Gordon region
192 (worst-case fit) contained 
the highest overall mean estimate within the
Dorsal-Attention network community for both the intercept
and linear 2-back accuracy term.
This result may indicate, for example, that while the (relatively
simple) model we have used for 2-back accuracy provides a reasonable
fit to the task contrast data across most of the right hemisphere, it
may fail to perfectly encapsulate complex task-related activation
patterns in this sample.

\section{Discussion}
\label{sec:discussion}

Here we propose a Bayesian spatial model for group-level 
image-on-scalar regression analyses, and illustrate several ways to
consider working with the model in practice. We also show how the
spatial Gaussian process prior formulation and related approximation
through conditional independence methods can enable flexible and
reasonably efficient computation with MCMC.
Critically, our approach allows us to work with 
numerically full-rank spatial processes, and does not rely on lossy
compression schemes like down-sampling or low-rank projection.
We have shown in simulation that these strategies can improve on
alternative methods in terms of estimation and inferential accuracy.
Finally, we have illustrated
use of our method on (n-back) task contrast data from the Adolescent
Brain Cognitive Development study.

With the exception of a white-noise component in our model error
process, we express (in Section \ref{sec:methods}) our model as a sum
of terms assigned stationary spatial priors. In general, spatial
stationarity is not considered a realistic assumption for imaging data
\cite[e.g.,][]{aja2015spatially}. 
While we use stationary priors throughout for simplicity, we note that
the posterior distribution of our model parameters can still
reflect non-stationary processes, given the data.
Visually, the fitted model intercept (see
Fig. \ref{fig:results-intercept}) in fact suggests a degree of
posterior mean field non-stationarity. Moreover, since our model on
the white-noise process is inherently non-stationary, our prior
hierarchy can lead to more data-adaptive smoothing in the regression
coefficients compared to standard analysis streams where stationary
spatial smoothing is applied to the data at some level prior to
analysis.

As we alluded to in Section \ref{sec:analysis:data}, it may be of
interest to build extensions to our method to incorporate additional
variance components for more complex or specific study designs. 
While our present model is technically capable
of estimating such effects by pooling the corresponding $\zeta^2_j$ 
across related terms, including a large number of random spatial
effects in the analysis can be extremely demanding
computationally. One workaround might be to omit modeling spatial
correlation structures for these terms, and treat them as pure
nuisance parameters. At the time of writing, we have not yet studied
the practical consequences of doing so.

%


\section*{Acknowledgments}

The authors would like to thank Mike Angstadt, Dr. Chandra Sripada,
and Dr. Mary Heitzeg, who oversaw preprocessing of the fMRI data we
used to illustrate our methods, and who provided helpful feedback on
our work. This work was partially supported by NIH R01 DA048993
(Kang and Johnson).

\section*{Supplementary Materials}

Appendices, Tables, Figures, and software referenced in Sections
\ref{sec:methods} and \ref{sec:analysis} are available below.
Software for our methods is also available online at
\url{https://github.com/asw221/gourd}.

\section*{Data Availability}

The data that support the findings in this paper originate from the 
Adolescent Brain Cognitive Development (ABCD) imaging collective
(second annual release; version 2.0.1).
The ABCD data repository grows and changes over time. The ABCD data
used in this report come from NDA study 2573. DOIs can be found online
\cite[][\url{https://nda.nih.gov/study.html?id=721}]{abcd201}.

\clearpage
\renewcommand{\appendixname}{Supplement}
\begin{appendices}

\section{Posterior computation}
\label{sec:supp:posterior-computation}

We outline our general approach to computation using the working model
in the main text as a running example since this
variant is the easiest to work with. Posterior
computation with the conditional and marginal models can be
accomplished in very similar fashion.

Since we typically work on a fixed spatial domain $\mathcal{S}$, let
$\bbeta_j$  (dropping the superscript $w$ for simplicity)
denote the random field $[\beta_j^{w}(\spc)]_{\spc \in \mathcal{S}}$ 
for $j = 0, \ldots, P-1$, and let 
$\bbeta = (\bbeta_0\trans, \ldots, \bbeta_{P-1}\trans)\trans$.
Let $\bC = [C\{d(\spc, \spc')\}]_{\spc, \spc' \in \mathcal{S}}$
represent the $(M \times M)$ spatial correlation matrix such that the
prior on each $\bbeta_j$ is a zero-mean Gaussian random field with
covariance $\zeta_j^2 \tau^2 \bC$.
Similarly, let $\bSigma$ represent the variance of
$\epsilon_i^{w}(\cdot)$, here an $(M \times M)$ diagonal matrix with
the $\sigma^2(\spc)$, $\spc \in \mathcal{S}$ on the diagonal;
let $\bX$ denote the $(N \times P)$ matrix of participant-level
covariates; let $\by_i = [y_i(\spc)]_{\spc \in \mathcal{S}}$ denote the
vectorized outcome image for participant $i$; and let
$\by = (\by_1\trans, \ldots, \by_N\trans)\trans$ represent the
$(NM \times 1)$ vector of concatenated subject outcomes.

With the data in this ``long'' format, the model can be conveniently
expressed in terms of Kronecker products.
With $\bZ = \diag(\zeta_0^2, \ldots, \zeta_{P-1}^2)$, the conditional
posterior variance of $\bbeta$ can be written,
\begin{equation}
  \label{eqn:posterior-variance-beta}
  \var(\bbeta \mid \by, \cdot) =
  \big( \bX\trans \bX \otimes \bSigma^{-1} +
  \bZ^{-1} \otimes \tau^{-2} \bC^{-1} \big)^{-1},
\end{equation}
using shorthand to express conditioning on $\bSigma$, $\bZ$,
$\btheta$, and $\tau^2$.
Since the dimension of $\bbeta$ grows rapidly with $P$, it can be
difficult or even impossible to work with
\eqref{eqn:posterior-variance-beta} directly. Instead, we outline two
strategies to enable efficient posterior computation at this
scale. The first strategy, as alluded to above, is to replace
$\bC^{-1}$ with a sparse approximation $\tilde{\bC}^{-1}$ such that
$\tilde{\bC} \approx \bC$. In doing so, we follow work on the so
called ``Nearest Neighbor Gaussian Process'' 
\cite[][]{datta2016hierarchical, finley2019efficient}, replacing the
idea of $k$-nearest neighbors with small neighborhoods of fixed
physical radius $r$. Briefly, we replace $\bC^{-1}$ with a conditional
independence approximation, enforcing that $\tilde{C}_{ij}^{-1} = 0$
if $d(\spc_i, \spc_j) > r$ for $\spc_i, \spc_j \in
\mathcal{S}$. Similar ideas have been alternately called Vecchia
approximation 
\cite[][]{vecchia1988estimation, katzfuss2021general},
composite likelihood \cite[][]{varin2011overview}, or
Markov random field approximation \cite[][]{rue2005gaussian}, but in
general can lead to highly accurate and scalable approximations of
full rank spatial models
\cite[see e.g.,][]{taylor2014inla, datta2016hierarchical,
  heaton2019case}. Working with such an approximation of course
introduces a hyperparameter, $r$, for the neighborhood radius size. In
practice we found that in a large data setting choice of $r$ had very
little effect on our analysis (see Appendix
\ref{sec:supp:abcd-sensitivity} for a sensitivity analysis).
In a small $N$ setting, however, when the prior
has more influence on the posterior, $r$ must generally be chosen
large enough to obtain a good approximation of the log
prior. Anecdotally, we found that taking $r \geq 6$ mm worked well in
simulation.

Although replacing $\bC^{-1}$ with $\tilde{\bC}^{-1}$ in
\eqref{eqn:posterior-variance-beta} above lends sparsity and
efficiency to computation in our setting, it can still be burdensome
to evaluate or decompose \eqref{eqn:posterior-variance-beta} even for
moderate $P$. To overcome this issue we propose an approximate
quasi-Newton Hamiltonian Monte Carlo (HMC) algorithm for sampling from
the posterior of $\bbeta$, conditional on the other model
parameters. HMC is a hybrid, gradient-based MCMC method that is 
often more efficient in high dimensions than other MCMC
algorithms \cite[][]{neal2011hmc}, and can be used to
avoid direct computation with the very high dimensional covariance
matrix \eqref{eqn:posterior-variance-beta} here. In the general HMC
algorithm, sampling can be improved by scaling the gradients by a
carefully chosen ``mass matrix,'' $\bM$. In their highly influential
paper, Girolami and Calderhead showed that the most efficient choice
updates $\bM$ to be proportional to the posterior 
Fisher information matrix of the updated parameter
\cite[][]{girolami2011riemann}. Instead, we can choose to use the
prior information matrix to ``estimate'' the posterior information in
the spirit of a quasi-Newton algorithm: doing so results in a
computationally tractable and efficient alternative.
Taking $\bM \propto (\bZ^{-1} \otimes \tau^{-2} \bC^{-1})$ and
plugging in a sparse approximation of $\bC^{-1}$ as above can result
in dramatic improvement in Markov chain mixing with minimal increase
in computation time. In practice, we found that we need not use the
same $\tilde{\bC}^{-1}$ in $\bM$ as in our approximation of the log
prior. In fact, we found it better to use smaller neighborhood radii
in our construction of $\bM$, and that keeping the neighborhood radius
within the 2--4 mm range here resulted in the best Markov chain
mixing.

\subsection{Computation for our working model}

To help stabilize our computational steps, we first compute a rank
revealing decomposition of the covariate matrix $\bX$. We will work
here with the singular value decomposition (SVD)
$\bX = \bU \bD \bV\trans$, though the QR decomposition and its rank
revealing variants, etc. would work in the same way.
In general, computing the SVD is an
$\Order(N P^2)$ operation when $P \leq N$; even for relatively large
$P$ computing the SVD of $\bX$ takes minimal time
compared to MCMC. For simplicity, we will assume here that $\bX$ is
full column rank. Let $\bgamma = (\bV\trans \otimes \bI_M) \bbeta$
denote our parameter of interest, rotated by $\bV$. The effective
prior on $\bgamma$ is simply,
\[ \bgamma \sim \Gaussian\big( \bm{0}, \bV\trans \bZ \bV \otimes
  \tau^2 \bC \big), \]
which, as noted in the above, can be efficiently approximated by
plugging in a sparse matrix $\tilde{\bC}^{-1}$ such that
$\tilde{\bC} \approx \bC$.
In turn, the log prior and its gradient can be approximated via,
\begin{equation}
  \ln \pr(\bgamma \mid \bZ, \btheta, \tau^2) \approx -\frac{1}{2}
  \bgamma\trans \big( 
  \bV\trans \bZ^{-1} \bV \otimes
  \tau^{-2} \tilde{\bC}^{-1} \big) \bgamma + K(\bZ, \btheta, \tau^2), 
\end{equation}
where $K(\bZ, \btheta, \tau^2)$ is the log normalization constant
and,
\begin{equation}
  \nabla_\bgamma \ln \pr(\bgamma \mid \bZ, \btheta, \tau^2) \approx 
  - \big( \bV\trans \bZ^{-1} \bV \otimes
  \tau^{-2} \tilde{\bC}^{-1} \big) \bgamma.
\end{equation}
Kronecker identities facilitate numerical evaluation of these
quantities.
Similarly, the log likelihood can be rewritten in terms of
$\bgamma$. Up to the integration constant, the log likelihood of our
working model can be written,
\begin{equation}
  \label{eqn:supp:working-model-loglikelihood}
  \ln \pr(\by \mid \bSigma, \bgamma) = -\frac{1}{2}
  \gamma\trans \big( \bD^2 \otimes \bSigma^{-1} \big) \bgamma +
  \bgamma\trans \big( \bD \bU\trans \otimes \bSigma^{-1} \big) \by
  -\frac{1}{2} \by\trans \big( \bI_N \otimes \bSigma^{-1} \big) \by.
\end{equation}
From this expression, it can be seen that the part of the log
likelihood that includes $\bgamma$
depends on the data only through the sufficient statistic
$(\bU\trans \otimes \bI_M) \by$.
This implies that, within our working model framework, gradients and
Metropolis-Hastings ratios can be computed efficiently with respect to
$\bgamma$. 
Similarly, it can be shown that the residual sum of squares depends on
the data only through 
$(\bU\trans \otimes \bI_M) \by$ and an additional sufficient
statistic, $\sum_i \by_i^{\circ 2}$, where we use
$\ba^{\circ b} = (a_i^b)$ to denote element-wise or Hadamard
exponentiation. This additional fact suggests that $\sigma^2(\cdot)$
can be easily updated without reference to the original data. With
these two pieces in hand, we write our posterior computation algorithm
to alternate updating $\bgamma$ through Hamiltonian Monte Carlo (as
discussed in the main text), and updating the variance parameters with
Gibbs sampling. 
Samples of $\bgamma$ can easily be rotated back into samples of
$\bbeta$ by applying the reverse transformation,
$\bbeta = (\bV \otimes I_M) \bgamma$.
Within each HMC iteration, we update the
algorithm's mass matrix via,
\begin{equation}
  \bM(\bZ, \tau^2) = \bV\trans \bZ^{-1} \bV \otimes 
  \tau^{-2} \tilde{\bC}_M^{-1},
\end{equation}
where $\tilde{\bC}_M^{-1}$ is a sparse matrix again constructed so
that $\tilde{\bC}_M \approx \bC$.

\subsection{Approximation for our ``Conditional'' model}

Our computational strategy for the conditional method relies on the
observation that the full conditional distribution of the $\bomega_i$
is relatively easy to work with. Although it is too burdensome to
fully sample the $\bomega_i$ at each iteration of an MCMC routine, it
takes only a modest amount of time to find a maximum a posteriori
(MAP) estimate of the $\bomega_i$ given an estimate of $\bbeta$. As we
have shown above, gradient-based updates are efficient to compute for
$\bbeta$ in our working model. We first obtain an approximate MAP
estimate of $\bbeta$ using our working model with the restriction that
$\sigma^2(\spc) \equiv \sigma^2$ for all locations
$\spc \in \mathcal{S}$. An estimate of this parameter can be computed
quite quickly using gradient ascent. With estimates of $\bbeta$,
$\tau^2$, and $\bSigma$ in hand, the $\bomega_i$ can be set to their
conditional posterior mode analytically, 
\[ \bomega_i \gets (\tau^{-2} \bC^{-1} + \bSigma^{-1})^{-1}
  \bSigma^{-1} \{\by_i - (\bx_i\trans \otimes \bI_m) \bbeta \}. \]
To do this, we again construct a sparse, Vecchia-type approximation of
the matrix $(\tau^{-2} \bC^{-1} + \bSigma^{-1})^{-1}$.
Maximizing with respect to $\bbeta$ and $\bomega_i$ can be iterated if
necessary for convergence. Once we have a satisfactory estimate of 
$\bomega_i$, we can easily subtract it from $\by_i$ and switch to our
working model HMC algorithm for inference on $\bbeta$ if desired.

\subsection{Approximation for our ``Marginal'' model}

Rather than fix $\bomega_i$ at a point estimate as above, our
strategy for the marginal model will be instead to obtain a fixed
estimate of the correlated error variance---$\bH = \tau^2 \bC +
\bSigma$ in the main text---and use this estimate in our general HMC
algorithm (described above).
To compute with the marginal method, we first obtain an initial
estimate of $\bbeta$ using gradient ascent in our working model
approximation as above. With this estimate in hand, we can estimate
the marginal or sill variance $(\tau^2 + \sigma^2(\spc))$ for each
location $\spc$ using the standard formula
$\sum_i \{y_i(\spc) - \bx_i\trans \bbeta(\spc)\}^2 / (N - 1)$.
Then, again following \cite[][]{finley2019efficient}, it is
straightforward to construct a Vecchia-type
approximation $\tilde{\bH}^{-1}$ such that
$\tilde{\bH} \approx \bH$, and so that $\tilde{\bH}$ contains our
estimates of the spatial sills on the diagonal. To work with
MCMC, $\tilde{\bH}$ can simply be substituted in place of $\bSigma$ in
our working model HMC outline above. For computational savings, we do
not update $\tilde{\bH}$ over MCMC iterations when we work with the
model in this way.

\section{Estimation of $\btheta$ through maximum marginal
  likelihood} 

In general spatial kriging applications, it is common to estimate
$\btheta$ by maximum marginal likelihood
\cite[e.g.,][]{mardia1984maximum, rasmussen2006ch5}. This can be done,
for example by integrating out the mean model parameters and
optimizing the resulting marginal likelihood with respect to the
covariance and correlation parameters.
Retaining the vector-based notation from our posterior computation
sections and integrating the $\bbeta_j$ and $\bomega_i$ out of
equation (1) in the main text, the marginal log likelihood (less the
integration constant) for our spatial regression model is,
\begin{equation}
  \label{eqn:supp:marginal-likelihood}
  f(\by \mid \btheta, \bSigma, \bZ, \tau^2) =
  - \frac{1}{2} \sum_i \ln \det \bOmega_i +
  \by_i\trans \bOmega^{-1}_i \by_i, 
\end{equation}
where
$\bOmega_i = \tau^2 \big(1 + \sum_j \zeta_j^2 x_{ij}^2 \big) \bC +
\bSigma$,
and $\bSigma$ is the $(M \times M)$ diagonal matrix with
$[ \sigma^2(\spc) ]_{\spc \in \mathcal{S}}$ on the diagonal.
Equation \eqref{eqn:supp:marginal-likelihood} can of course be maximized
directly, but at the cost of also solving for $M + P + 1$ additional
parameters in $\bSigma$, $\tau^2$, and the $\zeta_j^2$. Also, from a
practical point of view, it is somewhat undesirable that the marginal
variance of $\by_i$ depends on $\bx_i$, implying the need to
re-optimize \eqref{eqn:supp:marginal-likelihood} every time a covariate is
added to or removed from the model.

Instead of working with \eqref{eqn:supp:marginal-likelihood} directly, we
choose to estimate $\btheta$ by optimizing the marginal
log likelihood for a surrogate simpler model. To estimate $\btheta$,
we replace \eqref{eqn:supp:marginal-likelihood} above with,
\begin{equation}
  \label{eqn:supp:surrogate-likelihood}
  \tilde{f}(\by \mid \btheta, \bSigma, \tau^2) =
  -\frac{N}{2} \ln \det (\tau^2 \bC + \bSigma) +
  -\frac{1}{2} \sum_i \by_i\trans (\tau^2 \bC + \bSigma)^{-1} \by_i,
\end{equation}
which, incidentally, is the unnormalized marginal likelihood for our
working model with an intercept as the only predictor. Equation
\eqref{eqn:supp:surrogate-likelihood} can be evaluated approximately 
either through use of a Vecchia-type approximation of
the matrix $(\tau^2 \bC + \bSigma)^{-1}$, or by down-sampling the
$\by_i$ to a more manageable number of spatial locations. We chose the
former option in the present paper, and in practice mean-center each
image $\by_i$ prior to optimization.
While this approach can work well, we have
noticed anecdotally that it can also tend to underestimate the width
of the correlation function. Obtaining a good estimate of $\btheta$ in
more complex settings---as in \eqref{eqn:supp:marginal-likelihood}---remains
an open research question. We do not, however, expect inference on
$\bbeta(\cdot)$ or other model parameters, to be overly sensitive to
the choice of $\btheta$, given a reasonable amount of  data (e.g. see
our sensitivity analysis in Appendix
\ref{sec:supp:abcd-sensitivity}).

Finally, we have used the gradient-free
optimization routine BOBYQA \cite[][]{powell2009bobyqa} to maximize
\eqref{eqn:supp:surrogate-likelihood}, which, surprisingly, improved
performance over gradient-based optimizers (both run time and
stability). The BOBYQA algorithm works by iteratively constructing a
quadratic approximation to the objective function at a set of
interpolation points, which are themselves updated as a trust region
is progressively estimated \cite[][]{powell2009bobyqa}. The algorithm
may fail if, for example, \eqref{eqn:supp:surrogate-likelihood} exhibits
local behavior that cannot be well approximated by a quadratic
function.

\section{Additional simulation study results}
\label{sec:supp:simulation}

\begin{figure}[H]
  \centering
  \begin{tabular}{ c }
    SNR = 4\% \\
    \includegraphics[width=0.55\textwidth]{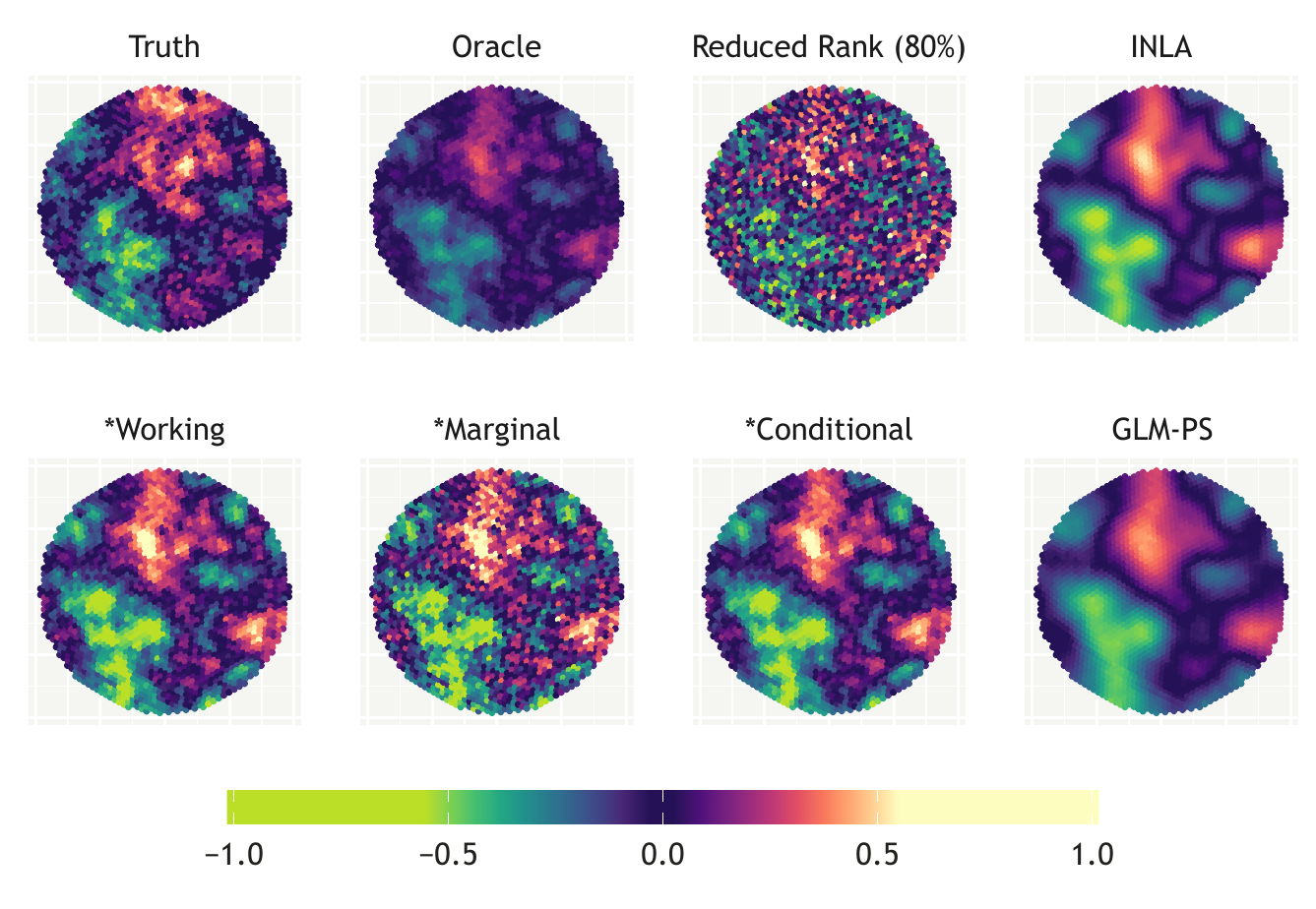}
    \\
    \hline
    \\
    SNR = 40\% \\
    \includegraphics[width=0.55\textwidth]{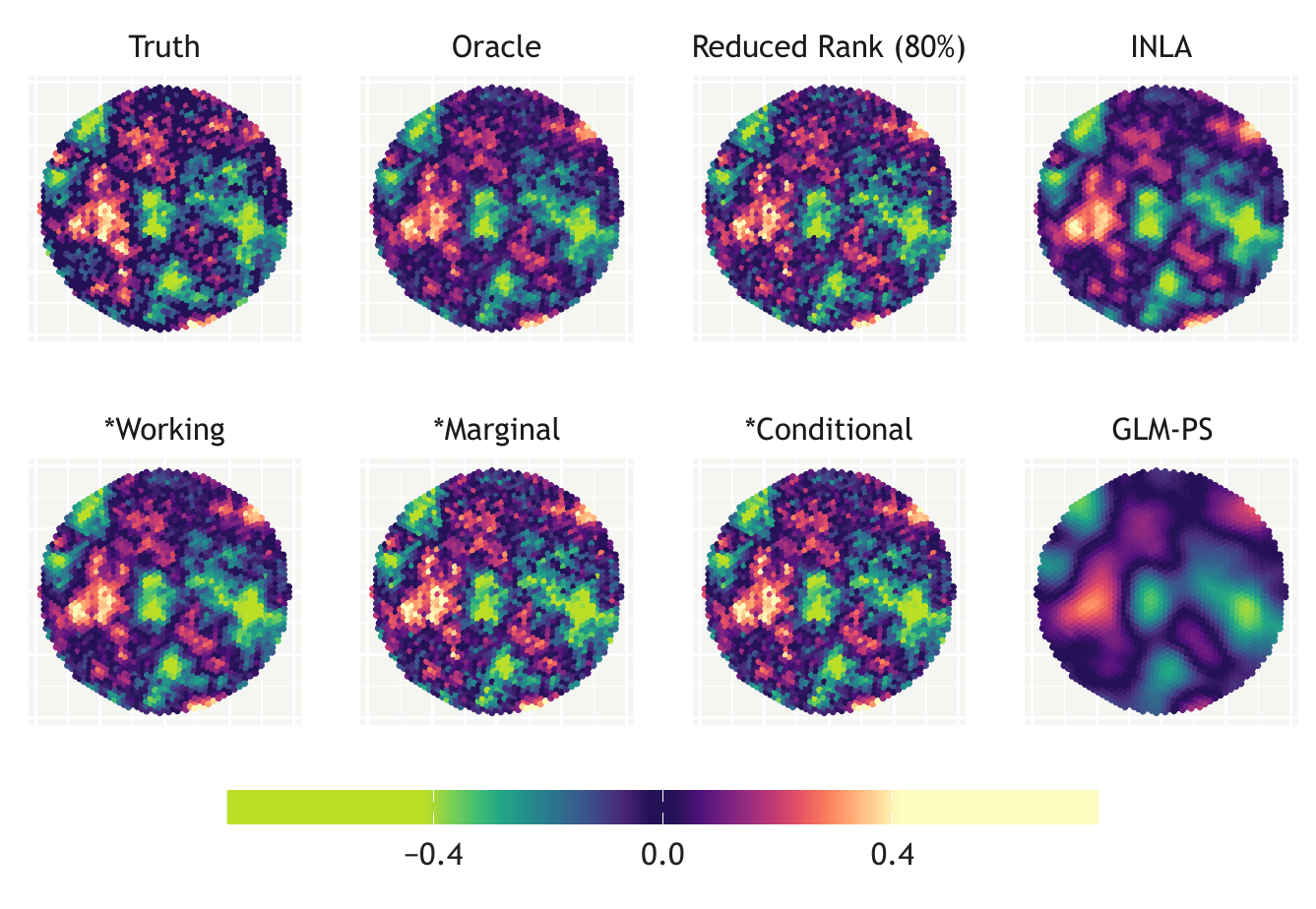}
  \end{tabular}
  \caption{Example posterior mean field estimation of example sparse,
    spatially varying intercepts in our low (\emph{top}) and high
    (\emph{bottom}) SNR settings. Estimates are from single
    representative simulation iterations, both with sample size
    $N = 50$. Note that color bar scales
    are different between the two sets of figures. Methods developed
    in the main text are indicated by asterisks.
    \protect{\label{fig:sim-mean-fields}}}
\end{figure}

\begin{figure}[H]
  \centering
  \begin{tabular}{ c c }
    \includegraphics[width=0.44\textwidth]{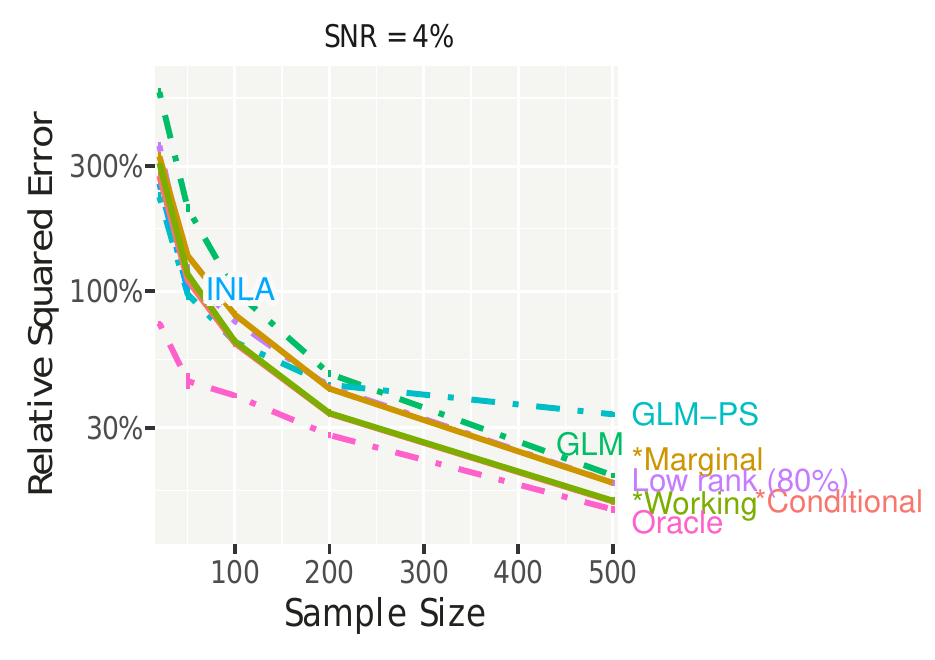}
    &
      \includegraphics[width=0.44\textwidth]{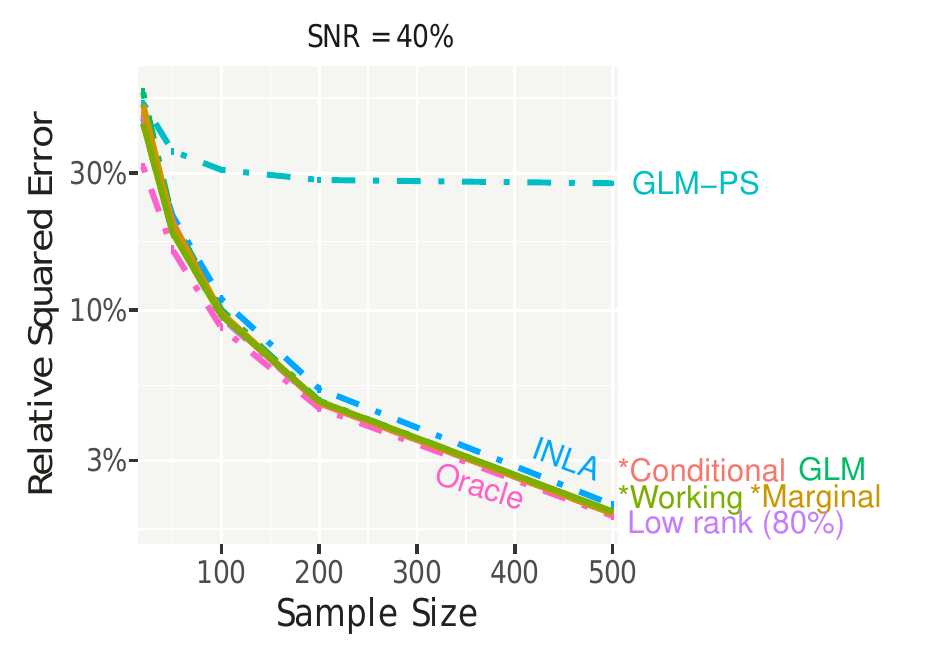}
    \\
    \includegraphics[width=0.44\textwidth]{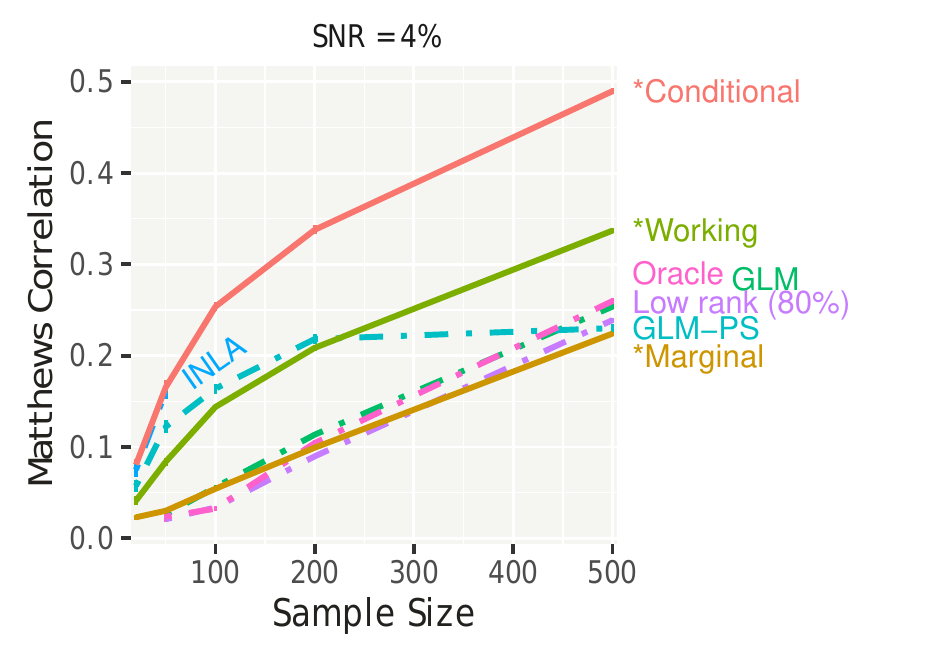}
    & \includegraphics[width=0.44\textwidth]{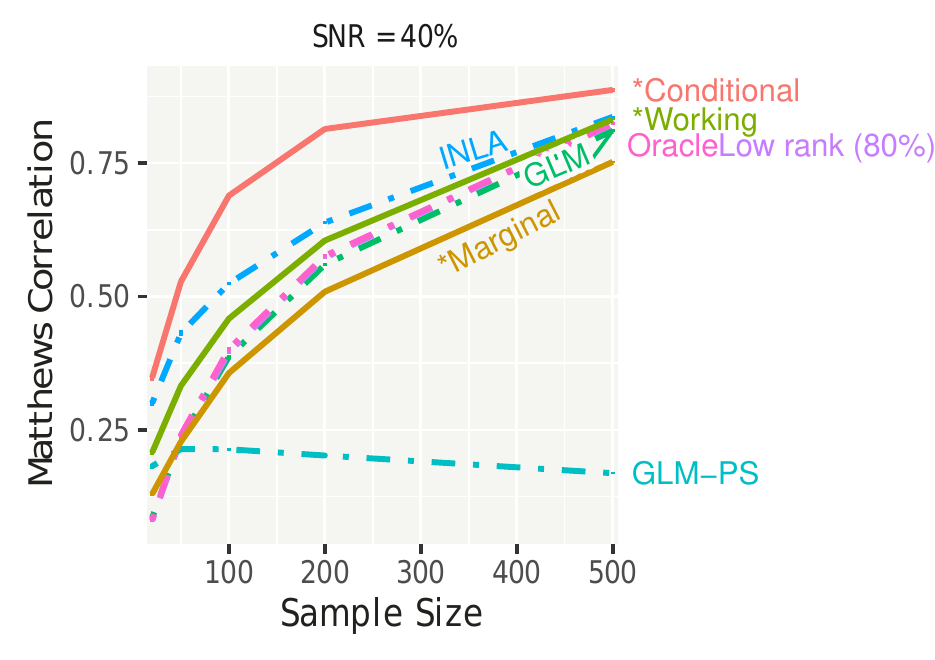}
    \\
    \includegraphics[width=0.44\textwidth]{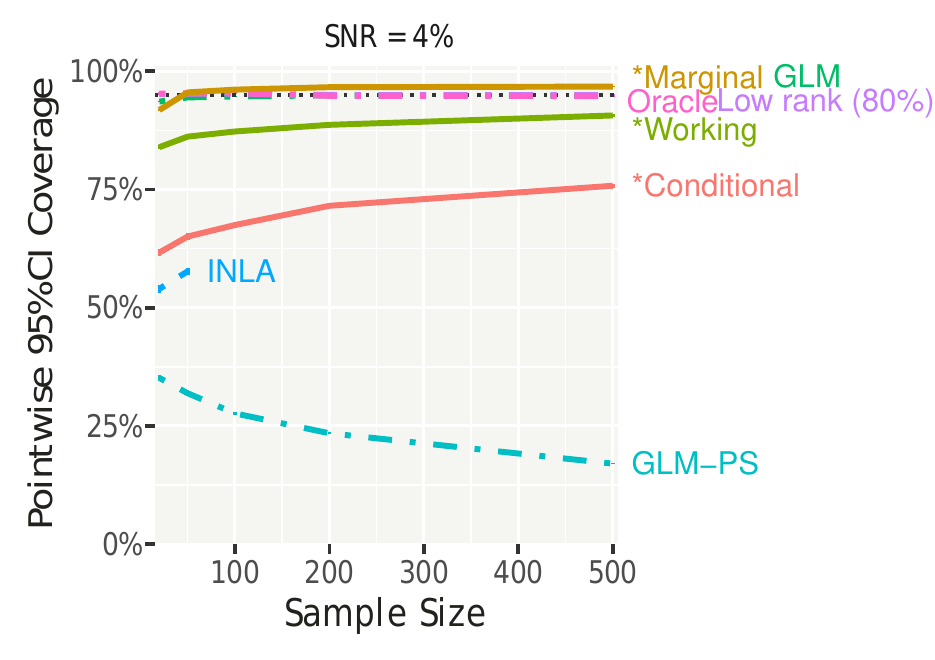}
    & \includegraphics[width=0.44\textwidth]{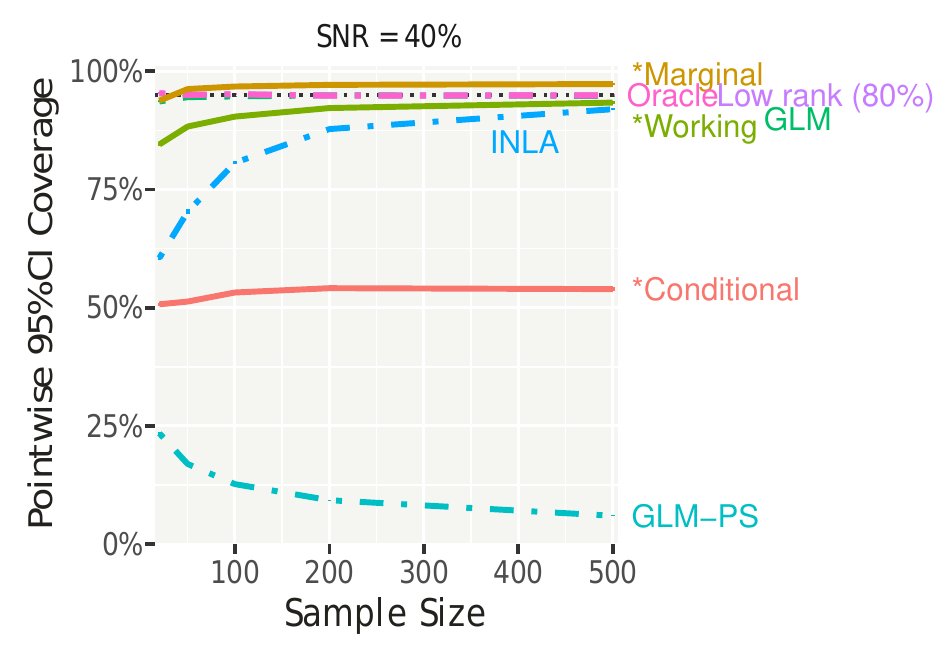}
  \end{tabular}
  \caption{Simulation results focusing on estimation and inferential
    accuracy. In the ``Relative Squared Error'' graphics, $y$-axes are
    logarithmically scaled. CI--posterior credible interval. Methods
    developed in the main text are indicated by
    asterisks. \protect{\label{fig:sim-full-results}}} 
\end{figure}

This appendix includes additional graphics that further unpack results
from the simulation study presented in the main
text. Fig. \ref{fig:sim-mean-fields} shows example posterior mean 
field estimates for all methods except the standard general linear
model analysis (GLM). Estimation differences are particularly apparent 
in the low signal-to-noise ratio setting (SNR = 4\%). Though it is
difficult to tell visually, the Oracle estimate is closest to the
Truth in a squared error sense in both
examples. Fig. \ref{fig:sim-full-results} shows extended
simulation results for select metrics.

\section{Additional ABCD study data description, analysis, and
  results}

Data collection and processing has been harmonized across 21 research
sites in the continental United States.
Details regarding study design and
recruitment \cite[][]{garavan2018recruiting},
neurocognitive assessment \cite[][]{luciana2018adolescent}, and
neuroimage acquisition \cite[][]{casey2018adolescent} are available in
published literature.

Data used in the preparation of this article were obtained from the
Adolescent Brain Cognitive Development (ABCD) Study
(\url{https://abcdstudy.org}), held in the NIMH Data Archive
(NDA). This is a multisite, longitudinal study designed to recruit
more than 10,000 children age 9-10 and follow them over 10 years into
early adulthood. The ABCD Study is supported by the National
Institutes of Health and additional federal partners under award
numbers
U01DA041048, U01DA050989, U01DA051016, U01DA041022, U01DA051018, 
U01DA051037, U01DA050987, U01DA041174, U01DA041106, U01DA041117,
U01DA041028, U01DA041134, U01DA050988, U01DA051039, U01DA041156,
U01DA041025, U01DA041120, U01DA051038, U01DA041148, U01DA041093,
U01DA041089, U24DA041123, U24DA041147. A full list of supporters is
available at \url{https://abcdstudy.org/federal-partners.html}. A
listing of participating sites and a complete listing of the study
investigators can be found at
\url{https://abcdstudy.org/consortium_members/}.
ABCD consortium investigators designed and implemented the study
and/or provided data but did not necessarily participate in analysis
or writing of this report. This manuscript reflects the views of the
authors and may not reflect the opinions or views of the NIH or ABCD
consortium investigators.

\subsection{fMRI preprocessing}

Preprocessing of the fMRI task data was accomplished through use of a
published standardized pipeline
\cite[see][``Supplemental FMRIPrep Methods'']{sripada2021brain}.
Briefly, time series acquisitions were filtered with a 0.005 Hz high
pass filter and spatially smoothed with a surface-based 2 mm
kernel. Within patient task-based modeling was accomplished using
tools from FSL \cite[][]{jenkinson2012fsl}, removing high motion time
points (frame-wise displacement $> 0.9$ mm) from the data. Additional
regressors of no interest included 24 total motion parameters (linear
and quadratic terms for each of six estimated motion
parameters---three rotational, three translational---and
their derivatives); five white matter principal components estimated
with CompCor software \cite[][]{behzadi2007component}; and five
cerebrospinal fluid principal components also estimated with CompCor.
Contrast images we analyzed in the present paper were derived from the
results of these first-level task-based models.

\subsection{Demographics and description of cohort}

\begin{table}[htb]
\centering
\caption{Demographic information for children in our
  sample. Continuous covariates are summarized by their mean, standard
  deviation and interquartile range; categorical covariates are
  summarized by percentage of the sample in the respective category.
  \protect{\label{tab:abcd-subset-demographics}}}
\vspace*{1em}
\begin{tabular}{ r c c c }
 \multicolumn{1}{l}{Descriptor} & Mean & SD & IQR \\
  \hline
  0-Back Accuracy       & 0.87 & 0.09 & 0.11 \\
  2-Back Accuracy       & 0.80 & 0.08 & 0.12 \\
  \{0 - 2\}-Back Difference & 0.07 & 0.09 & 0.12 \\  
  Age (yrs)             & 9.99 & 0.62 & 1.08 \\
  Fluid IQ              & 0.29 & 0.75 & 0.97 \\
 & Percentage & \\
  \hline
 \multicolumn{1}{l}{Child Gender} & \\
  Female & 50.8\% \\ 
  Male & 49.2\% \\
 \multicolumn{1}{l}{Child Race/Ethnicity} & \\
  Asian & 2.4\% \\ 
  Black & 8.8\% \\ 
  Hispanic & 17.1\% \\ 
  Other & 9.5\% \\ 
  White & 62.2\% \\ 
 \multicolumn{1}{l}{Household Income (US\$/yr)} & \\
  $<$ 50K & 22.4\% \\
  50K--100K & 30.7\% \\ 
  $\geq$ 100K & 46.9\% \\ 
 \multicolumn{1}{l}{Parental Education} & \\ 
  $<$ HS Diploma & 2.1\% \\
  HS Diploma/GED & 5.3\% \\
  Some College & 23.0\% \\ 
  Bachelor & 28.6\% \\ 
  Post Graduate Degree & 41.0\% \\ 
 \multicolumn{1}{l}{Parental Marital Status} & \\
  Married Household & 76.0\% \\ 
  Unmarried Household & 24.0\% \\
  \hline
\end{tabular}
\end{table}

For this illustration we will work exclusively with data from a subset
of 3,267 children in the baseline cohort that were scanned while
performing an n-back task
\cite[][]{barch2013function, casey2018adolescent} with pictures of
human faces expressing emotion as stimuli.
The n-back task has enjoyed wide use in the neuropsychological and
imaging community for its relationship with executive function
and as a correlate of working memory processes
\cite[e.g.,][]{jansma2000specific, owen2005nback,
  jaeggi2010concurrent}.
Our subsample of children is limited to those who scored
at or above 60\% correct on both 0-back and 2-back task conditions.

Table \ref{tab:abcd-subset-demographics} gives a summary of the
demographic information for this sample. 
As a final note,
the ABCD study more broadly contains imaging data acquired from
siblings. Around 20\% of families
in the ABCD release 2.0 baseline data have two or more children
enrolled in the  study. This might additionally suggest the need for
an analysis with random family effects.
We avoid this issue entirely here: the cohort that we analyze contains
data from only one child per family in our subset.
While our method is capable of estimating effects like this in
general, it would be very slow computationally to give a fully
Bayesian treatment to a large number of random spatial effects. 
A more specific tool could be built on top of the methods we present
here to include such variance components and/or treat them as nuisance
parameters.

\subsection{Comparison with standard imaging software}

\begin{figure}[H]
  \centering
  \includegraphics[width=0.8\textwidth]{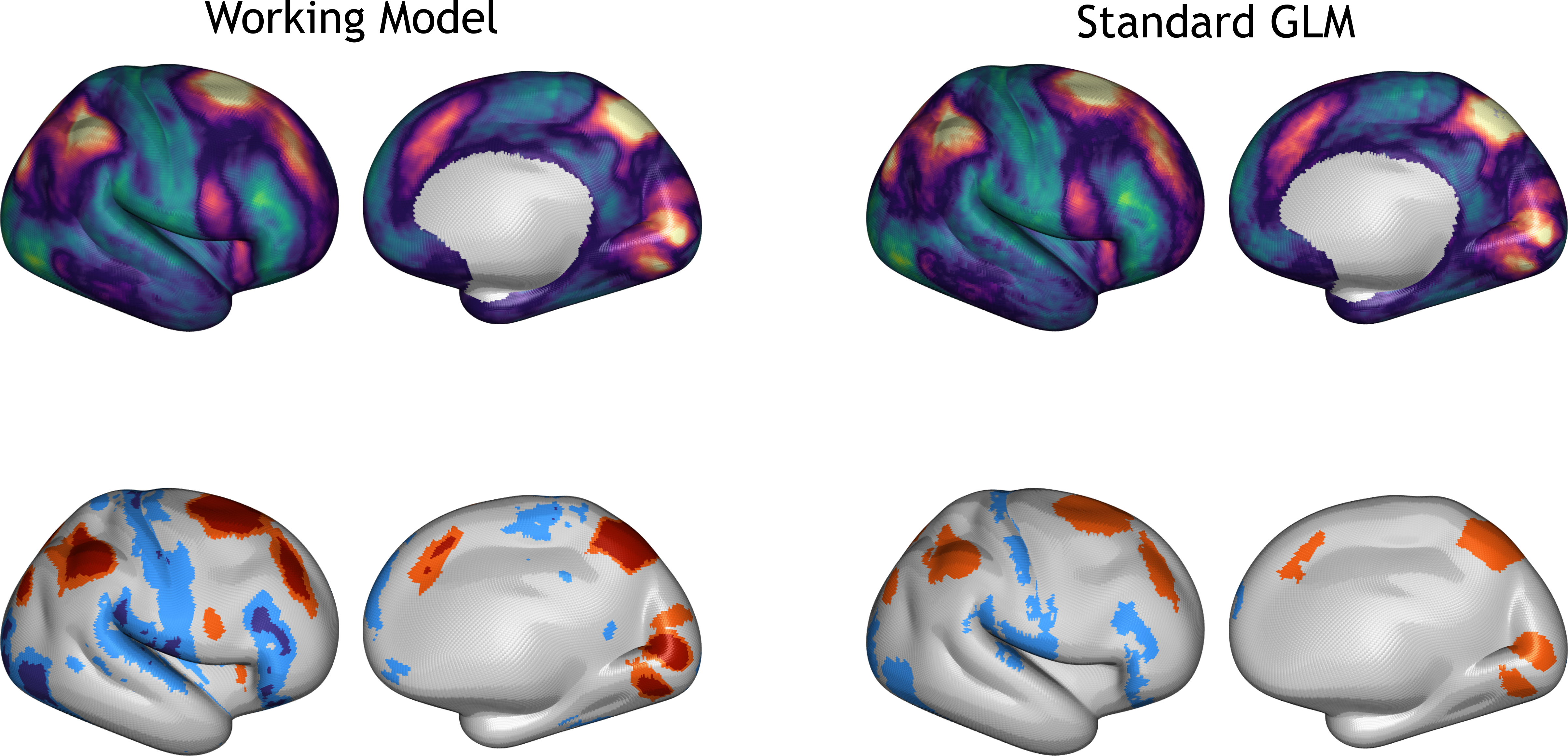}
  \caption{Comparison with AFNI software: spatial intercept estimation
    and inference. Surface images on the \emph{left} are reproduced from
    Figs. 3 and 4 in the main text; images on the \emph{right}
    parallel this analysis with AFNI software. The \emph{top} row
    shows the estimated spatially varying intercept, while the
    \emph{bottom} row shows example inference for regions where we might
    infer $|\beta_0(\spc)| > 0.4$ for at least some $\spc$. Reds
    correspond to functional activations and blues to functional
    deactivations.
    \protect{\label{fig:afni}}}
\end{figure}

We further compared results from our working model analysis of n-back
task contrast data (see Section 4 in the main text) with
paralleled results using the popular AFNI software package
\cite[][]{cox1996afni} and standard methods.
Fig. \ref{fig:afni} summarizes the results of this comparison. 
To restate, the bottom left panels of Fig. \ref{fig:afni} show
areas where we may infer $|\beta_0(\spc)| > 0.4$ using an 80\%
posterior credible band. Regions of darker color mark core areas where
our analysis suggests the probability that $|\beta_0(\spc)| > 0.4$ is
at least 80\% simultaneously for all vertices in those
regions. Regions of lighter color extend the core regions to areas
where the posterior mean estimate is that $|\beta_0(\spc)| > 0.4$. In
all cases, the inference is directly interpretable: it is spatially
and probabilistically precise conditional on the fitted model.

The bottom right panels of Fig. \ref{fig:afni} show regions
identified using AFNI software and tests of null hypotheses that
$|\beta_0(\spc)| \leq 0.4$. Regions were identified using contiguous
vertex cluster extent adjustment methods by applying sequential
``height'' and ``extent'' $p$-value thresholds. Briefly, a vertex-wise
$p < 0.05$ threshold was applied to a test statistic image (``height''
threshold). Conceptually, thresholding creates a binary labeled vertex
image where some labeled vertices form contiguous clusters on a
surface-defining mesh. Clusters were then thresholded based on their
area on the cortical surface (``extent'' threshold) to retain only
clusters with area at least 198 mm$^2$. The extent threshold was
chosen to control the expected cluster-wise false positive rate at a
$5\%$ level, as determined following output from AFNI program 
\texttt{slow\_surf\_clustsim.py} (with spatial smoothing parameter set
to 6 mm).
The null hypothesis testing probabilities are conditional on
the initial height threshold and other assumptions made by AFNI
programs internally.
Interpretation of resulting inferences are fundamentally different
from those based on posterior credible bands---all of the typical
arguments, pro or con, about posterior inference and null hypothesis
testing apply.
In addition, with cluster correction-based inference, there is little
in the way of implied spatial precision in the pattern of results.

\subsection{Additional brain region-level coefficient summaries}

\begin{figure}[!htb]
  \centering
  \includegraphics[width=0.8\textwidth]{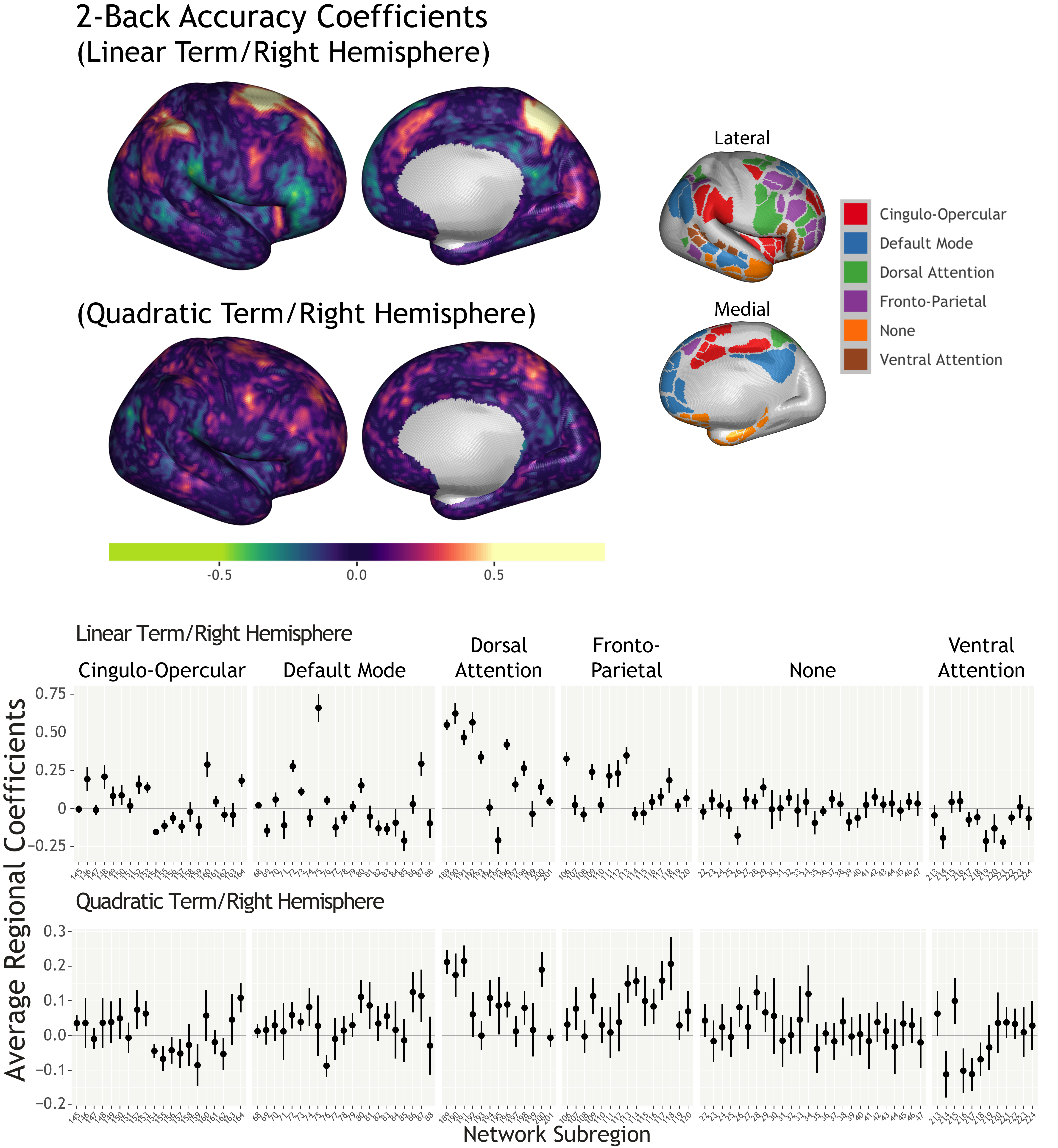}
  \caption{Coefficient summary for 2-back condition accuracy rate
    (linear and quadratic terms). The overall format of the figure is
    the same as in Fig. 4 in the main text.}
  \label{fig:results-c2back}
\end{figure}

Here we complete our report of demographic effects on the 2- vs 0-back
task contrast data from the ABCD study. We again summarize results
from the right hemisphere by averaging over all vertices within brain
regions from the Gordon 2016 cortical surface parcellation
\cite[][]{gordon2016generation}. Figures follow the same format as the
primary model intercept and 2-back accuracy rate results figures from
the main text. Results in the left hemisphere were generally highly
symmetric.

As noted in the main text, covariates were chosen on the basis of
known associations with n-back task accuracy
\cite[][]{pelegrina2015normative} and through preliminary exploratory
analyses.
Exploratory analyses served to help us
visualize and understand several important aspects of the data.
First, we observed modest but present nonlinear patterns in the
relationship between the contrast data and 2-back accuracy. Preferring
simplicity here, we found that these trends were reasonably well
characterized by a quadratic model for 2-back accuracy.
Including this term in the analysis resulted in a total of $P = 24$
predictors including the global intercept.
We scaled each continuous covariate by two standard deviations
\cite[][]{gelman2008scaling} so that resultant coefficient images are
more directly comparable with coefficient images for categorical
covariates.

Similarly to Fig. 4 in the main text,
Fig. \ref{fig:results-c2back} summarizes results for 
the effect of 2-back accuracy rate on the 2- vs 0-back
contrast. In the figure, coefficients for the linear and quadratic
accuracy terms reflect the expected change in activation between ten
year old female children scoring 96\% and 80\% correct on the 2-back
condition, respectively, holding all other demographic covariates
constant. Our analysis suggests high spatial overlap between the
intercept and areas where average activation increased linearly with
increasing 2-back accuracy (confer from main text
Fig. 4 and Fig. \ref{fig:results-c2back}).
Interestingly, however, the quadratic accuracy term largely seems to
reflect areas where average activation increased supra-linearly with
increasing 2-back accuracy. Based on our analysis, these areas are
more constrained to regions associated with the Dorsal-Attention and
Fronto-Parietal networks (Fig. \ref{fig:results-c2back}).

\begin{figure}[!h]
  \centering
  \includegraphics[width=0.75\textwidth]{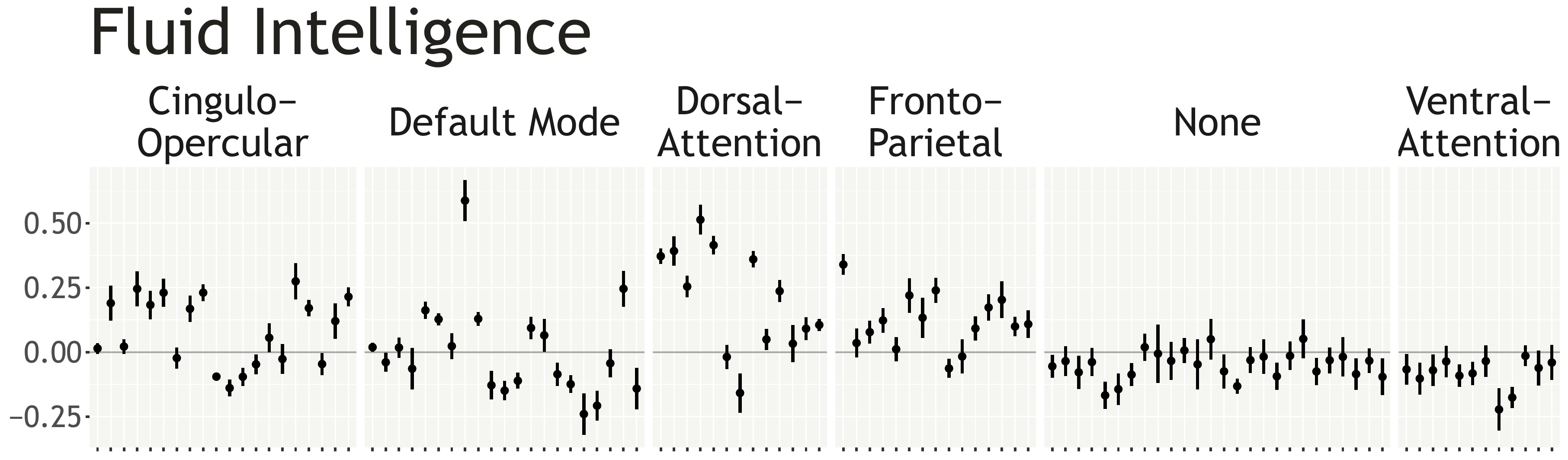}
  \caption{Regional average coefficients: fluid intelligence, linear
    term. Consistent with previous studies
    \cite[e.g.,][]{preusse2011fluid, li2021neural}, fluid intelligence
    is positively correlated with task-related activation in
    functionally relevant cingulo-opercular, dorsal-attention, and
    fronto-parietal network regions.}
\end{figure}

\begin{figure}[!h]
  \begin{minipage}{\textwidth}
  \centering
  \begin{tabular}{ c }
    \includegraphics[width=0.75\textwidth]{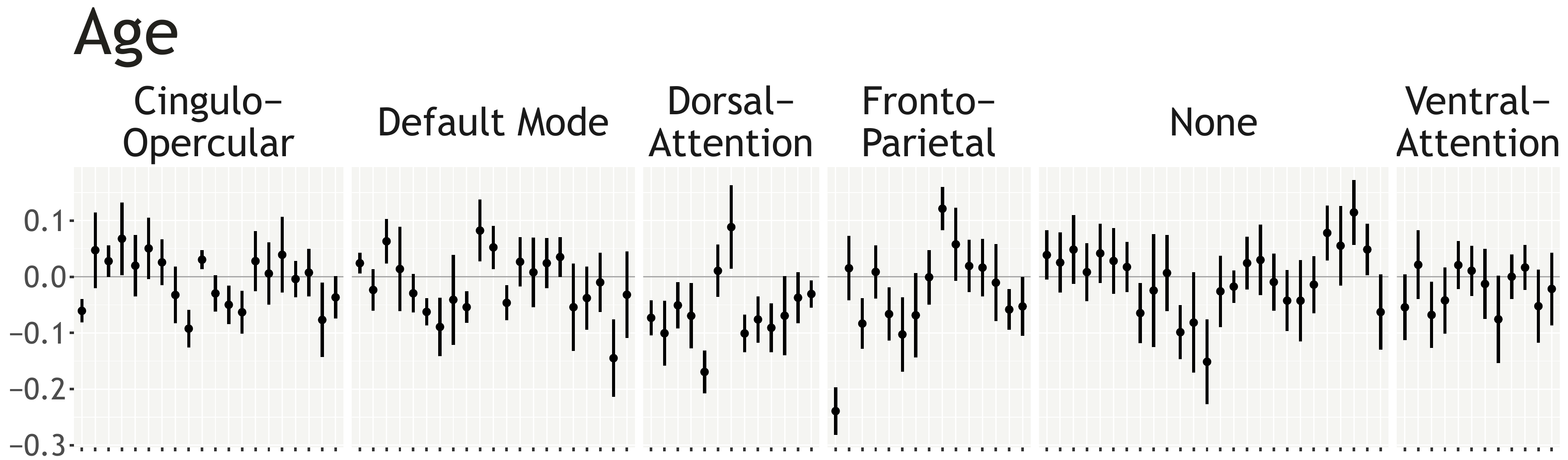} \\
    \includegraphics[width=0.75\textwidth]{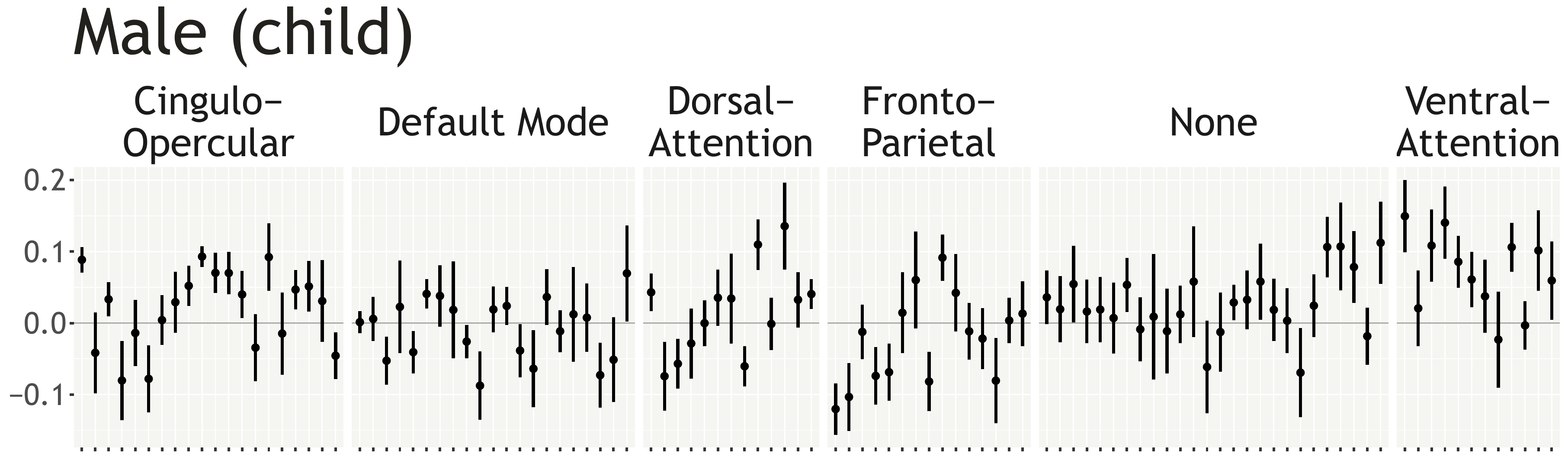} \\
    \includegraphics[width=0.75\textwidth]{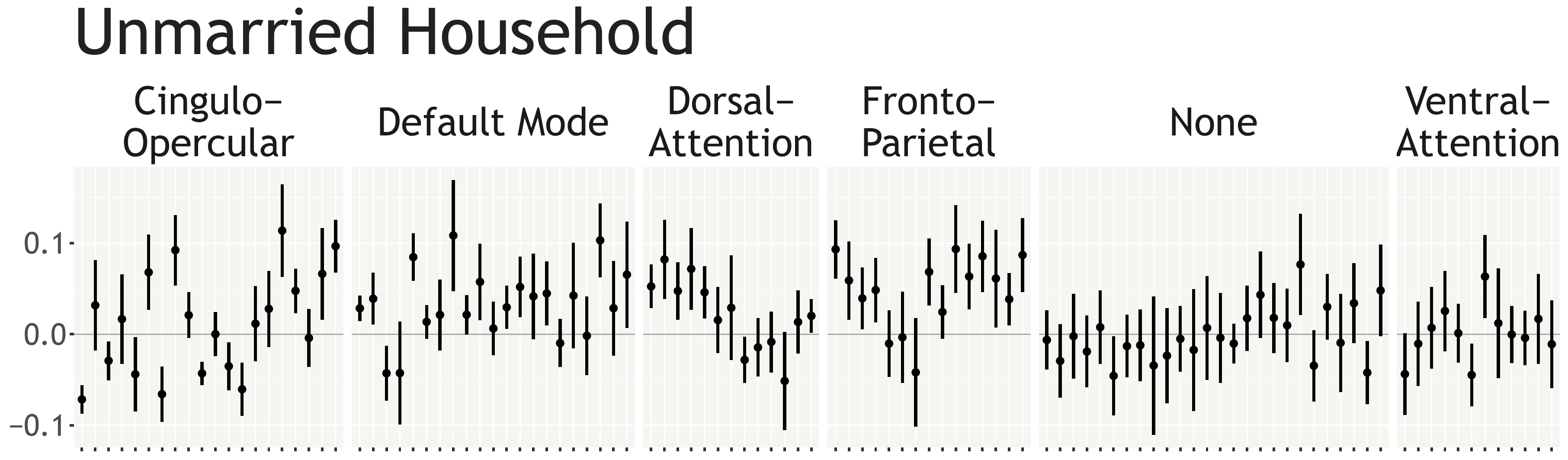} \\
    \includegraphics[width=0.75\textwidth]{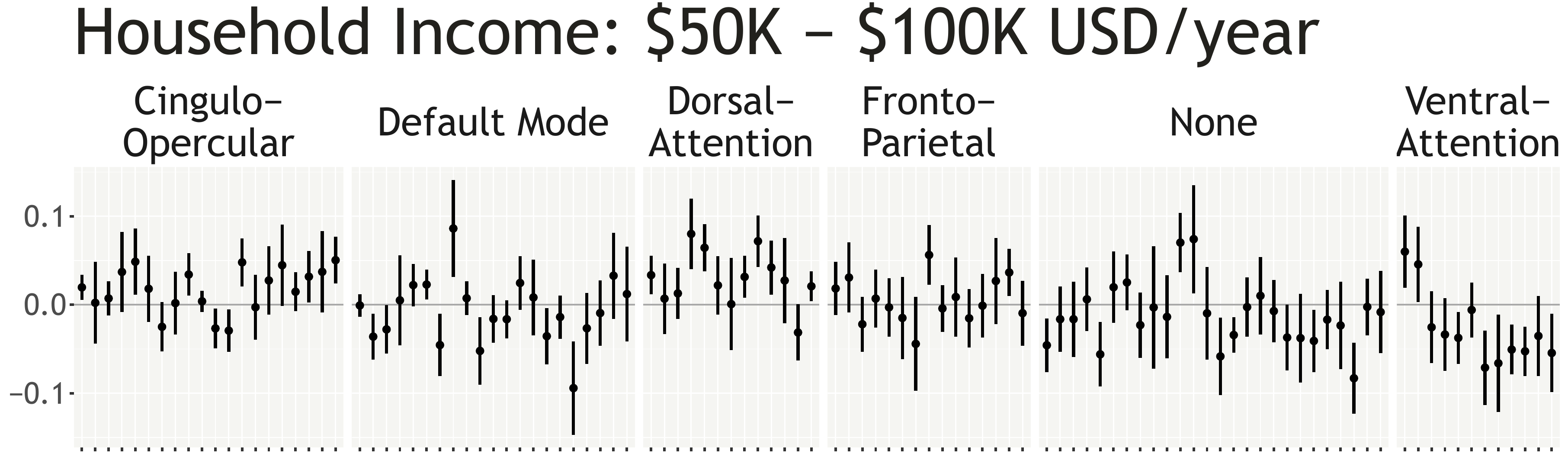} \\
    \includegraphics[width=0.75\textwidth]{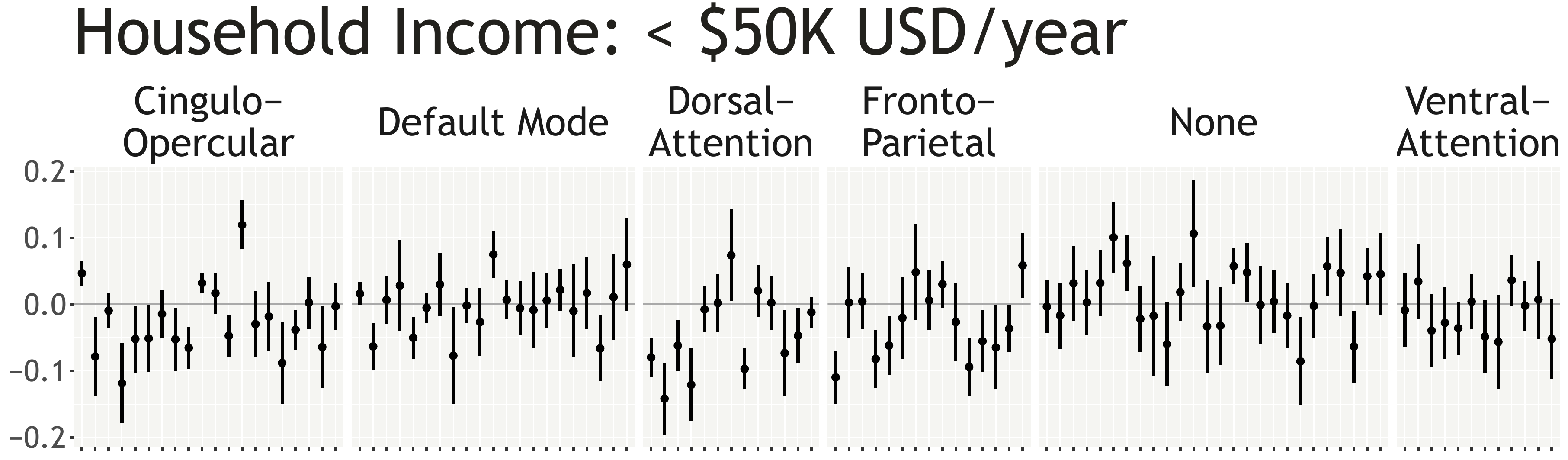} 
  \end{tabular}
  \end{minipage}
  \vspace*{1em}
  \caption{Regional average coefficients: additional demographic
    covariates. The majority of these effects are relatively small in
    magnitude with the notable exception of a negative association
    between child age and task-related activation in a functionally
    relevant fronto-parietal network region (region label: 106).}
\end{figure}

\begin{figure}[!h]
  \begin{minipage}{\textwidth}
  \centering
  \begin{tabular}{ c }
    \includegraphics[width=0.75\textwidth]{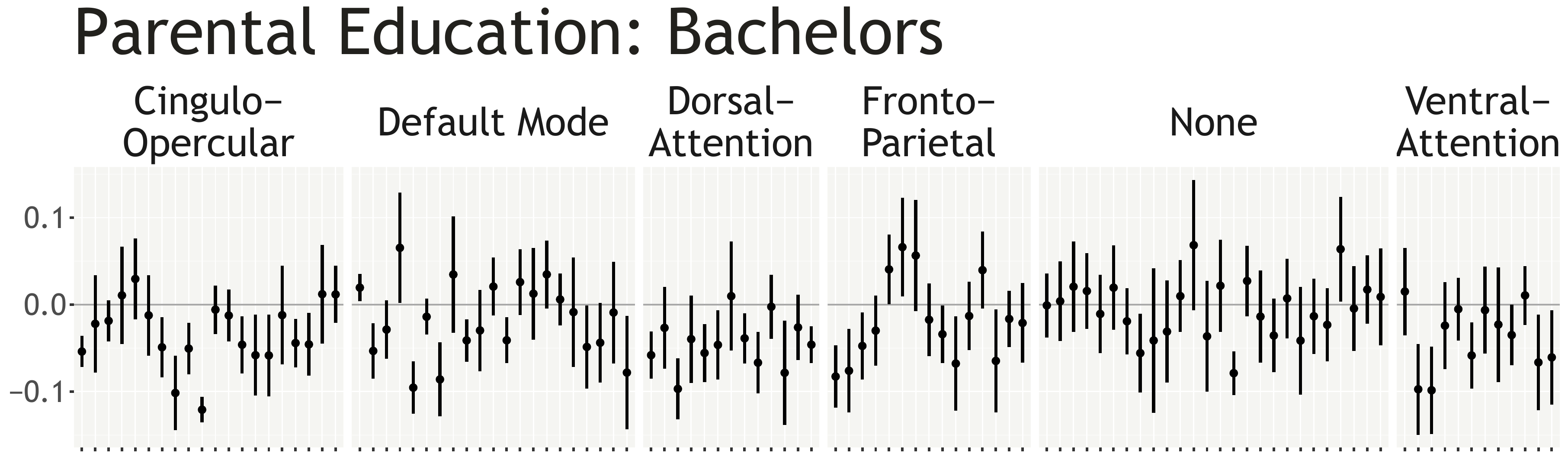} \\
    \includegraphics[width=0.75\textwidth]{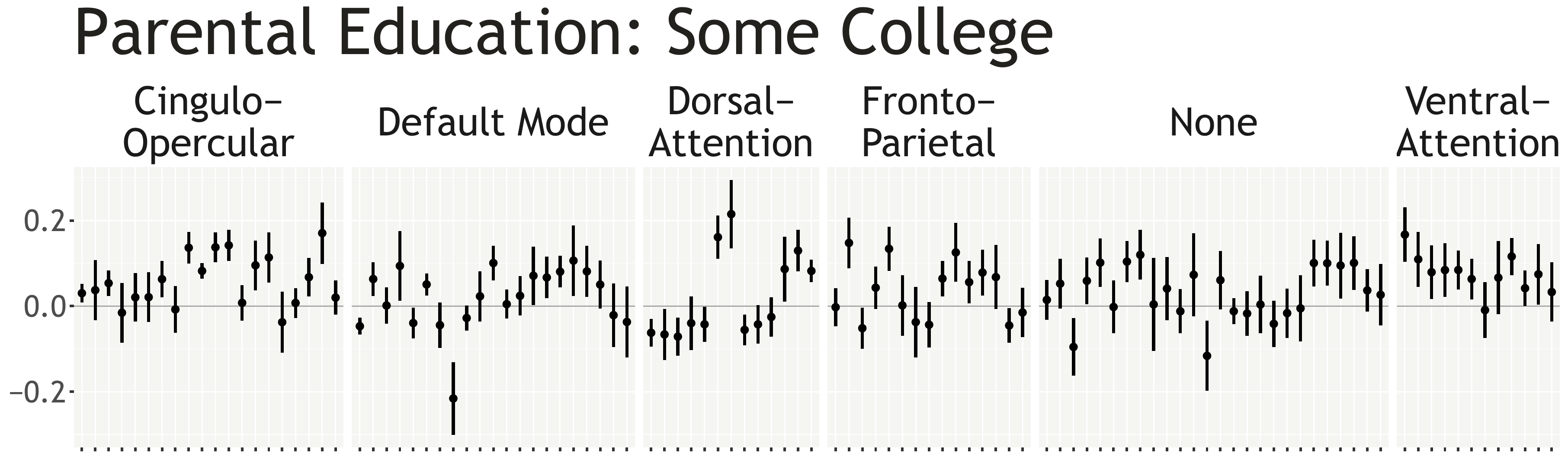} \\
    \includegraphics[width=0.75\textwidth]{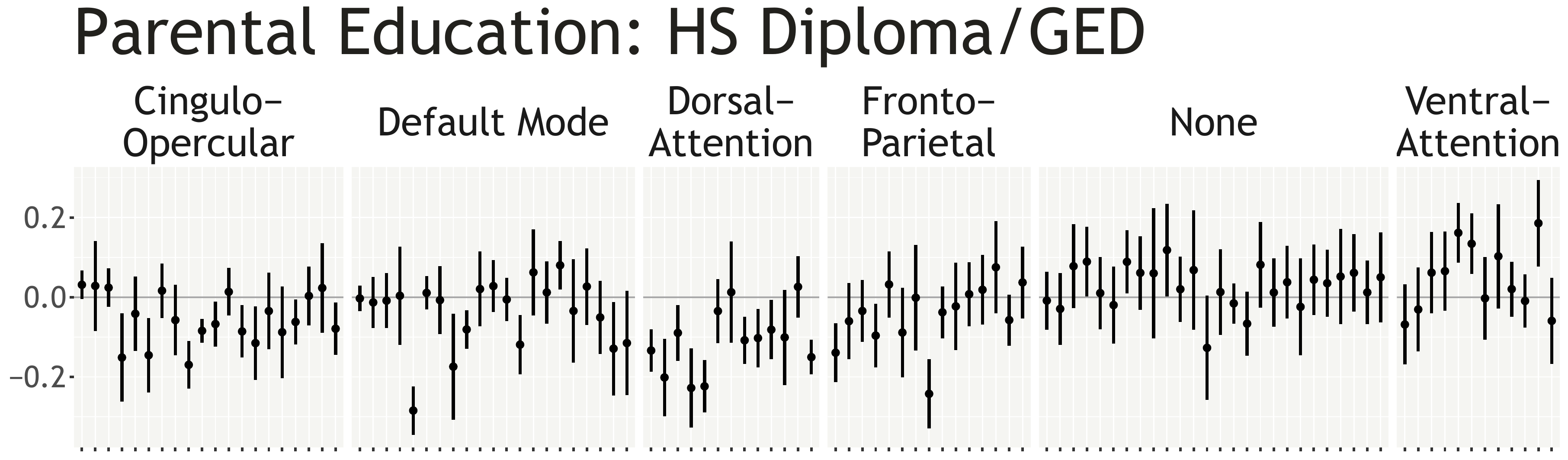} \\
    \includegraphics[width=0.75\textwidth]{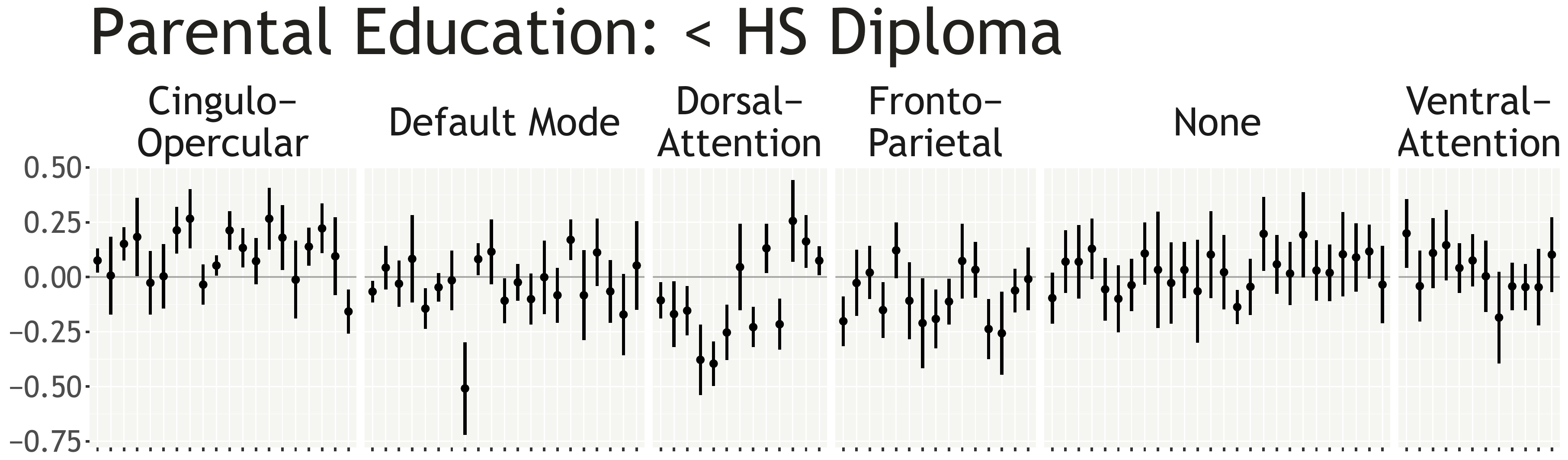}
  \end{tabular}
  \end{minipage}
  \vspace*{1em}
  \caption{Regional average coefficients: parental education (compared
    to a ``post-graduate degree'' reference group). The largest
    magnitude effects may suggest a pattern of decreased activation in
    functionally relevant dorsal-attention and fronto-parietal network
    regions in children of parents with less than ``some college''
    education.}
\end{figure}

\begin{figure}[!h]
  \begin{minipage}{\textwidth}
  \centering
  \begin{tabular}{ c }
    \includegraphics[width=0.75\textwidth]{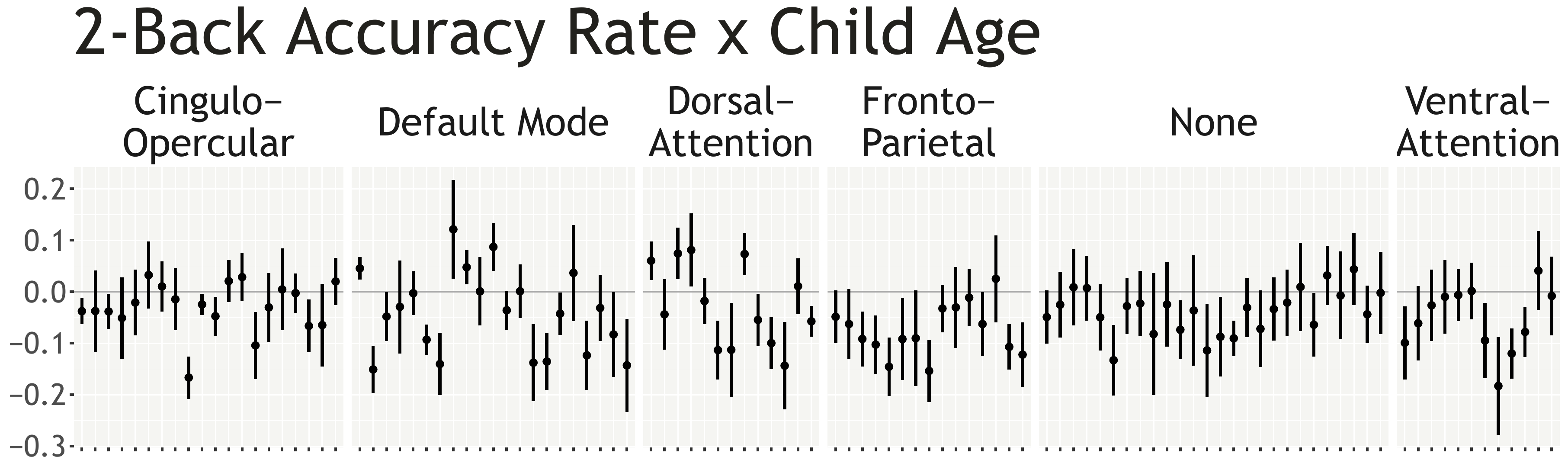} \\
    \includegraphics[width=0.75\textwidth]{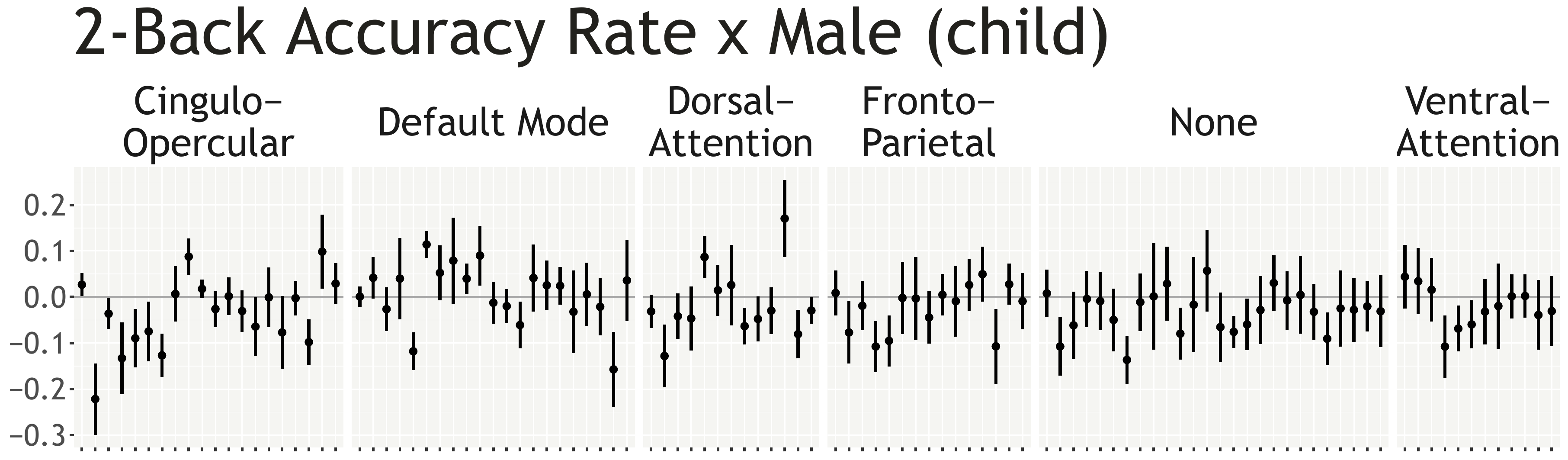} \\
    \includegraphics[width=0.75\textwidth]{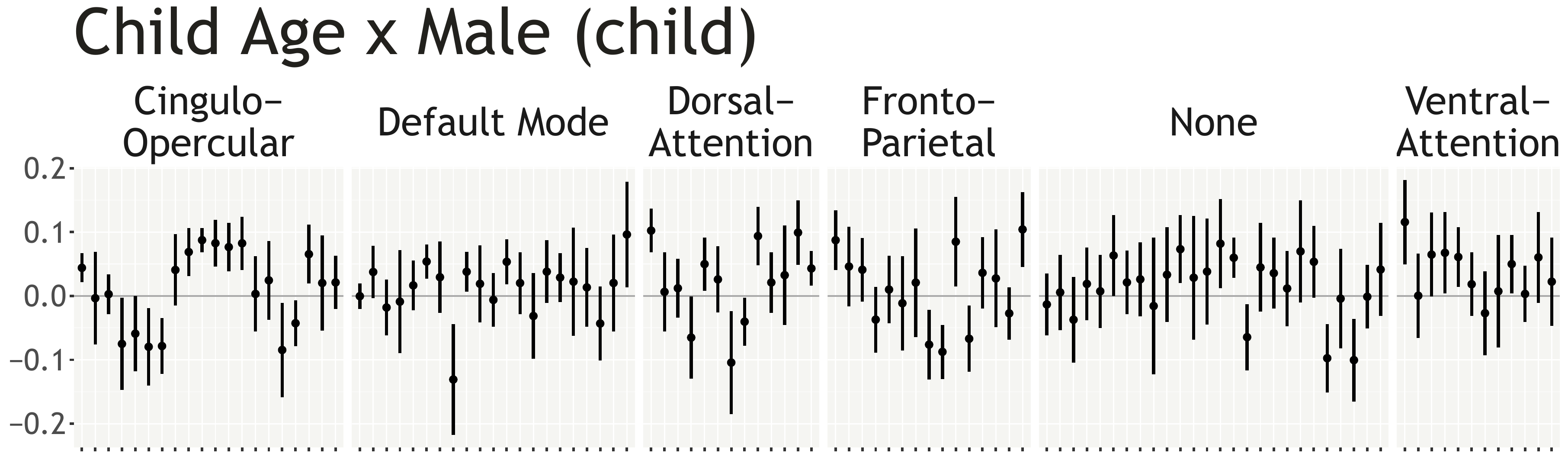}
  \end{tabular}
  \end{minipage}
  \vspace*{1em}
  \caption{Regional average coefficients: first-order interaction
    terms between 2-back accuracy, child age, and child sex. Most
    effects here are relatively small in magnitude.}
\end{figure}

\begin{figure}[!h]
  \begin{minipage}{\textwidth}
  \centering
  \begin{tabular}{ c }
    \includegraphics[width=0.75\textwidth]{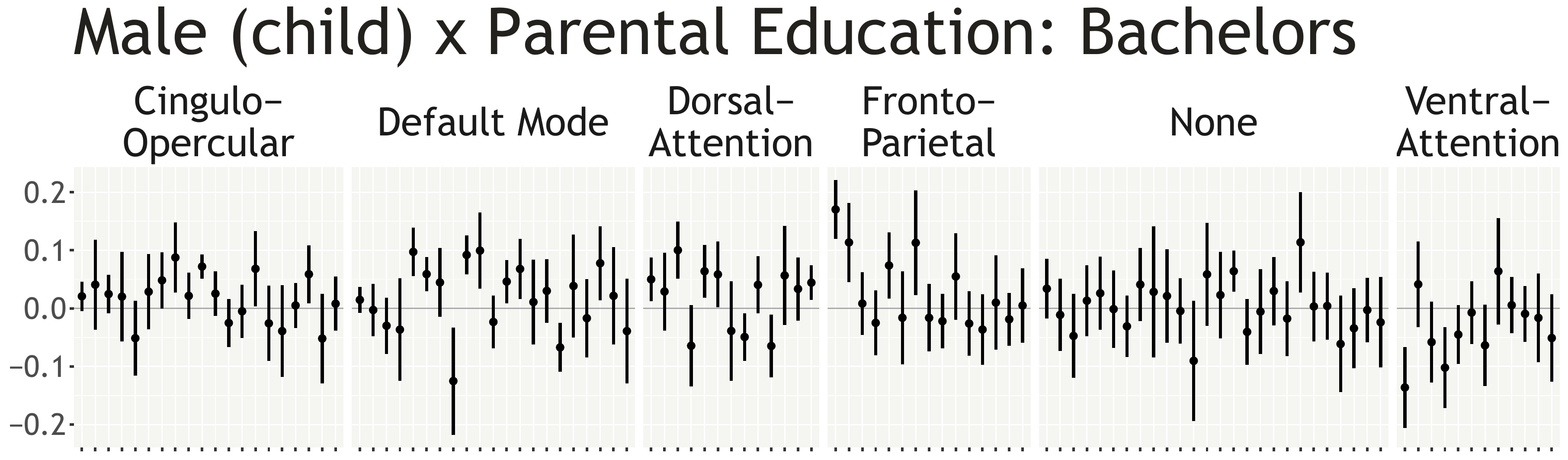} \\
    \includegraphics[width=0.75\textwidth]{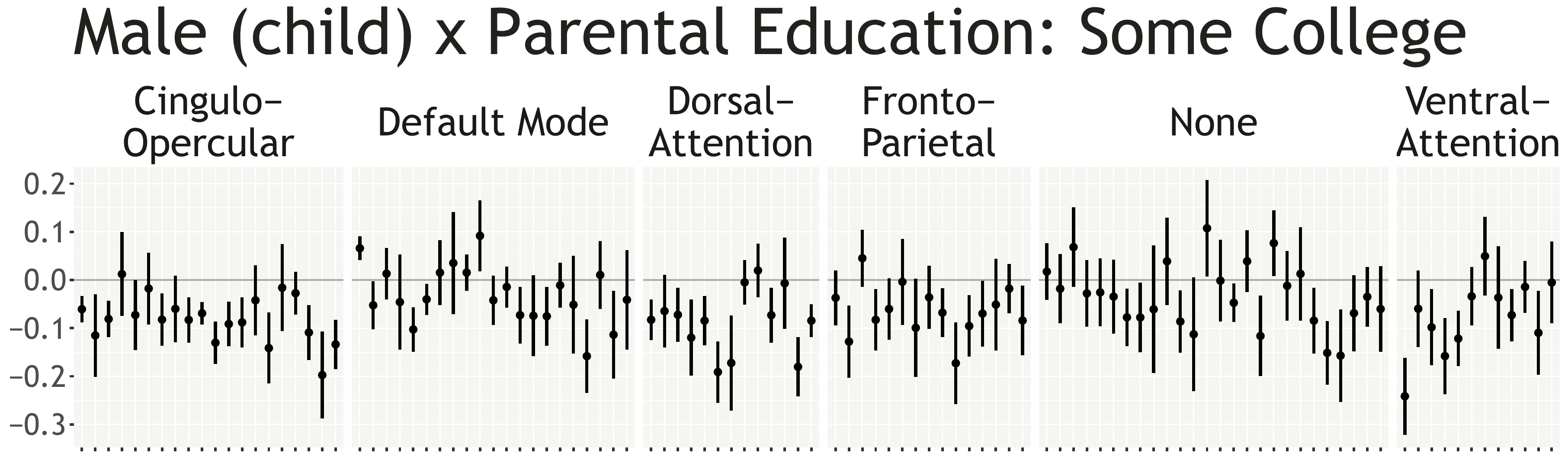} \\
    \includegraphics[width=0.75\textwidth]{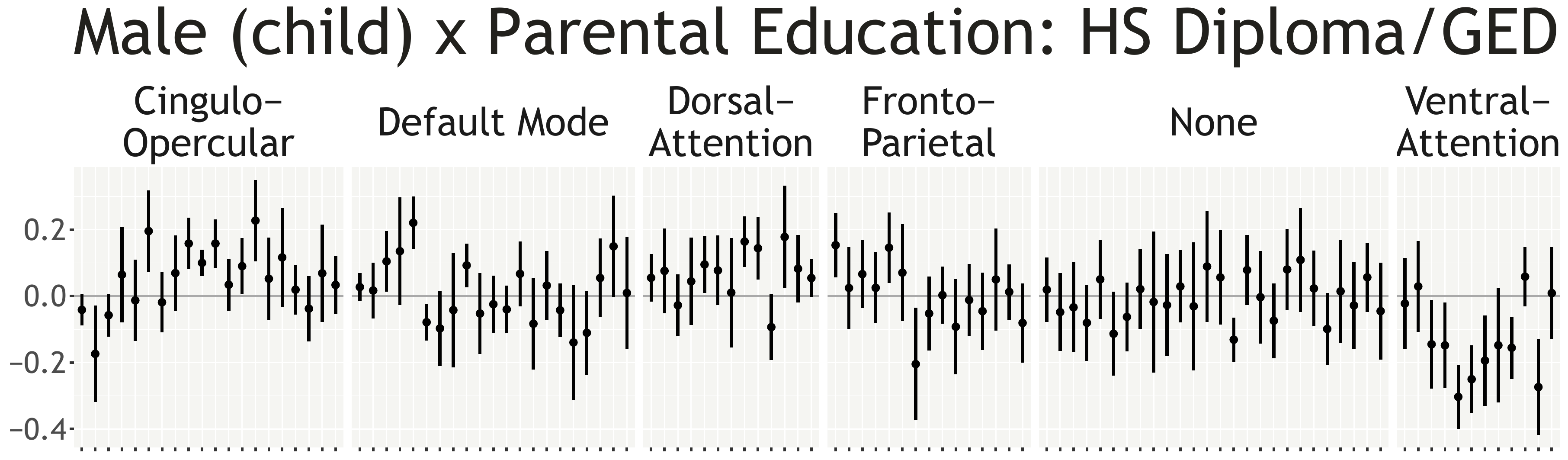} \\
    \includegraphics[width=0.75\textwidth]{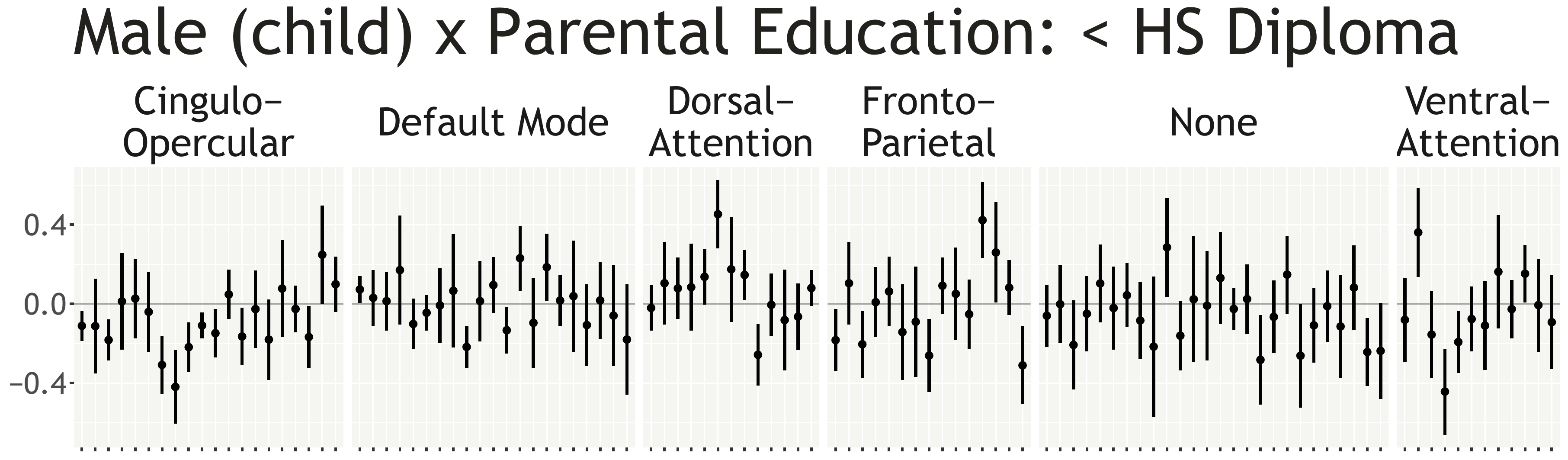}
  \end{tabular}
  \end{minipage}
  \vspace*{1em}
  \caption{Regional average coefficients: first-order interaction
    terms between child sex and parental education. No clear pattern
    of results is apparent here as with the parental education main
    effect terms.}
\end{figure}

\begin{figure}[!h]
  \begin{minipage}{\textwidth}
  \centering
  \begin{tabular}{ c }
    \includegraphics[width=0.75\textwidth]{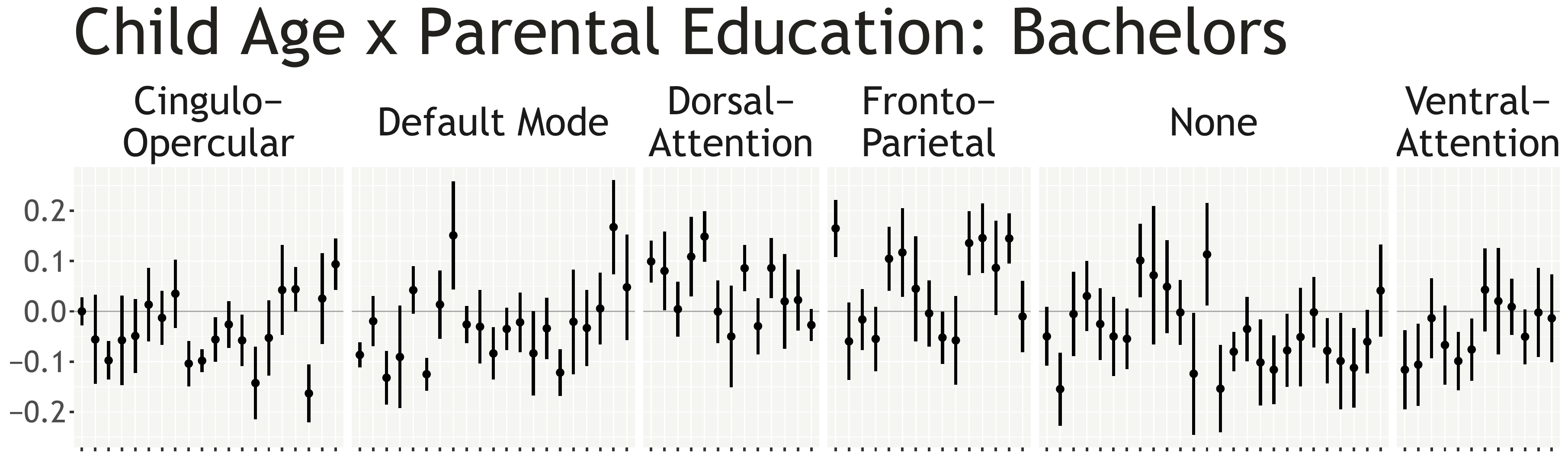} \\
    \includegraphics[width=0.75\textwidth]{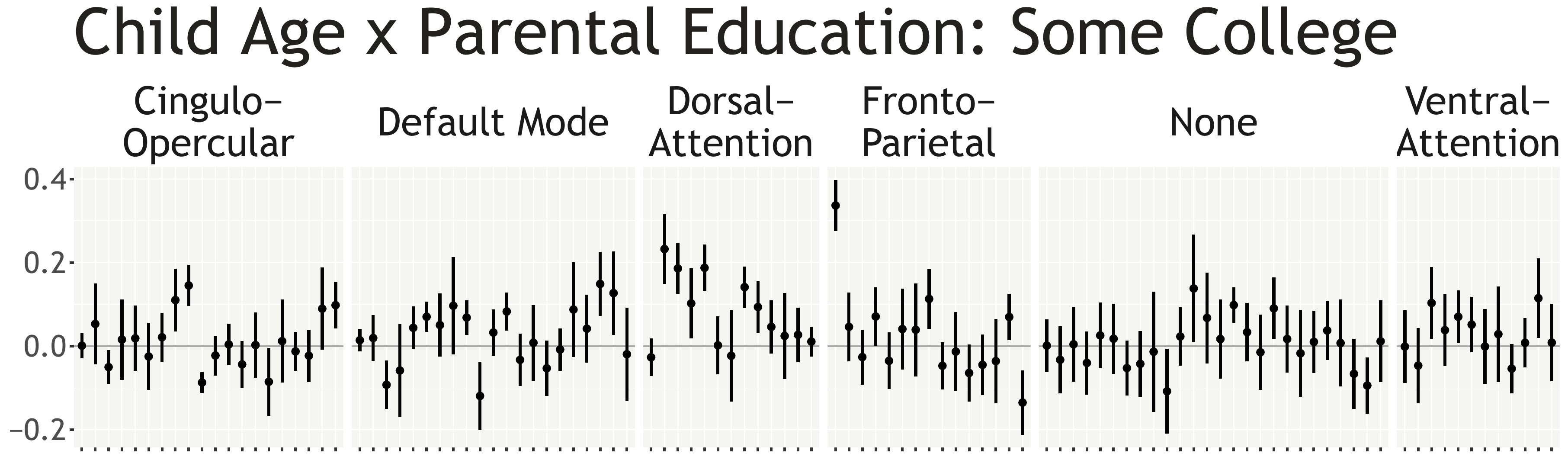} \\
    \includegraphics[width=0.75\textwidth]{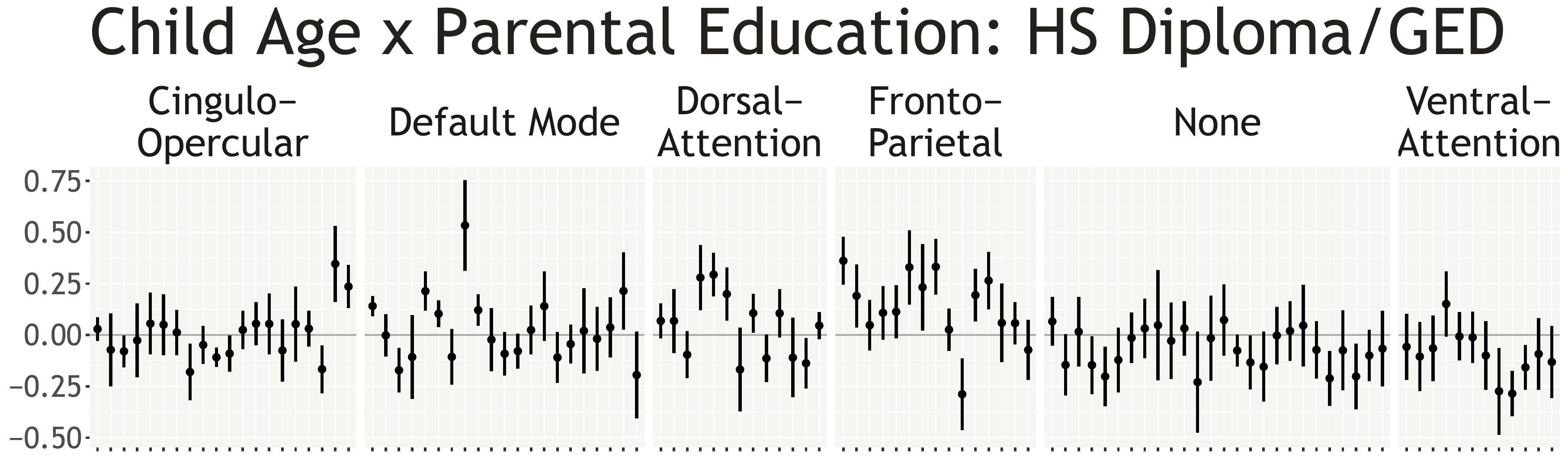} \\
    \includegraphics[width=0.75\textwidth]{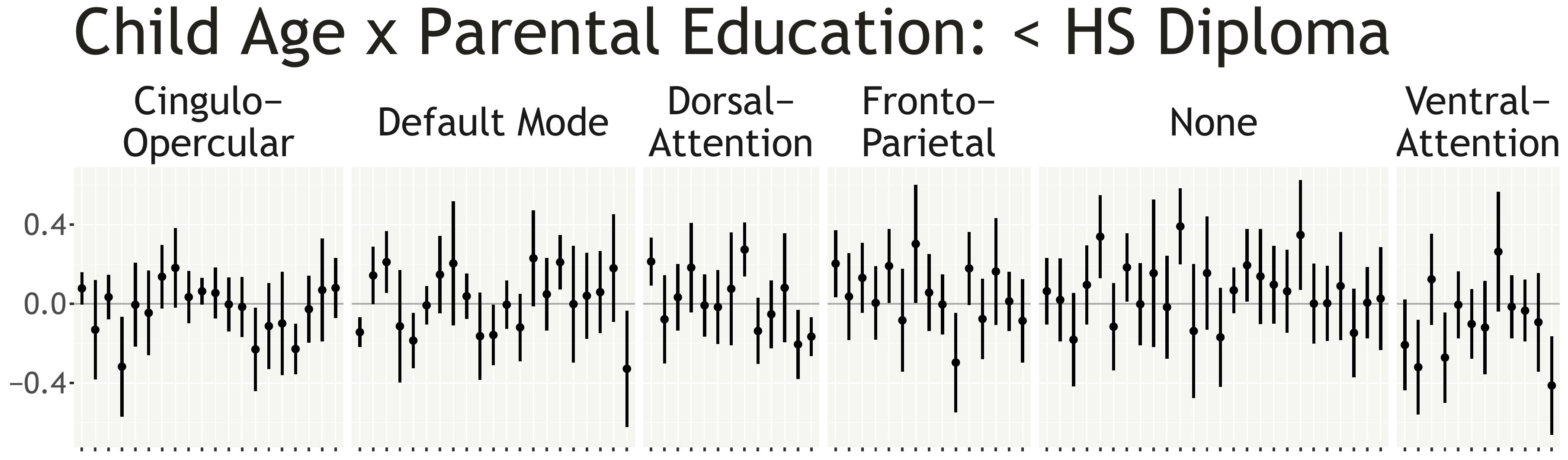}
  \end{tabular}
  \end{minipage}
  \vspace*{1em}
  \caption{Regional average coefficients: first-order interaction
    terms between child age and parental education. The uncertainty in
    many of these coefficients is relatively large, but there appears
    to be a consistent pattern of positive interactions in
    functionally relevant dorsal-attention network
    regions. Interpretation of this result is somewhat complicated by
    the general pattern of negative coefficients for the main effects
    of child age and parental education in these same regions.} 
\end{figure}

\begin{figure}[!htb]
  \begin{minipage}{\textwidth}
  \centering
  \begin{tabular}{ c c c c c }
    \makecell{Site 16 \\ ($n = 663$)} &
    \makecell{Site 14 \\ ($n = 319$)} &
    \makecell{Site 6 \\ ($n = 305$)} &
    \makecell{Site 2 \\ ($n = 304$)} &
    \makecell{Site 20 \\ ($n = 285$)} \\
    \includegraphics[width=0.15\textwidth]{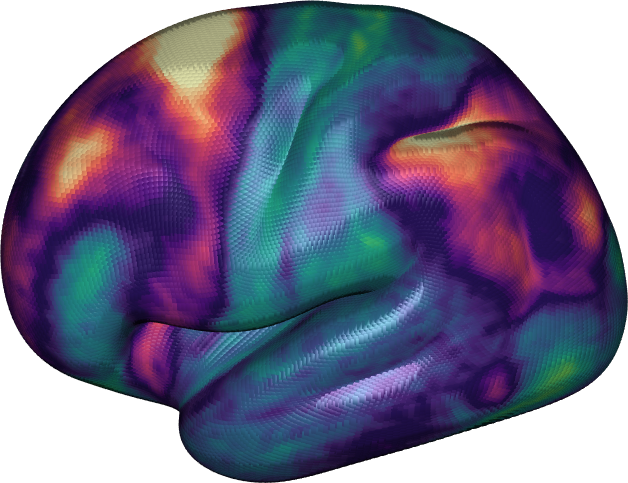}
    & 
    \includegraphics[width=0.15\textwidth]{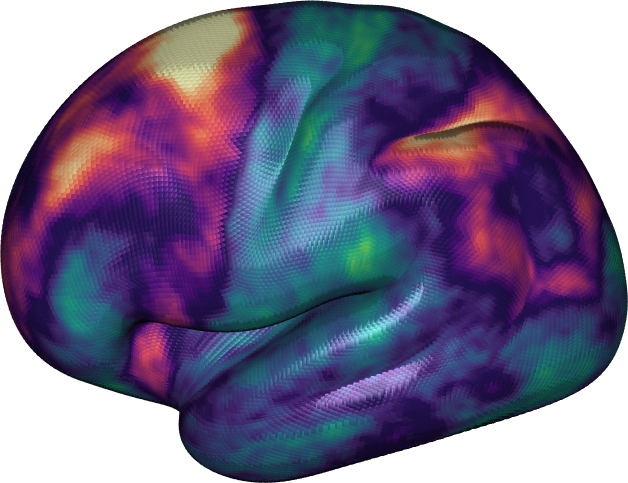}
    & 
    \includegraphics[width=0.15\textwidth]{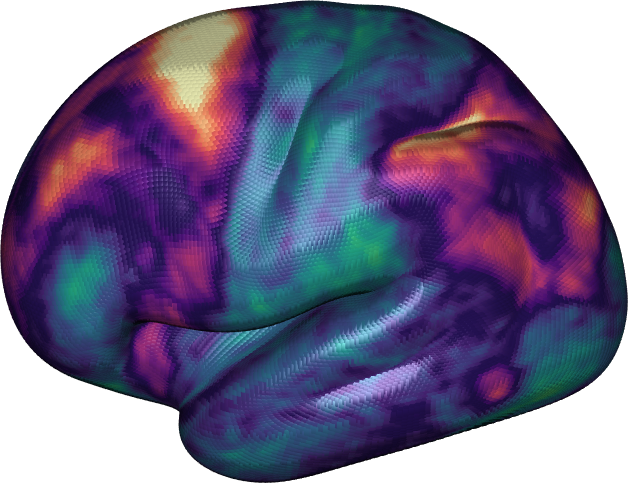}
    & 
      \includegraphics[width=0.15\textwidth]{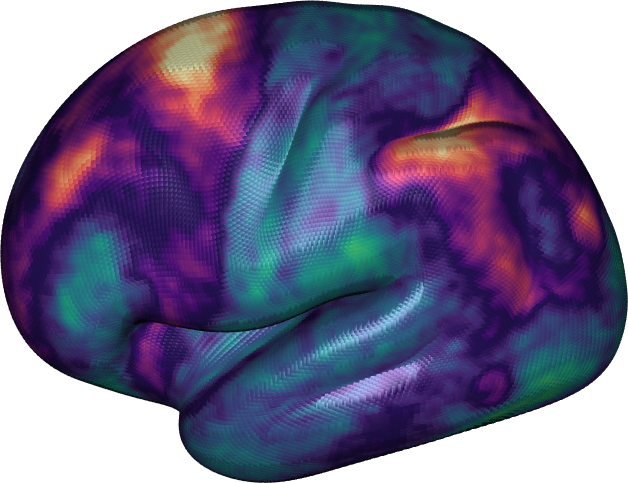}
      &
    \includegraphics[width=0.15\textwidth]{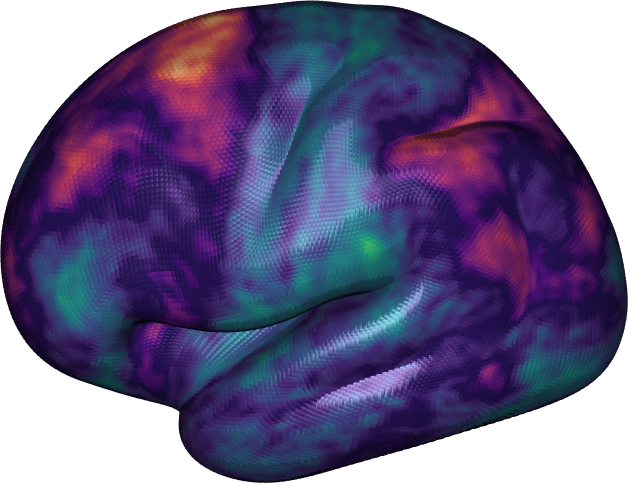}
    \\
    \makecell{Site 4 \\ ($n = 68$)} &
    \makecell{Site 8 \\ ($n = 68$)} &
    \makecell{Site 1 \\ ($n = 66$)} &
    \makecell{Site 12 \\ ($n = 65$)} &
    \makecell{Site 22 \\ ($n = 5$)} \\
    \includegraphics[width=0.15\textwidth]{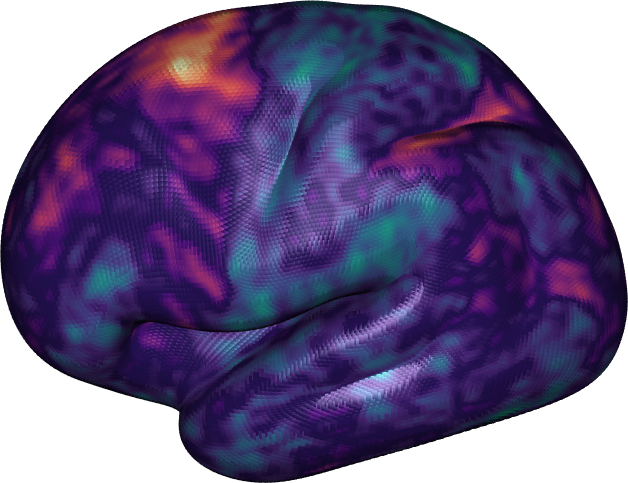}
    & 
    \includegraphics[width=0.15\textwidth]{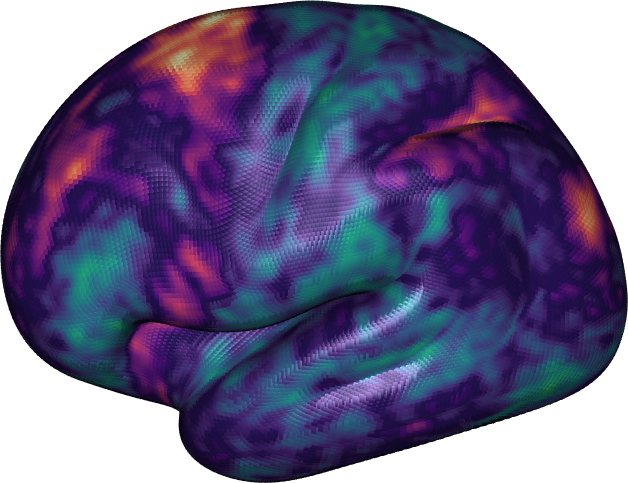}
    &  
    \includegraphics[width=0.15\textwidth]{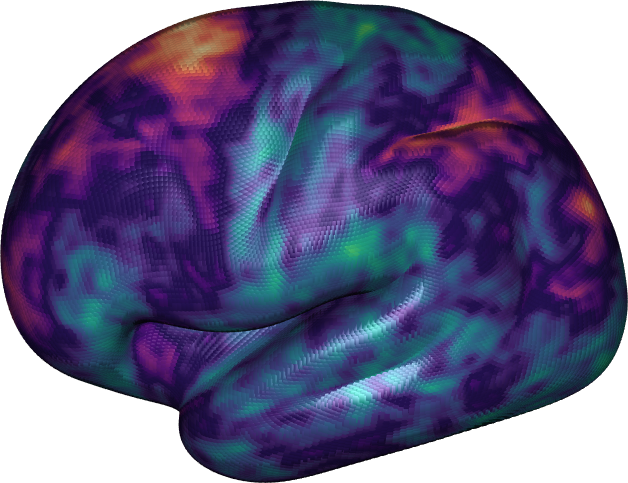}
    &
    \includegraphics[width=0.15\textwidth]{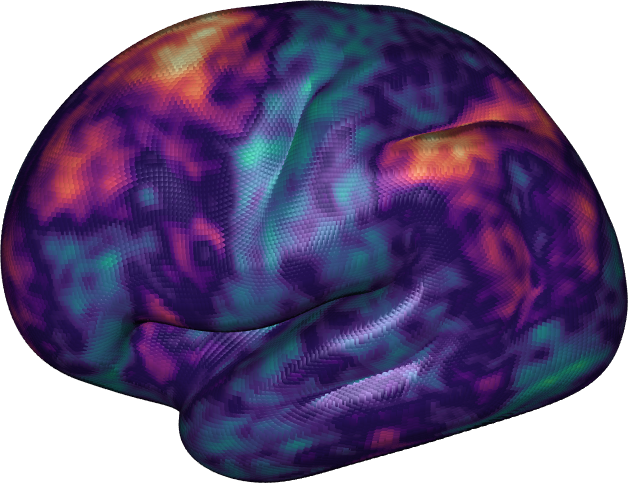}
    & 
    \includegraphics[width=0.15\textwidth]{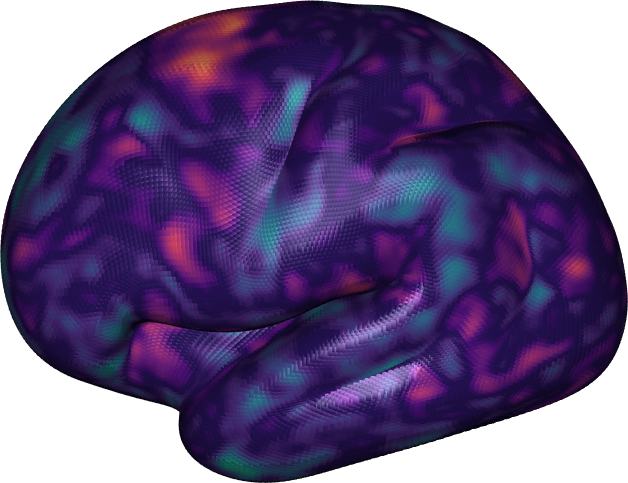}
    \\
    \multicolumn{5}{c}{\includegraphics[width=0.5\textwidth]{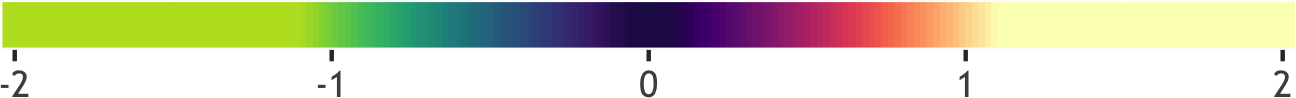}}
  \end{tabular}
  \end{minipage}
  \vspace*{1em}
  \caption{Site-specific effects for the five largest and five
    smallest sites in our ABCD study subset. We estimated the
    site-specific effects as random spatial intercepts using our
    working model framework. Site effects appear reasonably consistent
    across the 21 study locations, with of course smoother results 
    evident for the largest sites.}
  \label{fig:supp:random-site-effects}
\end{figure}

Since the ABCD data are naturally grouped by the study's
21 data collection sites, we explored the utility of including random
site effects. For these data, the
random site effects explained less than 1\% of the total variance in
over 97\% of vertices, and less than 0.1\% of the variance in nearly
half of vertices. We ultimately concluded that site-specific random
effects do not critically influence results here. Again preferring
simplicity, the results we show here and in the main text do not
include site effects as a variance component.
Fig. \ref{fig:supp:random-site-effects} displays posterior mean
estimates of site effects for the five largest and five smallest
collection sites.

\clearpage
\section{ABCD data analysis: MCMC diagnostics and sensitivity
  analyses} 
\label{sec:supp:abcd-sensitivity}

\begin{figure}
  \centering
  \includegraphics[width=0.65\textwidth]{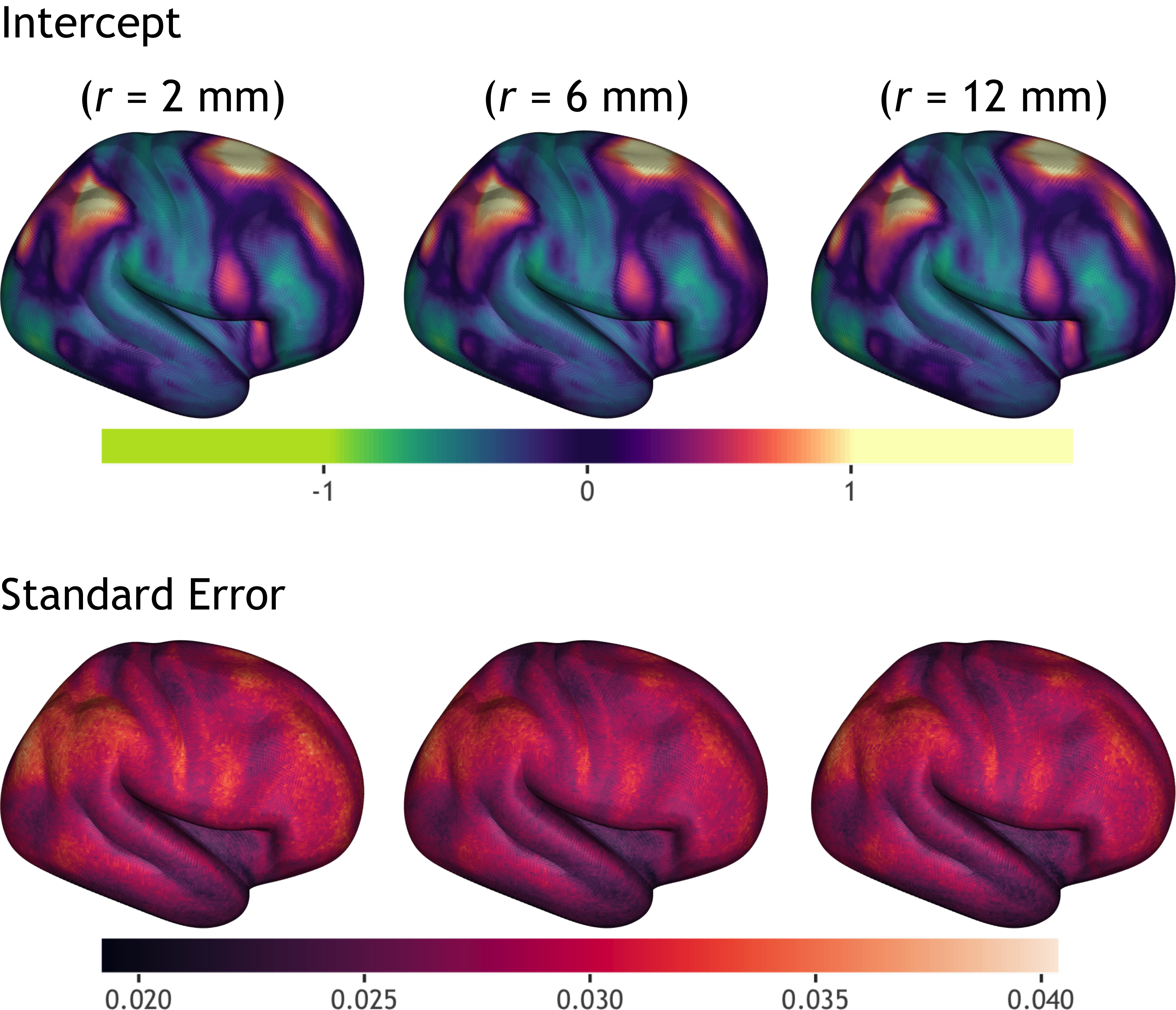}
  \caption{Sensitivity of model estimation to varying conditional
    independence neighborhood radii, $r$. Here, we explore the
    sensitivity of an intercept-only model for the ABCD
    study data at varying $r$.
    \protect{\label{fig:supp:sensitivity-r}}}
\end{figure}

In this section we describe additional sensitivity analyses and MCMC
diagnostics we have performed within the scope of the ABCD study
data.

We fit our model with Hamiltonian Monte Carlo (HMC) as noted in
the main text.
For this analysis, we ran eight chains of 7,000 iterations
each, discarding the first 5,000 as adaptation and burnin, and saving
200 samples from the final 2,000 iterations of each
chain. Convergence was assessed via univariate folded and non-folded
rank-normalized split $\hat{R}$ \cite[][]{vehtari2021rank} for each
parameter $\beta_j(\cdot)$, and by visual examination of trace plots
for subsets of these parameters. The folded split $\hat{R}$ statistic
was below the recommended threshold of 1.01 for over 99.9\% of the
$\beta_j(\cdot)$ (the worst case scenario was 1.02), indicating
reasonable convergence in the posterior spread and tail behavior for
these parameters. Similarly, the worst-case non-folded split $\hat{R}$
statistic was 1.04 across all $\beta_j(\cdot)$, indicating reasonable
convergence of the center of the posterior distribution for these
parameters. We set the neighborhood radius of the Vecchia
approximation of our prior precision to 8 mm, and the neighborhood
radius of our HMC mass matrix to 3 mm. While the algorithm can be
quite sensitive to the choice of mass matrix neighborhood radius,
values in the range 2--4 mm led to efficient and well-mixing chains
both here and in simulation. 
For readers familiar with Hamiltonian Monte Carlo:
Metropolis-Hastings rates were tuned during burnin to be
approximately 65\%; automatic tuning was achieved using the
dual-averaging method presented in \cite[][]{hoffman2014nuts}.
Additionally, we fixed the number of numerical integration steps in
our HMC to 35, which we noted produced well-mixing chains.

We noted (in the main text and in Appendix
\ref{sec:supp:posterior-computation}) that computationally we use a
specific sparse precision matrix approximation to induce conditional
independence between parameters at locations outside of an
$r$-neighborhood of each other. A natural question in this context is
how sensitive the analyses are to the choice of the neighborhood
radius $r$. We briefly explored this question by repeatedly fitting
our working model to the ABCD study data, using a spatial intercept as
the only predictor, and varying $r$ in the construction of our Vecchia
approximation to the prior. Fig. \ref{fig:supp:sensitivity-r} summarizes
the results of this sensitivity analysis. In the figure, the
posterior mean estimate (top row) is not visibly sensitive to the
choice of $r$ within a 2--12 mm range. The uncertainty in the spatial
intercept (bottom row), moreover, is at worst only modestly sensitive
to small $r$.

\begin{figure}
  \centering
  \includegraphics[width=0.65\textwidth]{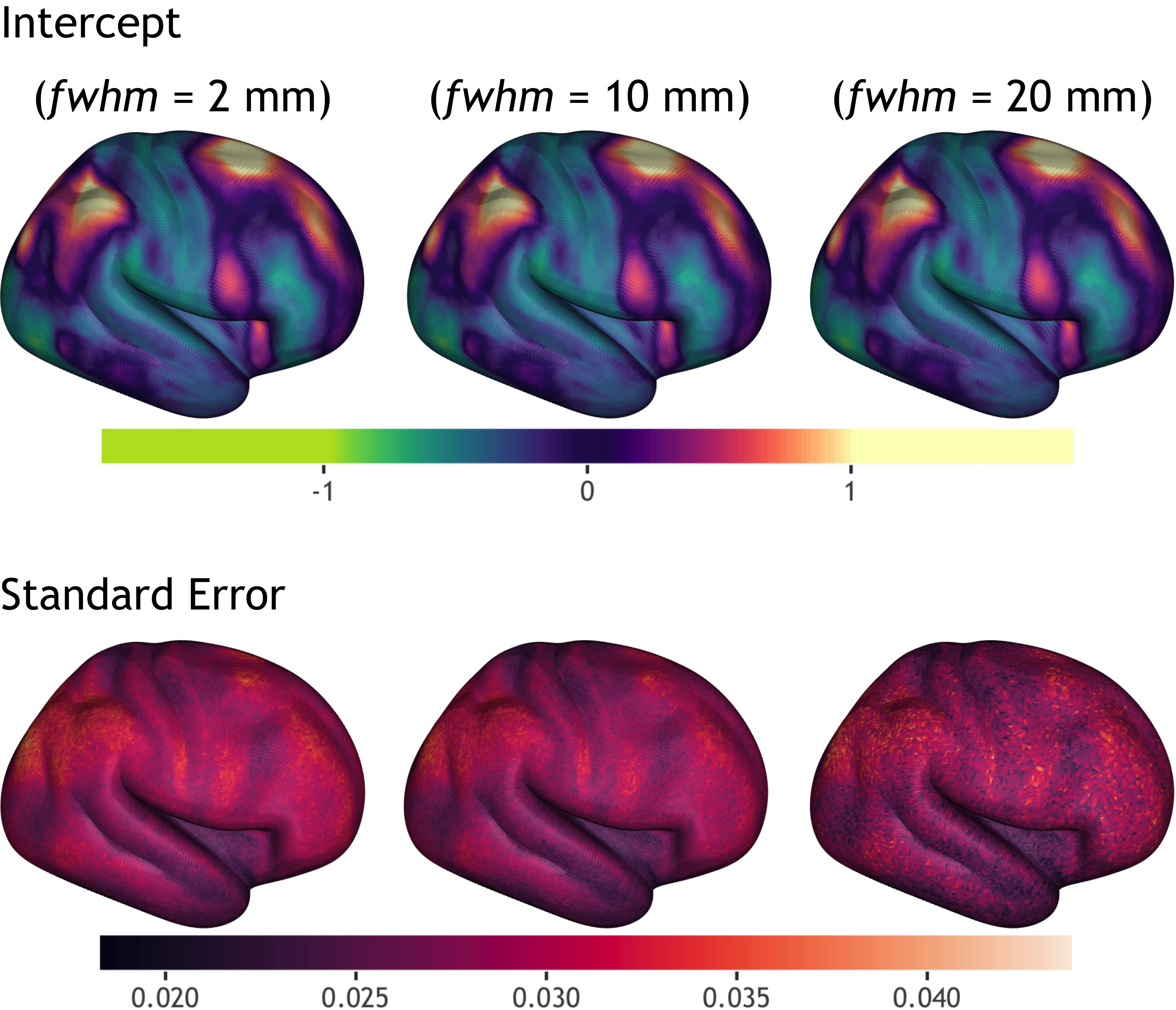}
  \caption{Sensitivity of model estimation to varying correlation
    function width. We again explored the sensitivity of an
    intercept-only model for the ABCD study data, this time for fixed
    $r$ and correlation function family. Here, we have varied the
    width of the correlation function to explore the effect on
    estimation.
    \protect{\label{fig:supp:sensitivity-fwhm}}} 
\end{figure}

A related question is how sensitive results are to the correlation
function parameters $\btheta$. As above, we repeatedly fit our working
model using a spatial intercept as the only predictor. For these
analyses, we fixed our conditional independence neighborhood radius
$r = 8$ mm and used radial basis correlation functions with exponent
parameter $1.38$ as in the main text. Here we varied only the width
of the correlation to probe for sensitivity in the
analysis. Fig. \ref{fig:supp:sensitivity-fwhm} summarizes the results of
this analysis across the varying correlation widths. As before, the
posterior mean (top row) is not visibly sensitive to the width of the
correlation within a 2--20 mm range. The uncertainty in the spatial
intercept (bottom row) is again modestly sensitive to the correlation
width. The estimate of the spatial standard error for the 20 mm
full-width-at-half-maximum correlation appears perhaps deteriorated
(bottom right panel).

\begin{figure}
  \begin{minipage}{\textwidth}
  \centering
  \begin{tabular}{ c c c }
    Best Case & Median Case & Worst Case \\
    \includegraphics[width=0.3\textwidth]{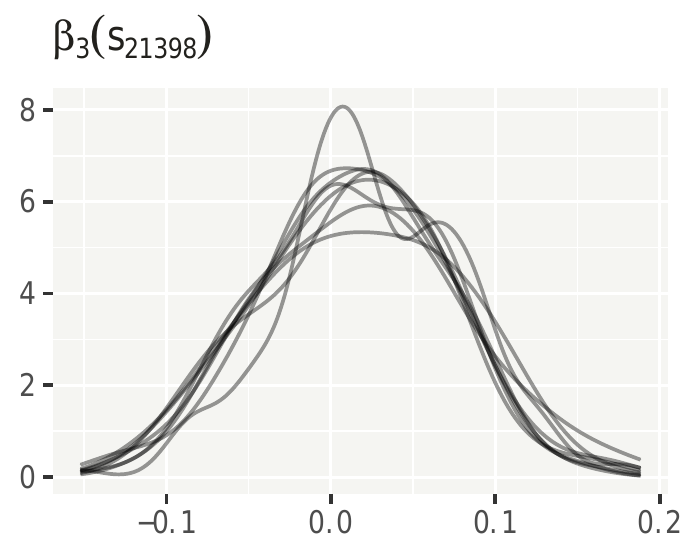} &
    \includegraphics[width=0.3\textwidth]{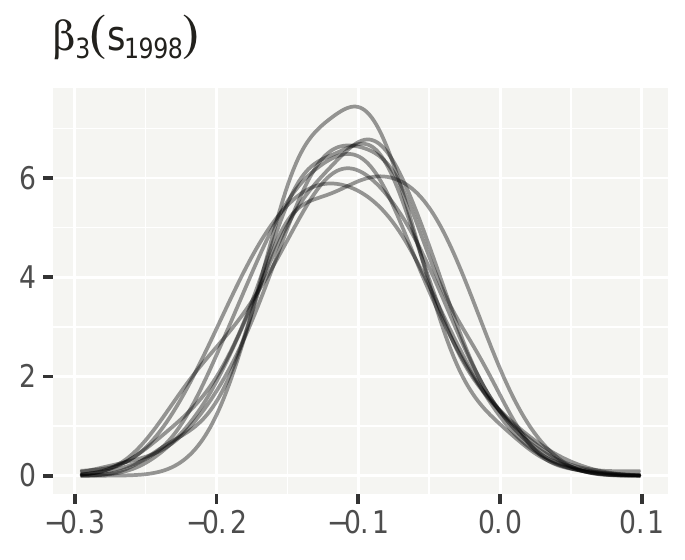} &
    \includegraphics[width=0.3\textwidth]{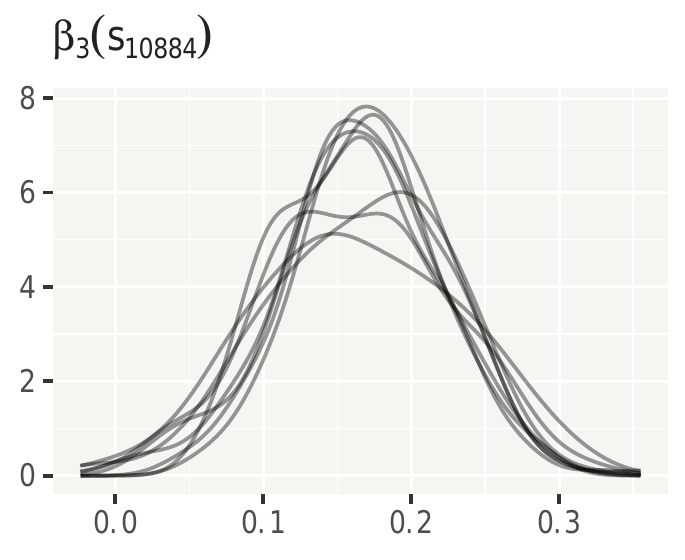}
  \end{tabular}
  \end{minipage}
  \vspace*{1em}
  \caption{Density estimates of the posterior distribution of
    $\beta_3(\cdot)$ for three different vertices and constructed from
    8 separate HMC chains. This diagnostic is for the analysis from
    the main text where $\beta_3(\cdot)$ represents the spatial
    coefficient function for the linear 2-back accuracy rate
    term. Selected vertices are rank-ordered from left to right by the
    corresponding split folded $\hat{R}$ statistic for diagnosing MCMC
    convergence. The posterior densities appear to have converged
    reasonably well across the different chains.
    \protect{\label{fig:supp:mcmc-densities}}}
\end{figure}

We also show an example MCMC convergence diagnostic for our
analysis of ABCD study data from the main text.
Fig. \ref{fig:supp:mcmc-densities} shows representative posterior density
estimates for the linear 2-back accuracy rate coefficient from three
vertices, constructed from 8 HMC chains. In the figure, we have
rank-ordered the selected vertices by the univariate split folded
$\hat{R}$ statistic \cite[][]{vehtari2021rank} for MCMC convergence
(left to right, $\hat{R} = 1$ to $\hat{R} = 1.01$). The posterior
densities show reasonable convergence across the MCMC chains.

\begin{figure}[ht]
  \begin{minipage}{\textwidth}
  \centering
  \begin{tabular}{ c c }
    \includegraphics[width=0.35\textwidth]{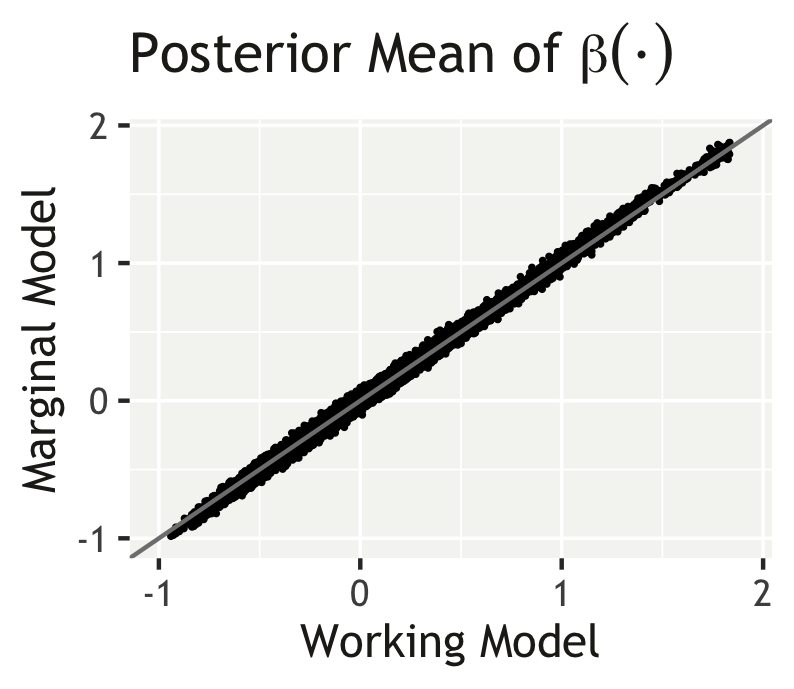} &
    \includegraphics[width=0.35\textwidth]{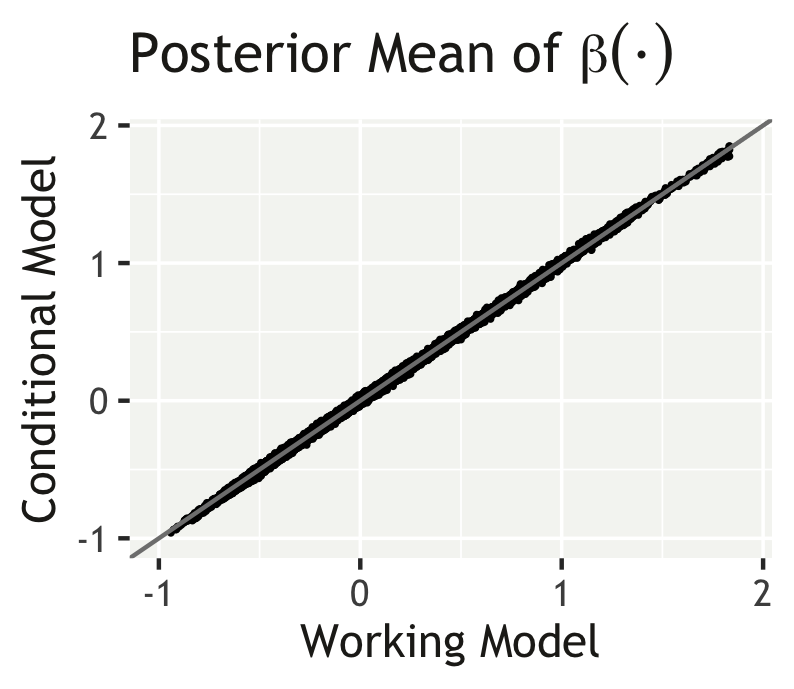}
  \end{tabular}
  \end{minipage}
  \vspace*{1em}
  \caption{Comparison of the posterior mean of $\bbeta(\cdot)$
    estimated from posterior samples drawn using each of our proposed
    conditional, marginal, and working model variants. Gray lines show
    identity relationships for reference.
    \protect{\label{fig:supp:compare-means}}
    }
\end{figure}

\begin{figure}[ht]
  \begin{minipage}{\textwidth}
  \centering
  \begin{tabular}{ c c }
    \includegraphics[width=0.4\textwidth]{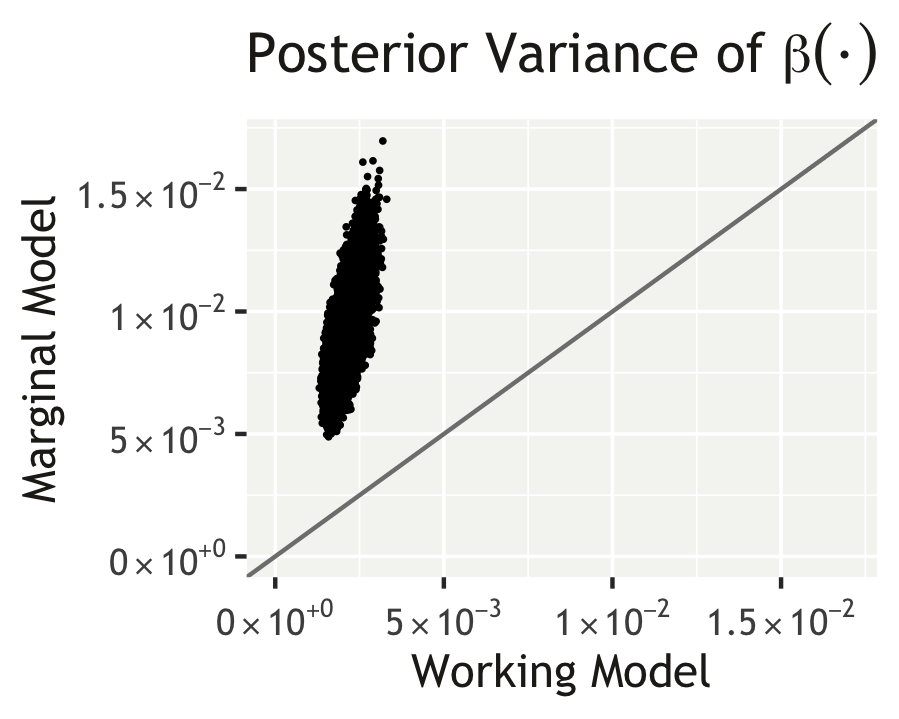} &
    \includegraphics[width=0.4\textwidth]{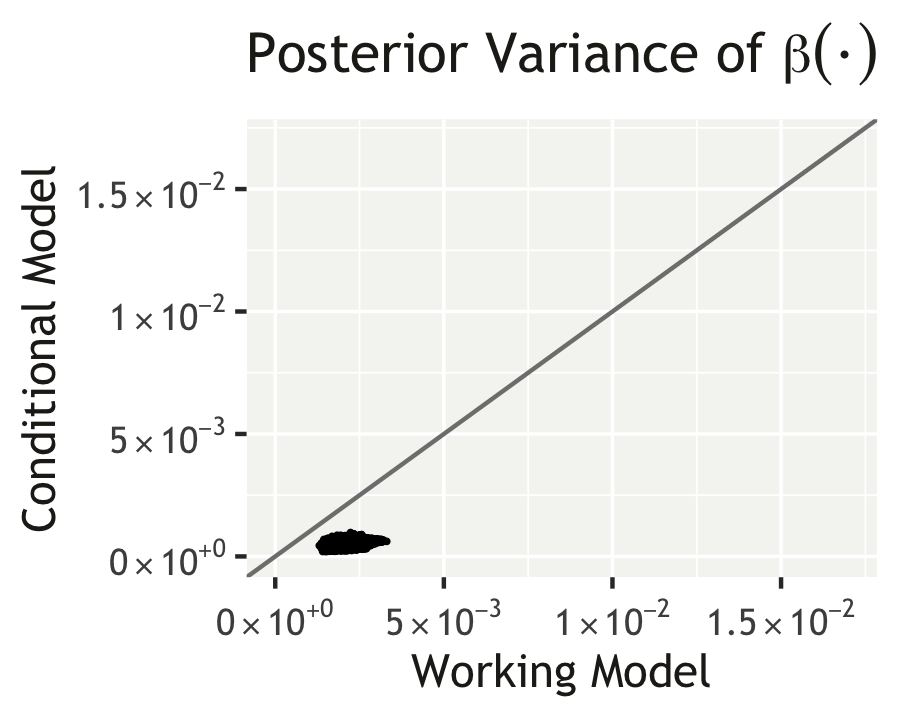}
  \end{tabular}
  \end{minipage}
  \caption{Comparison of the marginal posterior variances of each
    $\beta_j(\spc)$, $j \in 0, \ldots, 23$ and $\spc \in \mathcal{S}$,
    estimated from posterior samples drawn using each of our proposed
    conditional, marginal, and working model variants. Gray lines show
    identity relationships for reference.
    \protect{\label{fig:supp:compare-var}}
    }
\end{figure}

Finally, we give an informal comparison of realized estimation
differences arising from use of our conditional, marginal, and working
model variants in practice. For this comparison, we fit our various
models to the real ABCD study data following the protocol described in
the main text.
Figs. \ref{fig:supp:compare-means}
and \ref{fig:supp:compare-var} summarize the results of this
comparison due to both modeling and algorithmic differences between
the three methods. In particular, Fig. \ref{fig:supp:compare-means}
shows how the  posterior means of the $\beta_j(\spc)$ can be
quite similar across our proposed methods despite differences in
estimation strategy. Fig. \ref{fig:supp:compare-var} on the other hand
shows that, relative to our working model variant, marginal posterior
variances of the $\beta_j(\spc)$ were systematically larger for the
marginal model and smaller for the conditional model in these data. We
take these differences at face value here, and note only that in our
simulation studies, both the marginal and working models performed
quite well when data were generated directly from the conditional
model (see e.g. Table 1 in the main text).

\end{appendices}

\clearpage
\bibliography{references}

\label{lastpage}
\end{document}